\documentclass{article}%[12pt]{article}
\usepackage[top=2cm,bottom=2cm,left=2cm,right=2cm]{geometry}
\usepackage{amsmath, amsthm, amsfonts, bm}
\usepackage{amssymb}
\usepackage{mathtools}
\usepackage{float}
\usepackage{bbm} %for mathbbm

\usepackage{hyperref}
\hypersetup{colorlinks=true, citecolor=blue, linkcolor=blue, filecolor=blue}
\usepackage{graphicx,psfrag,epsf, float}
\usepackage{enumerate,titlesec, color}
\usepackage{natbib}
\usepackage{footnote}
\usepackage{url} % not crucial - just used below for the URL 
\usepackage{soul}
\usepackage{tikz}
\usepackage{algorithm, algorithmicx, algpseudocode}
\usepackage{rotating,tabularx,listings} % for sidewaystable
\usepackage[mathscr]{euscript}
\usepackage{enumitem} %for roman numeral enumeration
\usepackage{minitoc}
\usepackage{apalike}

\usepackage[font=small,labelfont=bf]{caption}
\usepackage{subcaption}

\usepackage{algorithm, algorithmicx, algpseudocode}
\usepackage{booktabs}
\usepackage{lineno}

\usepackage{soul}

\usepackage{setspace}
\renewcommand\footnotemark{}

\usepackage{booktabs}

\usepackage{array}
\usepackage{multirow}
\def\bSig\mathbf{\Sigma}

\newcommand{\tr}{\mbox{tr}}
\newcommand{\Var}{\text{Var}}
\newcommand{\Cov}{\text{Cov}}

\newcommand{\cdotb}{\boldsymbol{\cdot}}
\renewcommand \thepart{}

\begin{document}

\newcommand\mysubsection[1]{\noindent {\bf #1 \ }}

\newcommand\jiandel[1]{{\color{red} {\sout{#1}}}}
\newcommand\jian[1]{{\color{blue}{#1}}}
\newcommand\jianrp[2]{\jiandel{#1} \jian{#2}}
\newcommand\jiancm[1]{{\color{blue} {[[[***JK:  #1 ***]]]}}}
\newcommand{\cm}[1]{\ignorespaces}

%%%%%%%%%%%%%%%%%%%%%%
%bold small letters
\def\bfa{\mathbf a}
\def\bfb{\mathbf b}
\def\bfc{\mathbf c}
\def\bfd{\mathbf d}
\def\bfe{\mathbf e}
\def\bfp{\mathbf p}
\def\bfx{\mathbf x}
\def\bfy{\mathbf y}
\def\bfz{\mathbf z}
\def\bfs{\mathbf s}
\def\bfr{\mathbf r}
\def\bfv{\mathbf v}
\def\bfu{\mathbf u}
\def\bfw{\mathbf w}
\def\bfi{\mathbf i}
\def\bfq{\mathbf q}

\def\bw{\boldsymbol w}
%bold big letters
\def\bfA{\mathbf A}
\def\bfB{\mathbf B}
\def\bfC{\mathbf C}
\def\bfF{\mathbf F}
\def\bfS{\mathbf S}
\def\bfR{\mathbf R}
\def\bfI{\mathbf I}
\def\bfX{\mathbf X}
\def\bfY{\mathbf Y}
\def\bfZ{\mathbf Z}
\def\bfD{\mathbf D}
\def\bfL{\mathbf L}
\def\bfM{\mathbf M}
\def\bfN{\mathbf N}
\def\bfV{\mathbf V}
\def\bfU{\mathbf U}
\def\bfW{\mathbf W}
\def\bfH{\mathbf H}
\def\bfE{\mathbf E}
\def\bfeta{\mathbf \eta}
\def\bfP{\mathbf P}
\def\bfJ{\mathbf J}

%bold greek letters
\def\bfalpha{\boldsymbol \alpha}
\def\bfbeta{\boldsymbol \beta}
\def\bftheta{\boldsymbol \theta}
\def\bfepsilon{\boldsymbol \epsilon}
\def\bfdelta{\boldsymbol \delta} 
\def\bfgamma{\boldsymbol \gamma}
\def\bfGamma{\boldsymbol \Gamma}
\def\bfPhi{\boldsymbol \Phi}
\def\bfTau{\boldsymbol \Tau}
\def\bfmu{\boldsymbol\mu}
\def\bfSigma{\boldsymbol\Sigma}
\def\bfLambda{\boldsymbol\Lambda}
\def\bfpsi{\boldsymbol\psi}
\def\bfsigma{\boldsymbol\sigma}
\def\bfpi{\boldsymbol\pi}
\def\bfzeta{\boldsymbol\zeta}
%bold number
\def\bfzero{\boldsymbol 0}
\def\bfone{\boldsymbol 1}
\def\bfeta{\boldsymbol \eta}

\def\prob{{\mathrm prob}}

%%%%%%%%%%%%%%%%%%%%%%%%%%%%%%
% widetilde
\def\wtb{\widetilde b}
\def\wta{\widetilde a}
\def\wtbeta{\widetilde\beta}
\def\wtmu{\widetilde \mu}
\def\wtbfmu{\widetilde {\boldsymbol\mu}}
\def\wtbfSigma{\widetilde {\boldsymbol\Sigma}}

%%%%%%%%%%%%%%%%%%%%%%%%%%%%%%
%math calligraphic
\def\cA{\mathcal A}
\def\cB{\mathcal B}
\def\cN{\mathcal N}
\def\cT{\mathcal T}
\def\cG{\mathcal G}
\def\cP{\mathcal P}
\def\ctT{\widetilde{\mathcal T}}
\def\cbT{\overline{\mathcal T}}
\def\cR{\mathcal{R}}
\def\cD{\mathcal D}
\def\cX{\mathcal X}

\def\go{\rightarrow}
\def\up{\uparrow}
\def\down{\downarrow}

\def\cGP{\cG\cP}
\def\cF{\mathcal F}
\def\mCov{\mathrm{Cov}}
\def\mVar{\mathrm{Var}}
\def\mBer{\mathrm{Bernoulli}}
\def\mBin{\mathrm{Binomial}}
\def\mE{\mathrm{E}}
\def\mlogit{\mathrm{mlogit}}
\def\mdiag{\mathrm{diag}}
\def\mPr{\mathrm{Pr}}
\def\mDis{\mathrm{Discrete}}
\def\mDir{\mathrm{Dirichlet}}
\def\mBL{\mathrm{Beta\mbox{-}Laplace}}

\def\mP{\mathrm{P}}

%%%%%%%%%%%%%%%%%%%%%%%%%%%%%%
%mathbb
\def\mbZ{\mathbb Z}
\def\mbR{\mathbb R}
\def\mbP{\mathbb P}
\def\mbD{\mathbb D}
\def\mbI{\mathbb I}
\def\mbM{\mathbb M}

\def\md{\mathrm d}
\def\mx{\mathrm x}
\def\ms{\mathrm s}
\def\mm{\mathrm m}

%%%%%%%%%%%%%%%%%%%%%%%%%%%%%%
%mbox for distribution
\def\mLP{\mathrm{LP}}
\def\mN{\mathrm{N}}
\def\mIsing{\mathrm{Ising}}
\def\mBeta{\mathrm{Beta}}
\def\mG{\mathrm{G}}
\def\mU{\mathrm{U}}
\def\mIG{\mathrm{IG}}
\def\mv{\mathrm{v}}
\def\mw{\mathrm{w}}
\def\mg{\mathrm{g}}
\def\mh{\mathrm{h}}

\def\mGL{\mathrm{Gamma\mbox{-}Laplace}}
\def\arrowPhi{\overrightarrow \Phi}
\def\linePhi{\overline \Phi}
%%%%%%%%%%%%%%%%%%%%%%%%%%%%%%
% define operations

\def\mSign{\mathrm{Sign}}
\def\mPr{\mathrm{Pr}}
\def\mE{\mathrm{E}}
\def\iid{\scriptsize \mbox{iid}}

%%%%%%%%%%%%%%%%%%%%%%%%%%%%%%
%
\newcommand{\obeta}[1]{\widetilde\beta_{r_{#1}}}

%%additionals from FDA paper

\def\cGP{\mathcal{GP}}
\def\cK{\mathcal{K}}
\def\btheta{\boldsymbol \theta}
\def\bbeta{\boldsymbol \beta}
\def\bbetas{ \bm{\beta}_{\mathcal{S}}}
\def\bthetas{ \bm{\theta}_{\mathcal{S}}}
\def\wbthetaijk{\widetilde{\btheta_{ijk}}}
\def\wbthetaij{\widetilde{\btheta_{ij}}}
\def\wbthetai{\widetilde{\btheta_{i}}}
\def\wbbetaj{\widetilde{\bbeta_{j}}}

\def\bfOmega{\boldsymbol\Omega}
\def\bgamma{\boldsymbol \gamma}
\def\balpha{\boldsymbol \alpha}
\def\bsigma{\boldsymbol \sigma}
\def\bfTheta{{\ensuremath\boldsymbol{\Theta}}}
\def\bfrho{\boldsymbol{\rho}}

\def\bfT{\mathbf T}
\def\mt{\mathrm t}
\def\btau{\boldsymbol \tau}
\def\bs{\boldsymbol s}
\def\bS{\boldsymbol S}
\def\bvare{\boldsymbol \varepsilon}
\def\trit{\triangle t}
\def\btrit{\boldsymbol {\triangle t}}
\def\btriT{\boldsymbol {\triangle T}}
\def\diag{\mathrm{diag}}

\def\cA{\mathcal{A}}
\def\cGP{\mathcal{GP}}
\def\cK{\mathcal{K}}

\def\rT{\mathrm T}
\def\R{\mathbb R}
\def\go{\rightarrow}
\def\rank{\mathrm{rank}}
\def\vec{\mathrm{vec}}
\def\mCorr{\mathrm{Corr}}
\def\tr{\mathrm{tr}}

\newtheorem{theorem}{THEOREM}
\newtheorem{lemma}{LEMMA}
\newtheorem{remark}{REMARK}
\newtheorem{corollary}{COROLLARY}
\newtheorem{proposition}{PROPOSITION}[theorem]

	\bigskip
	\date{}

\title{Power calculation for cross-sectional stepped wedge cluster randomized trials with a time-to-event endpoint}
\author{Mary Ryan Baumann$^{1,2,*}$, 
Denise Esserman$^{3,4}$, Monica Taljaard$^{5,6}$ and 
Fan Li$^{3,4,7}$ \\
\small$^{1}$Department of Population Health Sciences, University of Wisconsin-Madison, Madison, WI U.S.A.\\
\small$^{2}$Department of Biostatistics and Medical Informatics, University of Wisconsin-Madison, Madison, WI U.S.A.\\
\small$^{3}$Department of Biostatistics, Yale School of Public Health, New Haven, CT, U.S.A.\\
\small$^{4}$Yale Center for Analytical Sciences, Yale School of Public Health, New Haven, CT, U.S.A.\\
\small$^{5}$Clinical Epidemiology Program, Ottawa Hospital Research Institute, Ottawa, Ontario, Canada.\\
\small$^{6}$School of Epidemiology and Public Health, University of Ottawa, Ottawa, Ontario, Canada.\\
\small$^{7}$Center for Methods in Implementation and Prevention Science, Yale University, New Haven, CT, U.S.A.}
\maketitle

\def\spacingset#1{\renewcommand{\baselinestretch}%
	{#1}\small\normalsize} \spacingset{1}

\begin{abstract}
Stepped wedge cluster randomized trials (SW-CRTs) are a form of randomized trial whereby clusters are progressively transitioned from control to intervention, with the timing of transition randomized for each cluster. An important task at the design stage is to ensure that the planned trial has sufficient power. While methods for determining power have been well-developed for SW-CRTs with continuous and binary outcomes, limited methods for power calculation are available for SW-CRTs with censored time-to-event outcomes. In this article, we propose a stratified marginal Cox model to analyze cross-sectional SW-CRTs and then derive an explicit expression of the robust sandwich variance to facilitate power calculations without the need for computationally intensive simulations. Power formulas based on both the Wald and robust score tests are developed, assuming constant within-period and between-period correlation parameters, and are further validated via simulation under different finite-sample scenarios. Finally, we illustrate our methods in the context of a SW-CRT testing the effect of a new electronic reminder system on time to catheter removal in hospital settings. We also offer an R Shiny application to facilitate sample size and power calculations using our proposed methods.\\
\end{abstract}

\noindent%
{\textbf{Keywords}: Generalized intracluster correlation coefficient; Kendall's tau; marginal Cox proportional hazards model; sample size estimation; small-sample corrections; survival analysis.}

\newpage
\spacingset{1.45}

\section{Introduction}
\label{sec:intro}
\begin{figure}
    \centering
    \includegraphics[width=\textwidth]{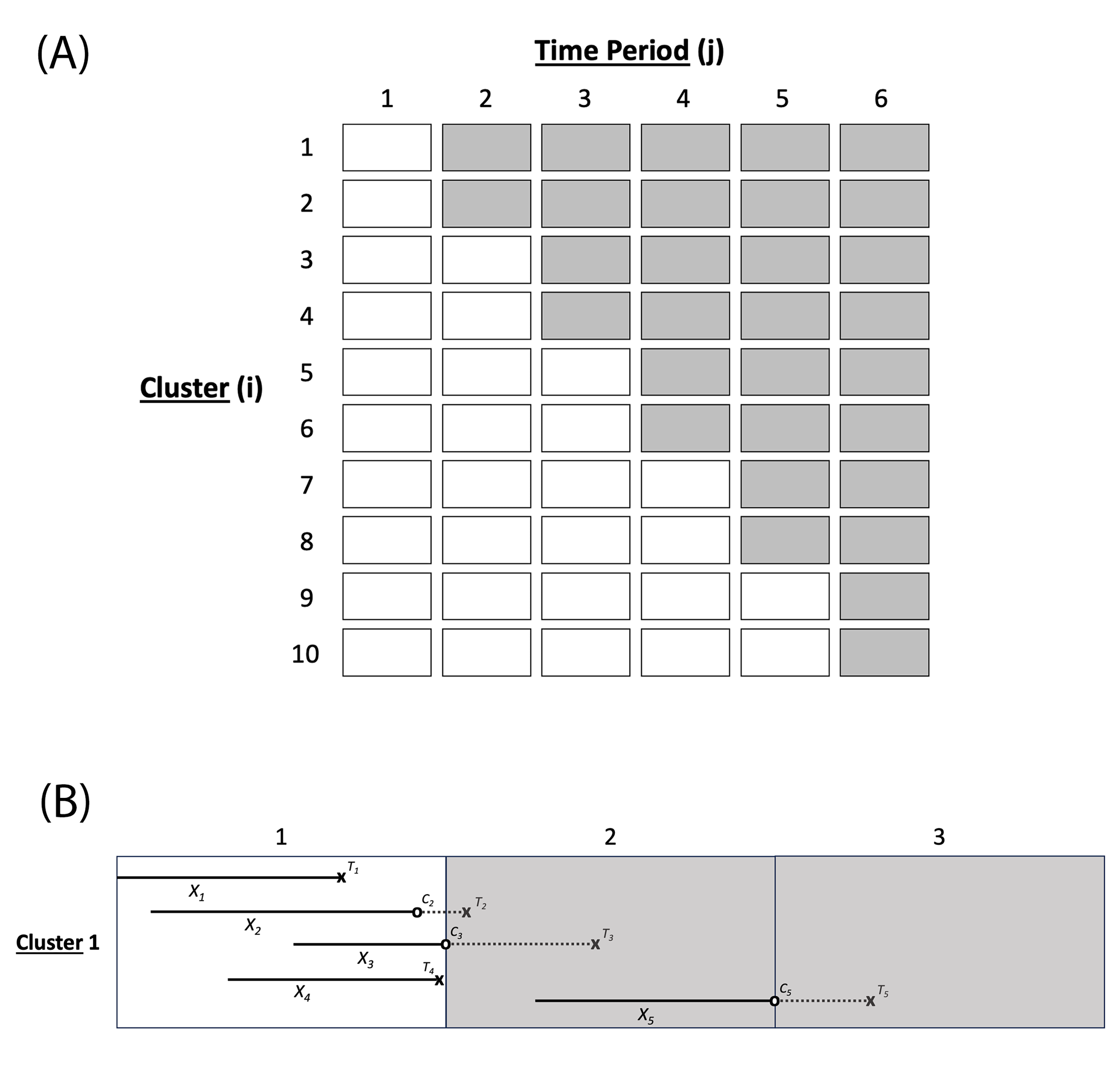}
    \caption{\label{fig:sw-crt}Panel (A): Example schematic of a stepped-wedge cluster randomized trial with $n=10$ clusters and $J=6$ time periods. White cells denote clusters under the control condition under a particular period, while gray cells denote cluster-periods under the intervention condition. Panel (B): Example schematic of observed event and censoring times for four individuals recruited during period $1$ and one individual recruited during period $2$ of a cross-sectional stepped-wedge cluster randomized trial. Cross symbols denote events and open circles denote censoring, while solid lines denote observed follow up time and dotted lines denote actual post-censoring time to event.}
\end{figure}

Cluster randomized trials (CRTs) are studies in which treatment is randomized at the cluster level. A popular class of these trials is the stepped wedge cluster randomized trial (SW-CRT), where all clusters begin on the control condition and are randomly switched to the treatment condition at staggered, pre-planned time points, until treatment is implemented in all clusters before the end of the study. An example SW-CRT with $10$ clusters observed over six time periods is illustrated in the top panel of Figure \ref{fig:sw-crt}. SW-CRTs can be classified into three types, depending on whether individuals within each cluster only contribute data to a single time period (cross-sectional), are followed longitudinally over multiple periods (closed-cohort), or may flexibly join or leave the study across time (open-cohort) \citep{copas_designing_2015}. 

To date, power calculation methods for SW-CRTs have primarily focused on continuous and binary outcomes; see, for example, \citet{hussey_design_2007}, \citet{li_sample_2018}, \citet{kasza2019impact}, \citet{wang_sample_2021} for methods with continuous outcomes, and \citet{harrison_power_2021}, \citet{davis-plourde_sample_2021} for methods with binary outcomes. A review of sample size formulas and software can be found in \citet{li_mixed-effects_2021} and \citet{ouyang2022sample}. However, there is a notable gap in the methods literature regarding SW-CRTs with time-to-event endpoints even though several published studies analyzing these endpoints have already been reported. For example, \citet{nevins_scoping_2023} reviewed $160$ SW-CRTs between 2016 and 2022 and identified at least nine health science cross-sectional SW-CRTs with time-to-event endpoints. While several sample size methods have been described for parallel-arm CRTs with a time-to-event outcome \citep{zhong_sample_2015,blaha_design_2022}, few methods are currently available to inform the planning of similar SW-CRTs. As a few exceptions, in an open-cohort SW-CRT, \citet{moulton_statistical_2007} used a log-rank type analysis to compare within-period incidence between arms where contributions were updated at the event level; power calculations were performed under a parallel-arm CRT framework with a simulation-based design effect to account for staggered randomization. In a closed-cohort SW-CRT, \citet{dombrowski_cluster_2018} investigated differences in time to viral suppression among HIV patients using a Cox proportional hazards model and a robust sandwich variance clustered at the provider level; power calculations were performed using the SW-CRT formula for binary outcomes. \citet{zhan_analysis_2016} assessed the use of discrete-time and continuous-time Cox proportional hazards models for the analysis of terminal endpoints with interval censoring in SW-CRTs via a simulation study, but noted that power formulas under their models were an area of future work. \citet{oyamada_comparison_2022} assessed the use of several recurrent event models and cluster stratification in open-cohort SW-CRTs, but did not address sample size considerations. Different from these previous studies, we focus on the planning of cross-sectional SW-CRTs based on a nested exchangeable type correlation structure with constant within-period and between-period correlations. We assume that the maximum follow-up time is pre-determined for individuals recruited within each period (Figure \ref{fig:sw-crt}B) and contribute novel non-simulation-based sample size formulas for time-to-event outcomes. Variations of designs concerning differing participant recruitment timelines and administrative censoring timing are presented in Web Appendix A.

This work is motivated in part by the CATH TAG trial \citep{mitchell_reducing_2019}, a study uncovered in the course of the \citet{nevins_scoping_2023} review. The CATH TAG trial aimed to evaluate whether attaching CATH TAG reminder devices to catheter bags reduced hospitalized patients' time on a catheter. Despite the primary analysis using a time-to-event outcome, power calculations were performed using the existing SW-CRT formula for binary outcomes (possibly due to limited methods available), resulting in the randomization of $10$ hospital wards to $5$ treatment sequences. To formally investigate more accurate sample size procedures for planning cross-sectional SW-CRTs with a right-censored time-to-event outcome, we first propose a period-stratified marginal Cox model, which is the analogue of marginal models developed to analyze non-censored outcomes in SW-CRTs \citep{li_sample_2018}. We consider both the Wald and robust score methods for testing the treatment effect, and leverage small-sample adjustments to combat inferential challenges that often arise with a limited number of clusters. For both tests, we then develop closed-form sample size formulas for study planning. A surprising finding of our work is that the associated sample size formulas share the same form as those developed for marginal analysis of continuous outcomes in cross-sectional SW-CRTs, with the exception that within-period and between-period correlations are now reformulated based on the martingale scores instead of the original outcomes. This insight provides a unification of the variance expression under marginal analyses of cross-sectional SW-CRTs. Simulations are carried out to validate our proposed methods in finite samples and the context of CATH-TAG is used to illustrate our methods. We also provide a free R Shiny application to implement the proposed sample size methodology, which can be found in the Supplementary Materials and at \url{https://mary-ryan.shinyapps.io/survival-SWD-app}.

\section{Period-stratified Cox proportional hazards model}\label{sec:background}
\subsection{Statistical model}\label{ss:model}
We consider a SW-CRT in which $n$ clusters are randomly assigned to $(J-1)$ treatment sequences to be stepped on to intervention across $J$ time periods; we assume each cluster includes $m$ individuals per period. Note that when the number of clusters is greater than the number of treatment sequences, $n > (J-1)$, at least one treatment sequence will be assigned multiple clusters. We assume the individual enrollment time is random within each period, and suppose we plan to follow individuals within time interval $(0,C^*]$ since enrollment. Here, $C^*$ is the maximum follow-up time (see Web Appendix A for design schematics with different specifications of $C^*$). We let $T_{ijk}$ and $C_{ijk}$ ($C_{ijk}\leq C^*$) denote the event and censoring times since enrollment, respectively, for the $k$th individual in cluster $i$ at period $j$, though we observe only $X_{ijk}=\min(T_{ijk}, C_{ijk})$. Define the observed event indicator $\Delta_{ijk} = \mathbb{I}(T_{ijk} \le C_{ijk})$, and at-risk indicators $Y_{ijk}(t) = \mathbb{I}(T_{ijk}\ge t)$, $Y^\dagger_{ijk}(t) = \mathbb{I}(C_{ijk}\ge t)$, and $\overline{Y}_{ijk}(t) = Y_{ijk}(t)Y^\dagger_{ijk}(t)$, where $\mathbb{I}(\cdotb)$ is an indicator function. We write $Z_{ij}$ as the treatment indicator for cluster $i$ at period $j$, where $Z_{ij} = 1$ indicates treatment and $Z_{ij} = 0$ indicates control. We also assume that $(C_{i11}, \dots, C_{iJm})' \perp\!\!\!\perp (T_{i11}, \dots, T_{iJm})' | Z_{ij}$.

We focus on the population-averaged hazard ratio as an effect measure, similar to the population-averaged effect that has been studied in SW-CRTs with non-censored outcomes \citep{li_sample_2018,li_marginal_2022}. To account for confounding by time, rather than including time periods as indicator variables and costing additional degrees of freedom, we propose a period-stratified marginal Cox model with separate baseline hazard functions for each time period:
\begin{equation}\label{eq:hazard}
    \lambda_{ijk}(t|Z_{ij}) = \lambda_{0j}(t)\exp(\beta Z_{ij})
\end{equation}
where $\beta$ is the treatment effect measured as a log hazard ratio and $\lambda_{0j}(t)$ is the period-specific baseline hazard. Stratifying the model by period allows us to adjust for underlying changes in baseline hazard functions over calendar periods (i.e., secular trend) in the marginal model without needing to specifically estimate each period effect. We pursue the independence estimating equations as the standard implementation in, for example, the \texttt{survival} R package \citep{survival-package}. Under working independence, the partial likelihood estimator in a stratified marginal Cox model solves: 
\begin{equation}\label{eq:score}
    U( \beta) 
    = \sum_{i=1}^n U_{i++}(\beta) 
    = \sum_{i=1}^n \sum_{j=1}^J \sum_{k=1}^m \int_0^{C^*} \overline{Y}_{ijk}(t) \left\{Z_{ij} - \frac{S^{(1)}_j(t;\beta)}{S^{(0)}_j(t;\beta)}\right\} dN_{ijk}(t)=0,
\end{equation}
where $N_{ijk}(t) = \mathbb{I}\left(T_{ijk} \le t\right)$ is the counting process for the survival time and $\overline{Y}_{ijk}(t)$ is the observed at-risk indicator, while $S_j^{(0)}(t;\beta) = n^{-1}\sum_{i=1}^n\sum_{k=1}^m\overline{Y}_{ijk}(t)\exp\left(\beta Z_{ij}\right)$ is akin to the cluster-averaged survival function among those at risk in period $j$, and $S_j^{(1)}(t;\beta) = n^{-1}\sum_{i=1}^n\sum_{k=1}^m\overline{Y}_{ijk}(t)Z_{ij}\exp\left(\beta Z_{ij}\right)$ is its derivative. Inference on $\widehat{\beta}$ proceeds with a robust sandwich variance estimator $\widehat{\Var}(\widehat{\beta}) = A^{-1}(\widehat{\beta})B(\widehat{\beta})A^{-1}(\widehat{\beta})$, where $A^{-1}(\beta) = E\left\{-{\partial U_{i++}(\beta)}/{\partial \beta}\right\}^{-1}$ and $B(\beta) = E\left\{U_{i++}(\beta)^2\right\}$ \citep{lin_cox_1994}.

\subsection{Generic power and sample size requirements}\label{ss:power}
Generally, the power to detect an effect size $\beta_1 \ne \beta_0$, given the number of clusters $n$, cluster-period size $m$, number of periods $J$, and $\beta_0=0$, using a two-sided $\alpha$-level Wald test is:
\begin{equation}\label{eq:power}
    \text{power} \approx \Phi_t\left({\left|\beta_1\right|}/{\sqrt{\Var(\widehat{\beta})}}-t_{\alpha/2, \text{DoF}}\right),
\end{equation}
where $t_{\alpha/2, \text{DoF}}$ is the upper $\alpha/2$th quantile of a central $t$-distribution with $\text{DoF}$ degrees of freedom, and $\Phi_t(\cdotb)$ is the cumulative $t$-distribution function. Following \citet{ford_maintaining_2020} and \citet{ouyang2023maintaining}, we consider the $t$-distribution with $\text{DoF} = n-2$ as a finite-sample correction. This empirical choice of degrees of freedom correction has proven effective in prior simulation studies for SW-CRTs with non-censored outcomes. To provide additional finite-sample improvement, we also examine several bias-corrected sandwich variance estimators in Section \ref{sec:sim}.

An alternative testing paradigm proceeds with the robust score statistic. Following \citet{self_powersample_1988}, the power for a two-sided $\alpha$-level robust score test is:
\begin{equation}\label{eq:power-score}
    \text{power} \approx \Phi\left({\left|E_{H_1}\left\{U_{i++}(\beta_0)\right\}\right|}/{\sqrt{\sigma^2_1}} - z_{\alpha/2}\right),
\end{equation}
where $\beta_0$ is the value of $\beta$ under the null hypothesis, $z_{\alpha/2}$ is the upper $\alpha/2$th quantile of the standard normal distribution, $\Phi(\cdotb)$ is the cumulative standard normal distribution function, and $E_{H_1}\left\{U_{i++}(\beta_0)\right\}$ is the expectation of the null score $U_{i++}(\beta_0)$ with data generated under $H_1$. Similarly, $\sigma^2_1 = \Var_{H_1}\left\{U_{i++}(\beta_0)\right\}$ is the variance of the null score with data generated under $H_1$. We will refer to equation \eqref{eq:power-score} as the S\&M method, which assumes $\sigma^2_1 = \Var_{H_1}\left\{U_{i++}(\beta_0)\right\} \approx \Var_{H_0}\left\{U_{i++}(\beta_0)\right\} = \sigma^2_0$ under contiguous alternatives \citep{self_powersample_1988}. For larger effect sizes, \citet{tang_improved_2021} suggested a correction method to more accurately estimate the power of a robust score test:
\begin{equation}\label{eq:power-tang}
    \text{power} \approx \Phi\left({\left|E_{H_1}\left\{U_{i++}(\beta_0)\right\}\right|}/{\sqrt{\sigma^2_1}} - z_{\alpha/2}\times \sqrt{{\sigma^2_0}/{\sigma^2_1}}\right).
\end{equation}
We will refer to equation \eqref{eq:power-tang} as the Tang method. Power formulas \eqref{eq:power}--\eqref{eq:power-tang} represent different paradigms (Wald versus robust score testing) within which we will propose analytic power procedures. An essential task is to characterize $\Var(\widehat{\beta})$ at the design stage to estimate power for the Wald test, and characterize $\sigma^2_0$, $\sigma^2_1$ to estimate power for the robust score test. Additional details about each testing procedure can be found in Web Appendix B.

\section{Power calculation for stepped wedge designs with a time-to-event endpoint}\label{sec:var}
\subsection{The Wald testing paradigm}\label{ss:beta-var}
Assuming model \eqref{eq:hazard} is correct and an absence of within-cluster dependence between survival times, $\widehat{\beta}$ is approximately normal with mean $\beta$ and variance given by \citep{lin_cox_1994}

\begin{equation}\label{eq:A}
    A^{-1}({\beta}) 
    = E\left\{\frac{-{\partial U_{i++}(\beta)}}{{\partial \beta}}\right\}^{-1} 
    =\left[\sum_{j=1}^JE_{Z_{ij}}\left\{\sum_{k=1}^m\nu(Z_{ij})\right\}\right]^{-1},
\end{equation}
where $\nu(Z_{ij})=\int_0^{C^*} \mathscr{G}(t) \mu_j(t)\left\{1 - \mu_j(t)\right\} f(t|Z_{ij})dt$, $E_{Z_{ij}}\{\cdotb\}$ is the expectation with respect to treatment assignment during study period $j$, $\mu_j(t) = {s^{(1)}_j(t;{\beta})}/{s^{(0)}_j(t;{\beta})}$,  $s_j^{(0)}(s;\beta)=E\left\{\sum_{k=1}^m \overline{Y}_{ijk}(s)\exp (\beta Z_{ij})\right\}$ and $s_j^{(1)}(s;\beta)=E\left\{\sum_{k=1}^m \overline{Y}_{ijk}(s)Z_{ij}\exp (\beta Z_{ij})\right\}$ are the almost sure limits of $S_j^{(0)}(s;\beta)$ and $S_j^{(1)}(s;\beta)$, $\mathscr{G}(t)$ is the censoring survival function for $C_{ijk}$, and $f(t|Z_{ij})$ is the conditional density of event time $T_{ijk}$ given the treatment status. The derivation of \eqref{eq:A} is found in Web Appendix C. To account for the within-cluster correlation and misspecification of the working independence assumption, the sandwich variance expression is required to reflect actual uncertainty of $\widehat{\beta}$, and is given by $A^{-1}(\beta)B(\beta)A^{-1}(\beta)$, where
\begin{align}
\begin{split}\label{eq:B}
    B(\beta) &= n^{-1}\sum_{i=1}^n\left[\sum_{j=1}^J\sum_{k=1}^m \Var\left\{U_{ijk}(\beta)\right\} + \sum_{j=1}^J\mathop{\sum_{k=1}^m\sum_{d=1}^m}_{k\ne d} \Cov\left\{U_{ijk}(\beta),U_{ijd}(\beta)\right\}\right.\\
    &\hspace{2.3cm}\left.+ \mathop{\sum_{j=1}^J\sum_{l=1}^J}_{j\ne l}\sum_{k=1}^m\sum_{d=1}^m \Cov\left\{U_{ijk}(\beta),U_{ild}(\beta)\right\}\right].
\end{split}
\end{align}
The first term in equation \eqref{eq:B} corresponds to the total marginal variance of the score for each individual, while the remaining two terms correspond to the total within-cluster-period covariance and the total within-cluster, between-period covariance, respectively. Power calculation for the Wald $t$-test requires the expression of $\text{Var}(\widehat{\beta})=A^{-1}(\beta)B(\beta)A^{-1}(\beta)$, while power calculation for the robust score test requires the expression of $B(\beta)=\Var \left\{U_{i++}(\beta)\right\}$, which we outline below; for full derivation details, see Web Appendix C.

In Web Appendix C, we provide an intermediate result on the variance and covariance expressions in equation \eqref{eq:B}. Through this intermediate result, we rewrite $B(\beta)$ as:
\begin{equation*}
    m\sum_{j=1}^J E_{Z_{ij}}\left\{q_0(Z_{ij})\right\} + m(m-1)\sum_{j=1}^J E_{Z_{ij}}\left\{\sum_{r=1}^4q_r(Z_{ij},Z_{ij})\right\} + m^2\mathop{\sum_{j=1}^J\sum_{l=1}^J}_{j \ne l} E_{Z_{ij}, Z_{il}}\left\{\sum_{r=1}^4q_r(Z_{ij},Z_{il})\right\},
\end{equation*}
where $E_{Z_{ij}}\{\cdotb\}$ is the expectation with respect to the marginal distribution of the treatment variable at period $j$, and $E_{Z_{ij},Z_{il}}\{\cdotb\}$ is the expectation with respect to joint distribution of the treatment variables at study periods $j$ and $l$. Furthermore, the function $q_0(Z_{ij}) = \int_0^{C^*}\mathscr{G}(s)\left\{Z_{ij} - \mu_j(s)\right\}^2 f(s|Z_{ij})ds$ is a single integral; each function $q_r(Z_{ij},Z_{il})$ is a double integral over $(0,C^*]^2$ with integrand defined as a function of the bivariate censoring distribution for $(C_{ijk},C_{ild})$, treatment assignments $(Z_{ijk},Z_{ild})$, limit functions $\mu_j(s)$ and $\mu_l(s)$, and the bivariate survival function for $(T_{ijk},T_{ild})$ given $(Z_{ijk},Z_{ild})$. Web Appendix C provides their explicit expressions.

Let $P(Z_{ij}=a)$ be the probability that cluster $i$ is in the treatment condition $a\in\{0,1\}$ during period $j$, and $P(Z_{ij}=a, Z_{il}=a')$ be the joint probability of the cluster in periods $j$ and $l$. We can explicitly write $E_{Z_{ij}}\{q_0(z_{ij})\}  = P(Z_{ij}=1) q_0(Z_{ij}=1)+ P(Z_{ij}=0)q_0(Z_{ij}=0)$. We then define $\Upsilon_0(j) = \sum_{a=0}^1P(Z_{ij}=a)q_0(Z_{ij}=a)$ and $\Upsilon_1(j,l) = \sum_{a=0}^1\sum_{a'=0}^1P(Z_{ij}=a, Z_{il}=a')\sum_{r=1}^4 q_r(Z_{ij}=a, Z_{il}=a')$. 
Thus we can succinctly write
\begin{equation}\label{eq:B_reexp2}
    B({\beta}) = m\sum_{j=1}^J\Upsilon_0(j) + m(m-1)\sum_{j=1}^J \Upsilon_1(j,j) + m^2\mathop{\sum_{j=1}^J \sum_{l=1}^J}_{j \ne l}\Upsilon_1(j,l),
\end{equation}
where $\sum_{j=1}^J\Upsilon_0(j)$ corresponds to the marginal variance of the individual score, $\sum_{j=1}^J \Upsilon_1(j,j)$ represents the within-period covariance, and $\mathop{\sum_{j=1}^J \sum_{l=1,j \ne l}^J}\Upsilon_1(j,l)$ represents the between-period covariance of two individual scores. Moreover, the model-based variance \eqref{eq:A} can be represented as 
$A^{-1}({\beta})=\left\{m\sum_{j=1}^J \sum_{a=0}^1P(Z_{ij}=a)\nu(Z_{ij}=a)\right\}^{-1}$. In Web Appendix C, we also show that when model \eqref{eq:hazard} is correctly specified, $\Upsilon_0(j)=\sum_{a=0}^1P(Z_{ij}=a)\nu(Z_{ij}=a)$. Based on these intermediate results, Theorem \ref{theorem:var_b} below provides a closed-form variance expression for $\widehat{\beta}$ estimated from the period-stratified marginal Cox regression.

%% theorem 1: explicit expression of var(beta) %%
\begin{theorem}\label{theorem:var_b}
Assuming known survival and censoring distributions and model \eqref{eq:hazard}, the variance of the treatment effect estimator based on a period-stratified marginal Cox model is
\begin{equation}\label{eq:var-robust}
    \Var(\widehat{\beta})=\left\{nm\sum_{j=1}^J \Upsilon_0(j)\right\}^{-1}\times \left\{1 + (m-1)\rho_w + m(J-1)\rho_b\right\},
\end{equation}
where $\rho_w = {\sum_{j=1}^J \Upsilon_1(j,j)}/{\sum_{j=1}^J \Upsilon_0(j)}$ and  $\rho_b = {\mathop{\sum_{j=1}^J \sum_{l=1}^J}_{j\ne l} \Upsilon_1(j,l)}/\{(J-1)\sum_{j=1}^J \Upsilon_0(j)\}$.
\end{theorem}

Several remarks are in order based on Theorem \ref{theorem:var_b}. First, although our primary context is cross-sectional SW-CRT, variance expression \eqref{eq:var-robust} is derived without restrictions on the design element $Z_{ij}$, and hence is general enough to accommodate all types of cross-sectional longitudinal CRTs, including the parallel-arm design and cluster randomized crossover design. The only difference in applying \eqref{eq:var-robust} is that the allocation probabilities $P(Z_{ij}=a)$ and $P(Z_{ij}=a,Z_{il}=a')$ will need to be modified according to the randomization schedule. 

Second, the two key parameters in variance expression \eqref{eq:var-robust} have an intuitive interpretation as the intracluster correlation coefficients (ICCs). Specifically, $\rho_w$ is the ratio of the average within-period covariance of the score over the average marginal variance of the score; we refer to $\rho_w$ as the \emph{within-period generalized ICC}. Similarly, we refer to $\rho_b$ as the \emph{between-period generalized ICC} (abbreviated as g-ICC hereafter). These two quantities are extensions of their counterparts in cross-sectional SW-CRTs with non-censored outcomes \citep{ouyang2023estimating}, and arise due to the specific features of censored survival outcomes.

When there is no covariation within or between periods (i.e., $\rho_w = \rho_b = 0$), such that there is an absence of any clustering, the data structure is akin to a period-stratified or period-blocked individually randomized trial. The variance of the treatment effect estimator will then simplify to $\Var(\widehat{\beta})=\left\{nm\sum_{j=1}^J \Upsilon_0(j)\right\}^{-1}.$ Thus, variance \eqref{eq:var-robust} consists of the variance without clustering, multiplied by a familiar design effect characterizing the nontrivial residual clustering: $\left\{1 + (m-1)\rho_w + m(J-1)\rho_b\right\}$. Furthermore, we explain in Web Appendix C that variance \eqref{eq:var-robust} also has a similar form to the treatment effect variance for marginal analyses of SW-CRTs with continuous outcomes \citep{wang_sample_2021,tian_information_2024}.

\subsection{The robust score testing paradigm}\label{ss:score-var}
We noted in Section \ref{ss:beta-var} that $B(\beta) = n^{-1} \sum_{i=1}^n \Var\left\{U_{i++}(\beta)\right\}$. The variance for the robust score statistic will follow a similar form. The major difference is that while $B(\beta)$ can be calculated at the design stage using the anticipated effect size $\beta = \beta_1$, $\sigma^2_1 = \Var_{H_1}\left\{U_{i++}(\beta_0)\right\}$ must be calculated such that the portions of the score concerning the observed data are generated under $\beta=\beta_1$ while the model-based portions are evaluated at $\beta=\beta_0$. For $\sigma^2_0 = \Var_{H_0}\left\{U_{i++}(\beta_0)\right\}$, all aspects of the calculation assume $\beta=\beta_0$. Thus, $\Upsilon^{H_c}_0(j)$ and $\Upsilon^{H_c}_1(j,l)$ are defined similarly as in Section \ref{ss:beta-var}, except that we introduce the superscript notation to denote that the data portions of the score are evaluated at $H_c: \beta=\beta_c$, $c\in\{0,1\}$. We summarize these modifications in Proposition \ref{coro:var_score}.

\vspace{-0.1in}
\begin{proposition}\label{coro:var_score}
Let $\Var_{H_c}\left\{U( \beta_0)\right\}$ be the variance of the score based on the period-stratified marginal Cox model, evaluated under $H_0: \beta=\beta_0$ and data generated under $H_c: \beta=\beta_c$, $c\in\{0,1\}$. Then we have the following 
\begin{equation}\label{eq:var-score}
Var_{H_c}\left\{U( \beta_0)\right\}=\left\{\frac{1}{nm}\sum_{j=1}^J\Upsilon^{H_c}_0(j)\right\}\times \{1 + (m-1)\kappa^{H_c}_w + m(J-1)\kappa^{H_c}_b\},
\end{equation}
where $\kappa^{H_c}_w = {\sum_{j=1}^J \Upsilon^{H_c}_1(j,j)}/{\sum_{j=1}^J \Upsilon^{H_c}_0(j)}$ and $\kappa^{H_c}_b = {\mathop{\sum_{j=1}^J \sum_{l=1}^J}_{j\ne l} \Upsilon^{H_c}_1(j,l)}/\{(J-1)\sum_{j=1}^J \Upsilon^{H_c}_0(j)\}$.
\end{proposition}

In \eqref{eq:var-score}, $\kappa^{H_c}_w$ is the ratio of the average within-period covariance of the score over the average marginal variance of the score, evaluated under $H_c$, $c\in\{0,1\}$, which we refer to as the within-period g-ICC at $H_c$. Similarly, we refer to $\kappa^{H_c}_b$ as the between-period g-ICC evaluated at $H_c$. We note that the score covariance components may be evaluated under different hypotheses, hence $\kappa^{H_0}_w$ and $\kappa^{H_0}_b$ are not necessarily equal to $\kappa^{H_1}_w$ and $\kappa^{H_1}_b$, respectively.

\subsection{Power calculation in practice}\label{sec:practical}
To use variance equations \eqref{eq:var-robust} and \eqref{eq:var-score} for power calculations, there are two options. First, one can directly assume specific values for the within-period and between-period g-ICCs and then use equation \eqref{eq:var-robust}. Operationally, this is no different than specifying the within-period and between-period ICC values for calculating power in SW-CRTs with non-censored outcomes, and therefore may be the preferred approach  for its simplicity. While convenient, a possible limitation of this approach is that it may be unclear how specific g-ICC values map to explicit features of the underlying within-cluster censoring and event outcome distributions. Therefore, a second approach is to consider a generative model for power calculation. In this approach, one directly specifies the survival distributions for the censoring and event times to calculate $\rho_w$ and $\rho_b$ for equation \eqref{eq:var-robust} in the Wald testing paradigm, or to directly calculate $\kappa^{H_c}_w$ and $\kappa^{H_c}_b$ via equation \eqref{eq:var-score} for the robust score paradigm. As a concrete example, we provide in Web Appendix D a nested Archimedean copula model \citep{mcneil_sampling_2008} with Gumbel transformations as a generative model for power calculation, and parameterize the dependency structure based on the within-period and between-period Kendall's tau (a type of rank correlation). Web Appendix E additionally explores the relationship between g-ICC and the Kendall's tau in specific scenarios, and our free R shiny application also allows one to explore their relationships more generally.

\section{Simulation study}\label{sec:sim}

We adopt the ADEMP (aims, data-generating mechanisms, estimands, methods, and performance measures) framework of \citet{morris_simulation_2019} to report our simulation studies. The R code to reproduce our simulations is available in the Supplementary Materials and at \url{https://github.com/maryryan/survivalSWCRT}.

\textbf{Aims}: We conduct a simulation study to (i) compare the type I error rate and empirical power of the Wald $t$-test and robust score test in SW-CRTs; (ii) assess the utility of finite-sample bias-correction methods (see Table \ref{tab:bias}) for maintaining the validity of tests with a small number of clusters; and (iii) examine the adequacy of our proposed sample size procedures among the valid tests that maintain the nominal type I error rate.

\textbf{Data-generating mechanisms:}
Simulation scenario combinations are enumerated in Web Table \ref{tab:simsettings} in Web Appendix G. We consider $J\in\{3, 4, 5, 6\}$ and $m\in\{15, 25, 40, 50\}$. We also vary the number of clusters, $n$, between $8$ and $30$ in multiples of $(J-1)$. These values are chosen to reflect study parameters typically reported for SW-CRTs \citep{nevins_scoping_2023}. We also vary the true treatment effect, $\beta$, between $0.25$ and $0.7$ in the non-null scenarios. In all simulations, we assume event times $T_{ijk}{\sim}\text{Exp}(\lambda_{ij})$ with independent loss to follow-up censoring such that $C_{ijk} {\sim} \text{Unif}(0,C^*)$ with administrative censoring time $C^*=1$. We also assume a baseline hazard that progressively increases with time, $\lambda_{0j}(t|Z_{ij}) = \lambda_0 + 0.2(j-1)$, to induce a non-zero period effect. Following \citet{zhong_sample_2015} and \citet{wang_improving_2023}, we set $\lambda_0$ as the solution to $P(T_{i1k} > C^*|Z_{i1}=0)=p_a$ in the first study period given a reference administrative censoring rate $p_a$; in these simulations, we consider $p_a=20$\%. With random loss to follow-up and an overall administrative censoring rate that changes with $\lambda_{0j}$, the total marginal censoring rate ranges from $38$\% to $42$\%. We use the algorithm in \citet{mcneil_sampling_2008} and \citet{li_sample_2022} to generate correlated survival times from a nested Gumbel copula model using $\theta_0=1/(1-\tau_b)$ and $\theta_{01}=1/(1-\tau_w)$, where $\tau_b$ and $\tau_w$ are the between-period and within-period Kendall's tau. We examined three sets of Kendall's tau: $(\tau_w, \tau_b) = \{(0.05, 0.01), (0.1, 0.01), (0.1, 0.05)\}$. The magnitude of correlation parameters were chosen to mimic, to the extent possible, the range of reported ICCs in the SW-CRT literature \citep{korevaar_intra-cluster_2021}. For presentation clarity, simulation combinations are chosen to ensure $80\%$--$95$\% empirical power based on two-sided Wald test. A step-by-step outline for generating correlated survival data for a single cluster $i$ is found in Algorithm \ref{algo:nestedGumbel}.

\begin{algorithm}
\caption{Generate correlated survival data from nested Gumbel copula in one cluster}\label{algo:nestedGumbel}
\begin{algorithmic}[1]
    \Require: $\theta_w = 1-\tau_w$; $\theta_b = 1-\tau_b$.
    
    \State Generate random variable $V_0$ from a stable distribution $S(\theta_b, 1, \cos{(\pi/(2\tau_b))},0)$ using the method described by \citet{j_nolan_stable_2003} or using \texttt{R} function \texttt{stabledist()}.
    
    \State Generate $J$ i.i.d. random variables $V_j$ from stable distribution $S(\theta_w/\theta_b, 1, \cos{(\pi\theta_w/2\theta_b)}, 0)$.
    
    \State Generate $J\times m$ independent random variables $Z_{i11}, \dots, Z_{iJm}$ from a standard Uniform distribution $U(0,1)$.
    
    \State Calculate $U_{ijk} = \exp{\left\{-[-\ln(Z_{ijk})/V_j]^{\theta_w/\theta_b}\right\}}$ for $j=1, \dots, J$ and $k=1, \dots, m$.
    
    \State Generate correlated failure times $T_{ijk} = [-\ln(U_{ijk})/V_0]^{\theta_b}/\lambda_{ijk}$ for $j=1,\dots, J$ and $k=1, \dots, m$.
\end{algorithmic}
\end{algorithm}

\textbf{Estimands:} Under the period-stratified marginal Cox model, the primary estimand is the treatment effect parameter, interpreted as a constant hazard ratio. 

\textbf{Methods:} Throughout, predicted power for the Wald $t$-test is based on equation \eqref{eq:power} and Theorem \ref{theorem:var_b}. For the robust score test, predicted power is based on equation \eqref{eq:power-score}, \eqref{eq:power-tang}, as well as Proposition \ref{coro:var_score}. As SW-CRTs often include a small number of clusters, we also explore several finite-sample corrections. In the Wald testing paradigm, to mitigate bias toward zero from the robust sandwich variance estimator, we adapt the methods of \citet{fay_small-sample_2001} (FG), \citet{kauermann_note_2001} (KC), and \citet{mancl_covariance_2001} (MD) to provide bias corrections, adapting the work of \citet{wang_improving_2023} from marginal Cox analysis of parallel CRTs to the period-stratified marginal Cox analysis of SW-CRTs. Finite-sample bias has also been reported for estimating $\sigma_1^2$ for robust score tests with a non-censored outcome \citep{guo_small-sample_2005}, resulting in conservative type I error rates. Therefore, we also apply the modified robust score test of \citet{guo_small-sample_2005} which weights $\sigma^2_1$ by $(n-1)/{n}$. In the ensuing simulation study, we compare the operating characteristics of these correction methods in finite-sample settings to identify valid tests. We also summarize the bias-correction methods under consideration in Table \ref{tab:bias}.

\begin{table}%\label{tab:bias}
    \caption{\label{tab:bias} Finite-sample bias-correction variance estimators under consideration. In the Wald $t$-test paradigm: robust sandwich variance estimator, \citet{fay_small-sample_2001} (FG), \citet{kauermann_note_2001} (KC), \citet{mancl_covariance_2001} (MD). In the robust score testing paradigm: robust score (SM), modified robust score \citet{guo_small-sample_2005}.}
    \vspace{0.25cm}
    \centering
    \begin{tabular}{ll}
    \hline
    Testing Paradigm \& Correction & Formula\\
    \hline\\[-2ex]
     $\boldsymbol{t}$\textbf{-test} & $A^{-1}(\widehat{\beta})\left(\sum_{j=1}^J\sum_{i=1}^n C_{ij} \widehat{U}_{ij}\widehat{U}_{ij}'C_{ij}'\right)A^{-1}(\widehat{\beta})$\\
     &\\[-1ex]
     \hspace{0.3cm}\textit{Lin's general variance} & \hspace{0.3cm} $C_{ij} = 1$\\
     \hspace{0.3cm}\textit{FG correction} & \hspace{0.3cm} $C_{ij} = \left(I - \frac{\partial U_{ij}(\widehat{\beta})}{\partial \widehat{\beta}}A^{-1}(\widehat{\beta})\right)^{-1/2}$\\
     \hspace{0.3cm}\textit{KC correction} & \hspace{0.3cm} $C_{ij} = \diag\left\{\left[1-\min\left(r,\left[\frac{\partial U_{ij}(\widehat{\beta})}{\partial \widehat{\beta}}A^{-1}(\widehat{\beta})\right]_{kk}\right)\right]^{-1/2}\right\}$\\
     \hspace{0.3cm}\textit{MD correction} & \hspace{0.3cm} $C_{ij} = \left(I - \frac{\partial U_{ij}(\widehat{\beta})}{\partial \widehat{\beta}}A^{-1}(\widehat{\beta})\right)^{-1}$\\
     \hline\\[-2ex]
     \textbf{Robust score} & c$\sum_{j=1}^J\sum_{i=1}^n \widehat{U}_{ij}\widehat{U}_{ij}'$\\
     &\\[-1ex]
     \hspace{0.3cm}\textit{SM general variance} & \hspace{0.3cm} $c=1$\\
     \hspace{0.3cm}\textit{Guo's modified correction} & \hspace{0.3cm} $c=(n-1)/n$\\
     \hline
    \end{tabular}
\end{table}

\textbf{Performance measures:} As the focus of the simulation study is on evaluating the performance of testing procedures rather than point estimation, we interpret ``estimands'' in the ADEMP framework as the nominal type I error rate—assessing the validity of each test—and the empirical power—assessing the accuracy of the predicted power based on analytical formulas. The empirical power of each test is calculated as the proportion of iterations that correctly rejected $H_0$ over 2,000 simulated SW-CRTs. Accuracy of predicted power is assessed by the difference in empirical power less predicted power. The empirical type I error rate is calculated as the proportion of iterations that incorrectly rejected $H_0$.

\subsection{Simulation results}\label{ss:results}

\begin{figure}
    \centering
    \includegraphics[width=\textwidth]{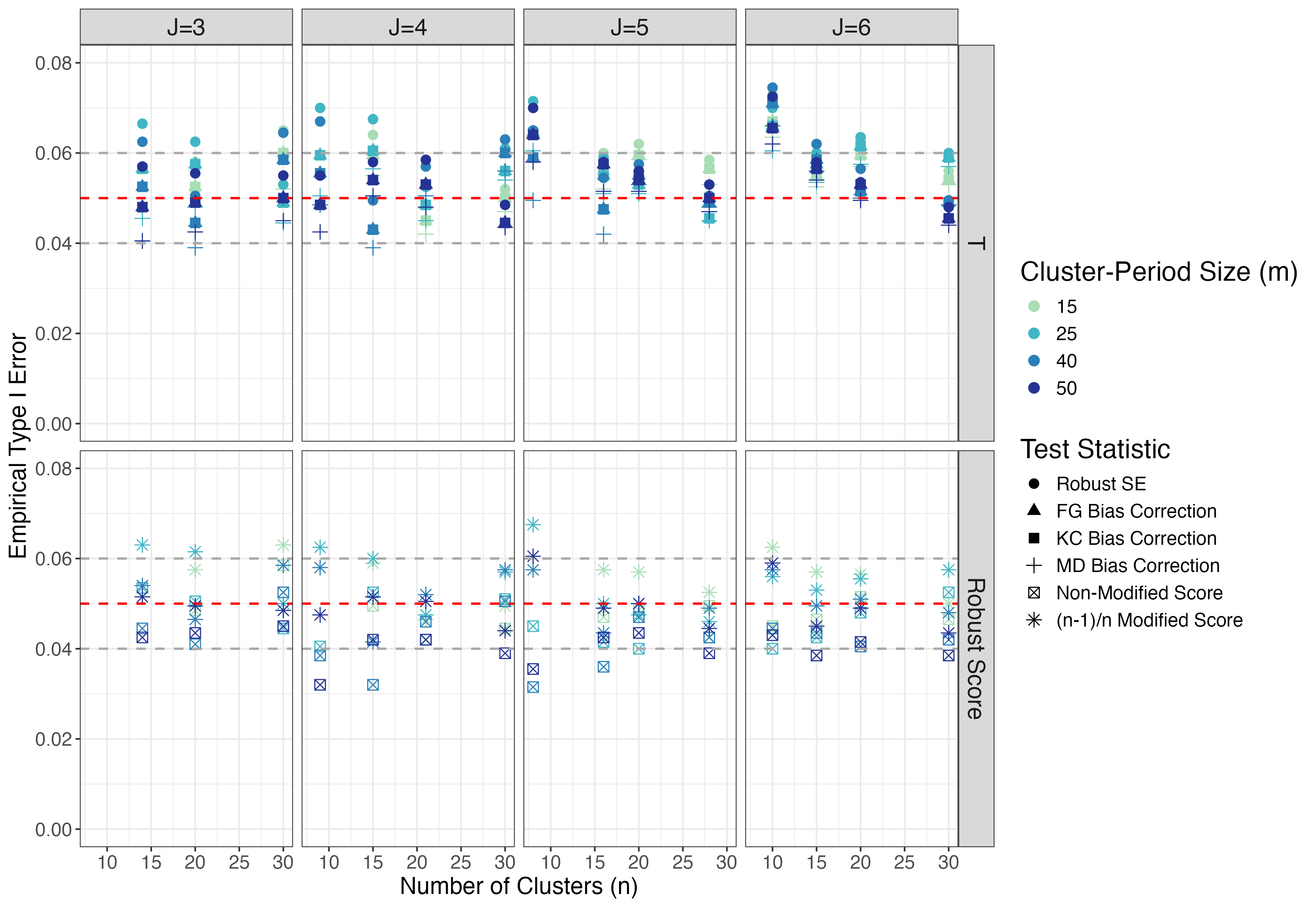}
    \caption{\label{fig:typeI}Empirical type I error rates for hypotheses testing paradigms when within-period Kendall's $\tau_w=0.05$ and between-period Kendall's $\tau_b=0.01$, given $n$ clusters of cluster-period size $m$ are transitioned onto intervention over $J$ periods (columns). The top row displays empirical type I error results for Wald $t$-tests using a robust sandwich variance (Robust SE) as well as \citet{fay_small-sample_2001} (FG), \citet{kauermann_note_2001} (KC), and \citet{mancl_covariance_2001} (MD) finite-sample adjusted variances (top row). The bottom row displays empirical type I error results for robust (Non-Modified Score) and modified robust score tests ($(n-1)/n$ Modified Score). The red dotted line represents the nominal 5\% error rate and gray dotted lines represent simulation 95\% confidence intervals.}
\end{figure}

\begin{figure}%[!ht]
    \centering
    \includegraphics[width=\textwidth]{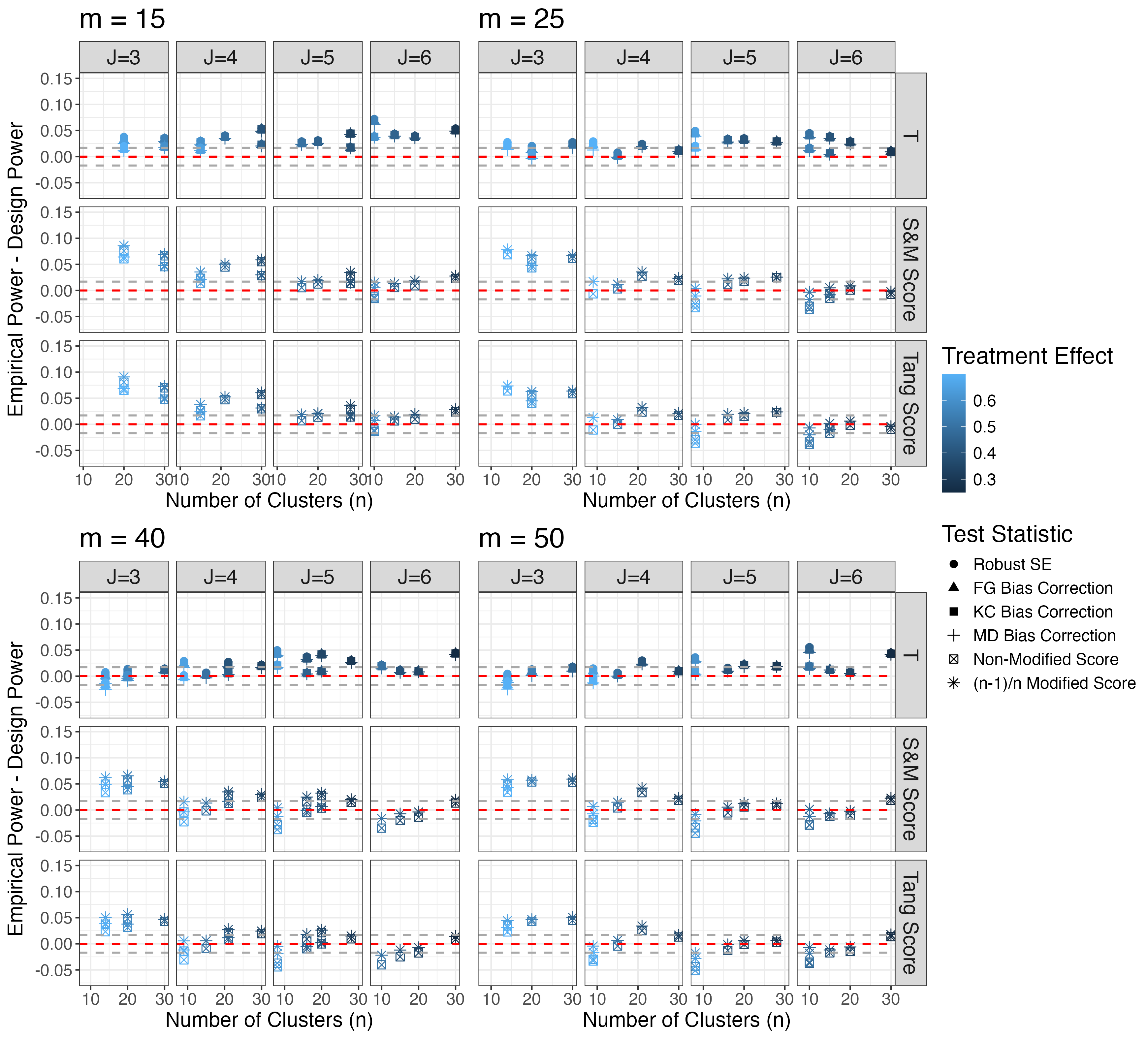}
    \caption{\label{fig:powerDiff}Difference between empirical and predicted power of hypothesis testing paradigms when within-period Kendall's $\tau_w=0.05$ and between-period Kendall's $\tau_b=0.01$, given $n$ clusters of cluster-period size $m$ are transitioned onto intervention over $J$ periods (columns) under a given treatment effect magnitude (color scale; lighter colors represent larger magnitude). The top row displays difference in power for Wald $t$-tests using a robust sandwich variance (Robust SE) as well as \citet{fay_small-sample_2001} (FG), \citet{kauermann_note_2001} (KC), and \citet{mancl_covariance_2001} (MD) finite-sample adjusted variances. The middle and bottom rows displays difference in power for robust (Non-Modified Score) and modified robust score tests ($(n-1)/n$ Modified Score) when power is predicted using the \citet{self_powersample_1988} methods (middle row) and the \citet{tang_improved_2021} methods (bottom row). The red dotted line represents a difference of $0$ and the gray dotted lines represent simulation 95\% confidence intervals.}
\end{figure}

The results of our simulation study under $(\tau_w, \tau_b)=(0.05, 0.01)$ are presented in Figures \ref{fig:typeI} and \ref{fig:powerDiff}; results for the remaining settings are qualitatively similar and presented in Web Appendix G. In general, the type I error rates (Figure \ref{fig:typeI}) under the uncorrected robust variance for the Wald $t$-test are almost always inflated unless $n$ or $m$ are of moderate size ($n\ge20$, $m\ge 40$), whereas the use of bias-correction variance estimators can maintain the nominal size if the number of clusters or cluster-period size are not especially small ($n > 10$, $m \ge 25$). More specifically, the Wald $t$-test coupled with the uncorrected robust sandwich variance estimator is the most liberal while the use of MD-corrected variance estimator is the most effective at controlling for type I error inflation. Furthermore, we also observe an inflation in type I error rate when the number of clusters per sequence is fewer than $3$, specifically, when $(J,n)\in\{(5,8), (6,10)\}$. We explored this issue in additional simulations with $J=9$ periods and $\{1, 2, 3, 4\}$ clusters per sequence (not presented) with similar findings. This suggests that in small sample settings, the number of clusters per sequence may be more important for test validity than the total number of observations. Without any finite-sample corrections, the robust score tests generally maintain the nominal test size, but may be occasionally conservative in the smallest sample size scenarios. However, the modified robust score test can sometimes carry a slightly inflated test size when $n\leq 15$ and $m\le 25$, suggesting that it may not be necessary to consider the finite-sample correction of \citet{guo_small-sample_2005} in small SW-CRTs.

Empirical power results for the same scenarios are shown in Web Figure \ref{fig:empPower0105} in Web Appendix G. Overall, the Wald $t$-test and robust score testing paradigms achieve similar levels of empirical power, though when the number of clusters and the number of periods are both not large ($n\le20$, $J\le 4$), the robust score test is frequently slightly more powerful. When $n\ge 20$, all tests generally carry the nominal size; the uncorrected robust sandwich variance estimator leads to the most powerful Wald $t$-test while the MD-corrected variance estimator corresponds to the least powerful test. Similarly, the modified robust score test is more powerful than the non-modified robust score test when $n$ increases.

Finally, Figure \ref{fig:powerDiff} presents the results for the difference between empirical and predicted power. The Wald $t$-tests generally tend to slightly under-predict power, though usually within $5$\%, while the robust score testing paradigm tends to over-predict power when the cluster-period size is moderate to large ($m >15$) and the number of clusters is small ($n\le10$). As $n$ and $m$ increase, the difference approaches $0$ approximately equally for both Wald and robust score methods. In addition, both the S\&M and Tang robust score methods tend to predict power similarly. Across all scenarios with valid tests, differences in empirical and predicted power for the Wald $t$-testing paradigm with an MD correction are between $-2.5$\% and $4.5$\%, whereas the differences for the robust score paradigm predicted using the S\&M and Tang methods are between $-4.4$\% and $5.4$\%, and $-5.2$\% and $4.5$\%, respectively. We observe that these results continue to hold with increasing $\tau_w$ (Web Appendix G: Web Figures \ref{fig:powerDiff101}-\ref{fig:powerDiff105}), though the robust score methods are more likely to under-predict the empirical power with increasing $\tau_w$.

\section{A data example with the CATH TAG stepped wedge trial}\label{sec:app}
We illustrate our analytic power methods in the context of a trial of the CATH TAG electronic reminder system \citep{mitchell_reducing_2019}. The study randomized $n=10$ wards of a large Australian hospital to transition to using CATH TAG devices over $J=6$ one-month periods ($5$ sequences; see Figure \ref{fig:sw-crt}A). Patients were censored only at transfer to another ward or hospital, not at the end of a period, meaning that patient follow up could in theory extend over multiple periods. However, as the mean catheter duration was short (approximately $5.51$ days in the control arm), we can assume minimal risk of treatment contamination. We assume a cluster-period size of $m=35$ patients for illustration. The original study protocol assumed a global ICC of $0.1$ but did not distinguish between within-period and between-period ICCs; for illustration, we assume the within-period and between-period Kendall's tau as $\tau_w=0.1$ and $\tau_b=0.05$, respectively, when predicting power under the generative procedure (Section \ref{sec:practical} and Web Appendix D). We will plan our hypothetical study to detect a hazard ratio of $1.5$ ($\beta=0.4$). Assuming uniformly-distributed loss to follow-up censoring, minimal administrative censoring ($p_a=5\%$), and a baseline hazard that increases by $5$\% with each additional period such that $\lambda_{0j}(t) = \lambda_0 + 0.05(j-1)$, our Wald approach predicts that $18$ wards would be required to to detect a HR=$1.5$ with $80$\% power with a within-period g-ICC of $0.1$ and a between-period g-ICC of $0.02$. Similarly, our robust score approach using the S\&M and Tang methods predict $18$ and $17$ clusters are needed, respectively, to detect the same effect size.

The above calculations assume the clusters are evenly distributed among the sequences, which could result in fractional clusters per sequence (e.g., when $n=18$). In such a case, one could either include additional clusters to ensure a balanced sequence assignment, or explore the power under a specific unbalanced assignment. Under the first strategy, if we increase our number of clusters to $n=20$ our Wald approach estimates $80.8$\% power to detect the treatment effect, while our robust score approach using the S\&M and Tang methods estimate $85.5$\% and $86.3$\% power, respectively. For the second strategy, our free R Shiny application allows one to upload a specific design matrix; a tutorial can be found in Web Appendix H.

\begin{figure}
    \centering
    \includegraphics[width=\textwidth]{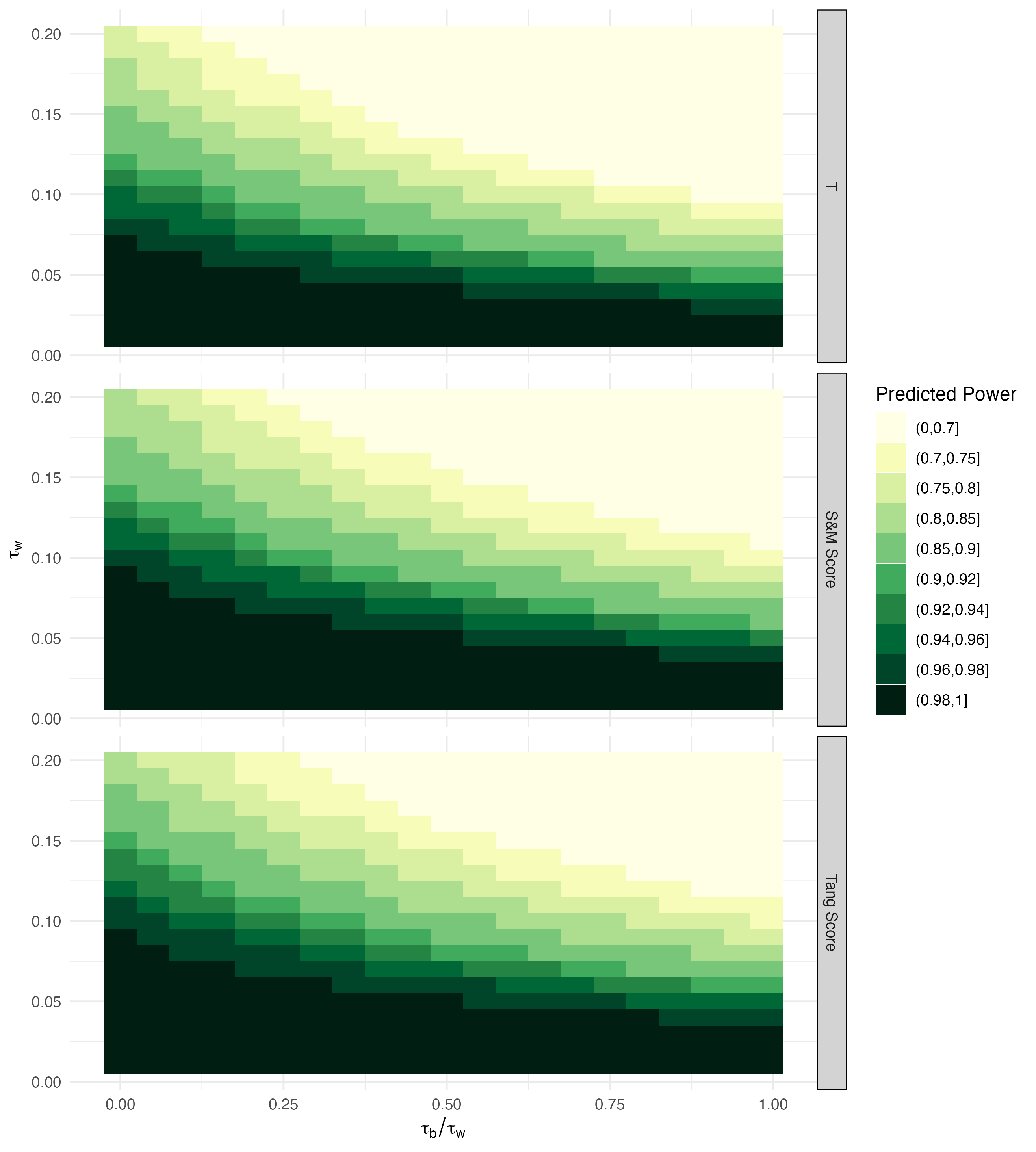}
    \caption{\label{fig:sensitivity} Contour plots of predicted power trends to detect $\beta=0.4$ (HR=$1.5$) across within-period Kendall's tau ($\tau_w$) and the ratio of between- and within-period Kendall's tau ($\tau_b/\tau_w$) within our application study of the CATH TAG trial, assuming a baseline hazard that increases by 5\% at each subsequent time period. The top row represents trends when power is predicted using the Wald $t$-test formula, the middle row when using the \citet{self_powersample_1988} robust score test formula, and the bottom row when using the \citet{tang_improved_2021} robust score test formula. Darker colors correspond to greater predicted power.}
\end{figure}

To assess the sensitivity of our sample size and power calculation to choice of Kendall's tau, we study how the predicted power for a balanced design may vary over $\tau_w\in [0, 0.2]$ with the ratio $\tau_b/\tau_w \in [0,1)$. Figure \ref{fig:sensitivity} presents results assuming $20$ clusters, and show that larger $\tau_w$ and $\tau_b$ result in smaller predicted power. Concordant with Section \ref{sec:sim}, the Wald $t$-test predicts the smallest power under all Kendall's tau combinations while the robust score power predictions using the Tang method return the highest, though the differences are slight. We can also see that when $\tau_w$ is below $0.05$, power under all paradigms is robust to changes in $\tau_b$; as $\tau_w$ increases and the range of values $\tau_b$ can take on grows, power predictions become more sensitive to $\tau_b$. For example, at $\tau_w=0.1$, predicted power ranges between $67$\% ($\tau_b=0.1$) and $98$\% ($\tau_b=0$). This speaks to the importance of differentiating the within-period and between-period correlations in power calculation, similar to SW-CRT settings with non-survival endpoints \citep{taljaard2016substantial}. Finally, to assess sensitivity to choice of baseline hazard, we also considered a constant baseline hazard, such that $\lambda_{0j}(t) = \lambda_0$, and decreasing baseline hazard, such that $\lambda_{0j}(t) = \lambda_0 - 0.05(j-1)$. The results and discussion of these analyses, along with an exploration of power under different g-ICC values, can be found in Web Appendix F; we generally find that power trends are largely robust to baseline hazard choice. Step-by-step R code to reproduce all calculations in Section 5 is available in the Supplementary Materials as well as at \url{https://github.com/maryryan/survivalSWCRT}; they may also be reproduced using our R Shiny application (Web Appendix H).

\section{Discussion}\label{sec:discuss}
In this article, we derived new analytic power calculation procedures for cross-sectional SW-CRTs with right-censored time-to-event outcomes, addressing an emerging scenario that has not been accommodated by current methods. In our numerical studies, the proposed Wald-based and score-based power formulas may under-predict power in finite samples (thus maybe considered conservative), though this improved as $n$ and $m$ increased.

We have based our power formulas on the period-stratified marginal Cox model, but this may not be the only choice of analytic model for cross-sectional SW-CRTs. For instance, an alternative approach is to account for the within-cluster correlation structures through a period-stratified frailty model with random effects and to develop variance formulas via the model-based variance, along the lines of \citet{hooper_sample_2016} and \citet{kasza2019impact}, and general mixed model formulation as in \citet{li_mixed-effects_2021}. While this approach might have higher power in some occasions by directly estimating the random-effects variance parameters, the model-based variance expression can also be sensitive to correlation misspecification \citep{kasza2019inference} and one could end up with an over-optimistic sample size estimate when the random-effects structure is incorrectly specified. Under a frailty model, it is generally challenging to obtain a closed-form variance expression and simulation-based power calculation can be used as a general and flexible approach for study planning. While simulation-based power calculations are usually an option, SW-CRTs with time-to-event outcomes often require more complicated data generating processes \citep{meng2023simulating} which can make power calculation computationally demanding, especially when considering many design scenarios with complex frailty models. From that standpoint, our approach serves as a complimentary yet computationally convenient alternative that exploits the sandwich variance expression under working independence to quickly provide insights into the key determinants of study power for SW-CRTs. We expect our formula to provide a conservative sample size estimate for cross-sectional SW-CRTs analyzed by frailty models, although a formal comparison merits future research.

%  The \backmatter command formats the subsequent headings so that they
%  are in the journal style.  Please keep this command in your document
%  in this position, right after the final section of the main part of 
%  the paper and right before the Acknowledgements, Supporting Information (Supplementary %  Materials),   and References sections. 

%\backmatter

%  This section is optional.  Here is where you will want to cite
%  grants, people who helped with the paper, etc.  But keep it short!

\section*{Acknowledgements}

Research in this article was supported by two Patient-Centered Outcomes Research Institute Awards\textsuperscript{\textregistered} (PCORI\textsuperscript{\textregistered} Awards ME-2020C3-21072, ME-2022C2-27676),  by CTSA Grant Number UL1 TR001863 from the National Center for Advancing Translational Science (NCATS), a component of the National Institutes of Health (NIH), and by Yale Claude D. Pepper Older Americans Independence Center (P30AG021342). The statements presented are solely the responsibility of the authors and do not necessarily represent the views of PCORI\textsuperscript{\textregistered}, its Board of Governors or Methodology Committee, or the National Institutes of Health.

\section*{Supplementary Materials}

Web Appendices and Figures referenced in Sections 2-5 are available with this paper on arXiv. R code for predicting power and for conducting the simulation and application studies described in Sections 4-5, and source code for the online R Shiny application, are available at \url{https://github.com/maryryan/survivalSWCRT}. The R Shiny application can be accessed at \url{https://mary-ryan.shinyapps.io/survival-SWD-app}.

\section*{Data Availability}

The illustrative data example in this paper only concerns sample size and power estimation in a real study context, and does not involve analysis of actual data sets. Further, no new primary individual-level data are generated in support of this paper.

\bibliographystyle{biom}

\newpage

\section*{Supplementary Materials for ``Power calculation for cross-sectional stepped wedge cluster randomized trials with a time-to-event endpoint" by Ryan Baumann, Esserman, Taljaard and Li}

%\maketitle
%%\vspace{-1cm}
%\part{Appendix} % Start the appendix part
%\parttoc % Insert the appendix TOC
%\doublespacing
%setstackgap{L}{.6\baselineskip}
\addcontentsline{toc}{section}{Web Appendix A: Variations of Study Timing and Censoring}
\section*{Web Appendix A: Variations of Study Timing and Censoring}\label{A}
\renewcommand\thefigure{A.\arabic{figure}}
\setcounter{figure}{0}

In the general setting, SW-CRTs can be classified into three types, depending on whether individuals within each cluster only contribute data to a single time period (cross-sectional design), are followed longitudinally and contribute information to multiple periods (closed-cohort design), or may flexibly join or leave the study across time (open-cohort design) \citep{copas_designing_2015}. These broad classifications help us identify appropriate correlation structures for observed data points, but also give us bounds around the time an individual study participant may be observed. In the context of time-to-event endpoints, however, where ``observation time'' is inherently part of the outcome and is not necessarily defined by study time periods, these definitions can be more complex.\\

\begin{figure}[ht!]
    \centering
    \includegraphics[width=0.75\textwidth]{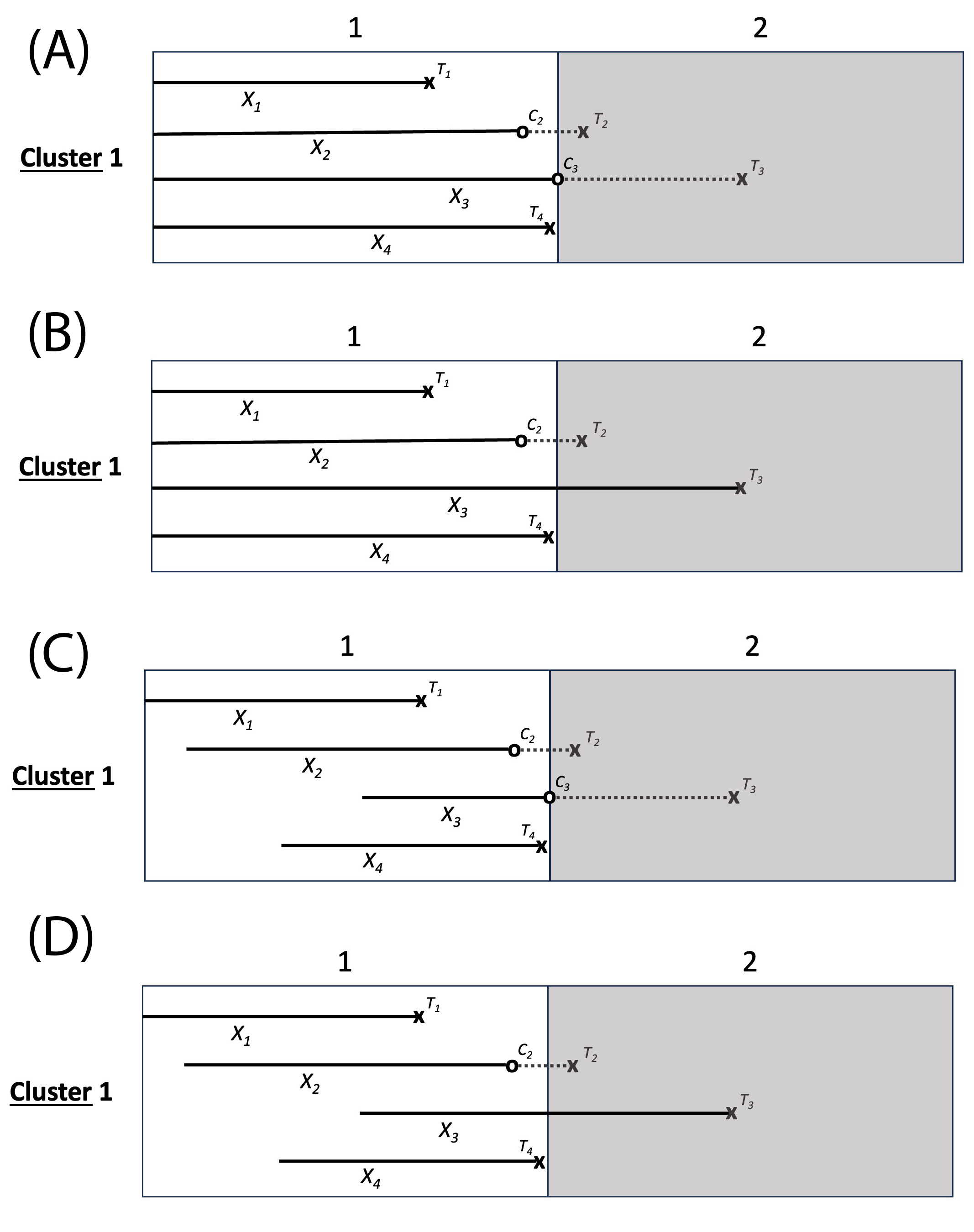}
    \caption{\label{fig:sw-timing}Example schematics of observed event and censoring times for four individuals recruited simultaneously (panels (A) and (B)) or continuously (panels (C) and (D)) during period 1 of a cross-sectional stepped-wedge cluster randomized trial. Panels (A) and (C) depict designs with strict study time period-end administrative censoring, while panels (B) and (D) illustrate flexible follow-up beyond the end of the study period. Cross symbols denote events and open circles denote censoring, while solid lines denote observed follow up time and dotted lines denote actual post-censoring time to event.}
\end{figure}

\noindent In particularly, the meaning of ``cross-sectional'' in time-to-event settings can refer to several different observation structures depending on the nature of participant recruitment and the rigidity of administrative censoring; four examples are shown in Figure \ref{fig:sw-timing}.\\

\noindent First, in Figure \ref{fig:sw-timing}(A), study participants are recruited simultaneously at the beginning of a study period (``fixed recruitment'') and are administratively censored at the end of the period, even if the participant was not lost to follow-up and did not experience an event. In this setting, maximum follow-up time is standardized to the length of the period and ensures no within-cluster treatment contamination. This setting may be appropriate when all participants eligible for a study period are present or can be identified at once, when the length of the study period is of clinical importance (e.g., survival up to $28$ days), or when the participant is in continuous contact with the trial condition under investigation. An example of this might be time to discharge under a new intensive care unit observation protocol.\\

\noindent As a variation, study participants may instead be followed-up past the end of the calendar time defining the study period in which they were recruited (Figure \ref{fig:sw-timing}(B)). This setting would allow for variations in maximum follow-up time and would result in fewer participants being administratively censored, though not necessarily fewer with random loss to follow-up, depending on the event of interest. This setting may be appropriate when there is not a major concern of within-cluster treatment contamination, such as when participants have a single point of interaction with the trial condition so that their follow-up past the end of the period will not be contaminated by interaction with the intervention condition. This extended follow-up may exacerbate confounding by time across the entire study, however, as participants who were recruited and treated earlier in the study timeline are then permitted longer follow-up than participants recruited at later periods.\\

\noindent It may be more realistic, though, that participants are not all identifiable at the beginning of the study period and will instead present themselves to the cluster randomly throughout the period (``continuous recruitment'', also see \citet{hooper2021key}). Depending on the nature of the intervention, there is still a choice in how participants are administratively censored. Censoring participants at the end of the period (Figure \ref{fig:sw-timing}(C)) would be appropriate for interventions with continuous participant contact, such as in scenario (A); the difference here is that the continuous recruitment of scenario (C) prevents a standardized maximum follow-up time like in scenario (A). On the other hand, participants could be followed-up beyond the end of the period (Figure \ref{fig:sw-timing}(D)), such as in scenario (B); this would be appropriate for interventions with a single point of contact with the participant.\\

\noindent Further variations on these schemes are also possible. For example, an adaptation may be made for situations when participants cannot all be readily identified at the start of the period but standardizing the maximum follow-up time is necessary. It is important to consider which observation timing scenario is most applicable when designing a cross-sectional SW-CRT as this will affect (i) administrative censoring rates and (ii) the possibility for within-cluster treatment contamination. While (i) will primarily impact study power and the presence of time confounding, (ii) may bias the treatment effect estimate and jeopardize the validity of the trial results.
\addcontentsline{toc}{section}{Web Appendix B}
\section*{Web Appendix B}\label{B}

\addcontentsline{toc}{subsection}{Wald Testing Procedure}
\subsection*{Wald Testing Procedure}
To test $H_0: \beta = \beta_0$ versus $H_1: \beta \ne \beta_0$, the Wald $t$-test statistic $t = |\hat{\beta} - \beta_0|/\sqrt{Var(\hat{\beta})}$ has a $t$ distribution with $DoF$ degrees of freedom under $H_0$. Thus, if $|\hat{\beta} - \beta_0|/\sqrt{Var(\hat{\beta})} \ge t_{1-\alpha/2; DoF}$, the null hypothesis is rejected, where $t_{p; DoF}$ is the $p$th percentile of a $t$ distribution with $DoF$ degrees of freedom.\\

\noindent To calculate the test statistic, $Var(\hat{\beta})$ is calculated according to the sandwich variance estimator under a working assumption of independent correlation between event times, $A^{-1}(\hat{\beta})\left(\sum_{i=1}^n\sum_{j=1}^J U_{ij+}(\hat{\beta})U^T_{ij+}(\hat{\beta})\right)A^{-1}(\hat{\beta})$. $\hat{\beta}$ is estimated according to the usual maximum partial likelihood estimator.\\

\noindent To accommodate finite-sample bias corrections, a cluster period-specific weight $C_{ij}$ can be applied to $U_{ij+}(\hat{\beta})$ such that the sandwich variance estimator takes the form $A^{-1}(\hat{\beta})\left(\sum_{i=1}^n\sum_{j=1}^J C_{ij}U_{ij+}(t; \hat{\beta})U^T_{ij+}(t; \hat{\beta})C^T_{ij}\right)A^{-1}(\hat{\beta})$

\addcontentsline{toc}{subsection}{Robust Score Testing Procedure}
\subsection*{Robust Score Testing Procedure}
To test $H_0: \beta = \beta_0$ versus $H_1: \beta \ne \beta_0$, the robust score test statistic $\left|U(\beta_0)\right|/\sqrt{\sigma^2}$ has a standard Normal distribution. Thus, if $\left|U(\beta_0)\right|/\sqrt{\sigma^2} \ge z_{1-\alpha/2}$, the null hypothesis is rejected, where $z_{p}$ is the $p$th percentile of a standard Normal distribution.\\

\noindent To calculate the test statistic, the score equation evaluated under $\beta_0$ takes the form
$$U(\beta_0) = \sum_{i=1}^n \sum_{j=1}^J \sum_{k=1}^m \int_0^{C^*} \overline{Y}_{ijk}(t) \left\{Z_{ij} - \frac{S^{(1)}_j(t;\beta_0)}{S^{(0)}_j(t;\beta_0)}\right\} dN_{ijk}(t)=0,$$
where $S_j^{(0)}(t;\beta_0) = n^{-1}\sum_{i=1}^n\sum_{k=1}^m\overline{Y}_{ijk}(t)\exp\left(\beta_0 Z_{ij}\right)$ and $S_j^{(1)}(t;\beta_0) = n^{-1}\sum_{i=1}^n\sum_{k=1}^m\overline{Y}_{ijk}(t)Z_{ij}\exp\left(\beta_0 Z_{ij}\right)$. If $H_0: \beta=\beta_0$ is true, the data $X_{ijk}$ should be consistent with the Cox survival model at $\beta=\beta_0$, and $U(\beta_0)$ should be close to $0$; if the data are inconsistent with $\beta=\beta_0$, however, $U(\beta_0)$ often consistently deviates from $0$. The variance $\sigma^2$ may be calculated as $n^{-1}\sum_{i=1}^n\{U_{i++}(\beta_0)U^T_{i++}(\beta_0)\}$.
\addcontentsline{toc}{section}{Web Appendix C}
\section*{Web Appendix C}\label{C}

\addcontentsline{toc}{subsection}{Derivation of $A^{-1}(\beta)$}
\subsection*{Derivation of $A^{-1}(\beta)$}
Under a working independence assumption, $\hat{\beta}$ is asymptotically normal with mean $\beta$ and covariance matrix $A^{-1}(\beta) = E\left\{-\partial U_{i++}(\beta)/\partial \beta\right\}^{-1}$. This can be estimated as
$$E\left\{-\frac{\partial U_{i++}(\hat{\beta})}{\partial \hat{\beta}}\right\}^{-1}= E\left\{n^{-1} \sum_{i=1}^n \sum_{j=1}^J \sum_{k=1}^m \int_0^{C^*} \overline{Y}_{ijk}(t) \mu_j(t)\left[1 - \mu_j(t)\right]dN_{ijk}(t)\right\}^{-1},$$
where $\mu_j(t) = s_j^{(1)}(t;\hat{\beta})/s_j^{(0)}(t;\hat{\beta})$ is the ratio of the almost sure limits of $S_j^{(0)}(t;\beta)$ and $S_j^{(1)}(t;\beta)$. We may expand this as:
\begin{align*}
    &E\left\{n^{-1} \sum_{i=1}^n \sum_{j=1}^J \sum_{k=1}^m \int_0^{C^*} \overline{Y}_{ijk}(t) \mu_j(t)\left[1 - \mu_j(t)\right]dN_{ijk}(t)\right\}^{-1}\\
    &= \left\{n^{-1} \sum_{i=1}^n \sum_{j=1}^JE_{Z_{ij}}\left[ E_{Y_{ijk}(t)|Z_{ij}}\left(E_{Y^{\dagger}_{ijk}(t)|Y_{ijk}(t), Z_{ij}}\right.\right.\right.\\
    &\hspace{1cm}\left.\left.\left.\left\{\sum_{k=1}^m \int_0^{C^*} \overline{Y}_{ijk}(t) \mu_j(t)\left[1 - \mu_j(t)\right]\lambda_{ijk}(t)dt\right\}\right)\right]\right\}^{-1}\\
    &= \left\{n^{-1} \sum_{i=1}^n \sum_{j=1}^JE_{Z_{ij}}\left[ E_{Y_{ijk}(t)|Z_{ij}}\left(\sum_{k=1}^m \int_0^{C^*} \mathscr{G}(t)Y_{ijk}(t) \mu_j(t)\left[1 - \mu_j(t)\right]\lambda_{ijk}(t)dt\right)\right]\right\}^{-1}\\
    &= \left\{n^{-1} \sum_{i=1}^n \sum_{j=1}^JE_{Z_{ij}}\left[ \sum_{k=1}^m \int_0^{C^*} \mathscr{G}(t)P(T_{ijk}\ge t|Z_{ij}) \mu_j(t)\left[1 - \mu_j(t)\right]\lambda_{ijk}(t)dt\right]\right\}^{-1}\\
    &= \left\{n^{-1} \sum_{i=1}^n \sum_{j=1}^JE_{Z_{ij}}\left[ \sum_{k=1}^m \int_0^{C^*} \mathscr{G}(t) \mu_j(t)\left[1 - \mu_j(t)\right]f(t|Z_{ij})dt\right]\right\}^{-1},
\end{align*}
where $E_{Z_{ij}}\{\cdot\}$ is the expectation with respect to treatment at study period $j$, $\mathscr{G}(t)$ is the marginal survival function for the censoring time $C_{ijk}$, and $f(t|Z_{ij})$ is the conditional density of event time $T_{ijk}$.

%% LEMMA 1 %%
\addcontentsline{toc}{subsection}{Intermediate Result to Equation (7)}
\subsection*{Intermediate Result to Equation (7)}
Power calculation for the Wald $t$-test requires the expression of $\text{Var}(\hat{\beta})$, while power calculation for the robust score test requires the expression of $B(\beta)=\Var \left\{U_{i++}(\beta)\right\}$. To facilitate the derivation, we first provide an intermediate result on the variance and covariance expressions in equation (7) to simplify the expression of $\Var\left\{U_{i++}(\beta)\right\}$. This will involve deriving and simplifying three components: $\Var\left\{U_{ijk}(\beta)\right\}$, $\Cov\left\{U_{ijk}(\beta),U_{ijd}(\beta)\right\}$ when $k \ne d$, and $\Cov\left\{U_{ijk}(\beta),U_{ild}(\beta)\right\}$ when $j \ne l$ but $k$ may be equal to $d$.\\

\noindent We will begin with the derivation of $\Var\left\{U_{ijk}(\beta)\right\}$. Given $E\left\{U_{i++}(\beta)\right\} = 0$, then $\Var\left\{U_{ijk}(\beta)\right\} = E\left\{U_{ijk}(\beta)^2\right\}$. Let $\mu_j(s) = s_j^{(1)}(s;\beta)/s_j^{(0)}(s;\beta)$ be the ratio of the almost sure limits of $S_j^{(0)}(t;\beta)$ and $S_j^{(1)}(t;\beta)$, and $S_j^{(r)}(s;\beta) = n^{-1}\sum_{i=1}^n\sum_{k=1}^m \overline{Y}_{ijk}(s) Z_{ij}^r\exp\left(\beta Z_{ij}\right)$. Also let $M_{ijk}(s) = N_{ijk}(s) - \int_0^s \overline{Y}_{ijk}(u)\exp(\beta Z_{ij})\lambda_0(u)du$ be a martingale. Notice that this is the martingale with respect to the marginal filtration defined based on individual $k$ in cluster $i$ during period $j$, but not a martingale for the joint filtration due to the intracluster correlations. Given these definitions and utilizing iterated expectations, we may expand the scalar variance expression as:
\begin{align*}
    E\left\{U_{ijk}(\beta)^2\right\} &= E\left\{\left[\int_0^{C^*} \overline{Y}_{ijk}(s) \left\{Z_{ij} - \mu_j(s)\right\}dM_{ijk}(s)\right]^2\right\}\\
    &= E\left\{\int_0^{C^*} \overline{Y}_{ijk}(s) \left[Z_{ij} - \mu_j(s)\right]^2 \lambda_{ijk}(s) ds\right\}\\
    &= E_{Z_{ij}}\left\{E_{Y_{ijk}(s)|Z_{ij}}\left(E_{Y^{\dagger}_{ijk}(s) | Y_{ijk}(s), Z_{ij}}\left[\int_0^{C^*} \overline{Y}_{ijk}(s) \left[Z_{ij} - \mu_j(s)\right]^2 \lambda_{ijk}(s) ds\right]\right)\right\}\\
    &= E_{Z_{ij}}\left\{E_{Y_{ijk}(s)|Z_{ij}}\left(\int_0^{C^*} \mathscr{G}(s)Y_{ijk}(s) \left[Z_{ij} - \mu_j(s)\right]^2 \lambda_{ijk}(s) ds\right)\right\}\\
    &= E_{Z_{ij}}\left\{\int_0^{C^*} \mathscr{G}(s)P\left(T_{ijk} \ge s | Z_{ij}\right) \left[Z_{ij} - \mu_j(s)\right]^2 \lambda_{ijk}(s) ds\right\}\\
    &= E_{Z_{ij}}\left\{\int_0^{C^*} \mathscr{G}(s) \left[Z_{ij} - \mu_j(s)\right]^2 f(s|Z_{ij}) ds\right\},
\end{align*}
where $E_{Z_{ij}}\{\cdot\}$ is the expectation with respect to treatment at study period $j$, $\mathscr{G}(s)$ is the marginal survival function for the censoring time $C_{ijk}$, and $f(s|Z_{ij})$ is the conditional density of event time $T_{ijk}$.\\

\noindent We may now derive $\Cov\left\{U_{ijk}(\beta),U_{ijd}(\beta)\right\}$ when $k \ne d$. Again, as $E\left\{U_{i++}(\beta)\right\} = 0$, we may write $\Cov\left\{U_{ijk}(\beta),U_{ijd}(\beta)\right\} = E\left\{U_{ijk}(\beta)U_{ijd}(\beta)\right\}$. We may expand this as:
    $$E\left\{U_{ijk}(\beta)U_{ijd}(\beta)\right\} = E\left\{\int\int_{(0,C^*]^2} \overline{Y}_{ijk}(s)\overline{Y}_{ijd}(t) \left[Z_{ij} - \mu_j(s)\right]\left[Z_{ij} - \mu_j(t)\right]dM_{ijk}(s)dM_{ijd}(t)\right\}.$$
As the derivatives of the martingales are not squared, they do not simplify as in the univariate variance. Thus,
\begin{align*}
dM_{ijk}(s)dM_{ijd}(t) = &dN_{ijk}(s)dN_{ijd}(t) - dN_{ijk}(s)\overline{Y}_{ijd}(t)d\Lambda_{ijd}(t)\\
&- \overline{Y}_{ijk}(s)d\Lambda_{ijk}(s)dN_{ijd}(t) - \overline{Y}_{ijk}(s)\overline{Y}_{ijd}(t)d\Lambda_{ijk}(s)d\Lambda_{ijd}(t).
\end{align*}
Therefore, we can express the expectation as
\begin{align}
\begin{split}\label{app_eq:cov_breakdown}
   E\left\{U_{ijk}(\beta)U_{ijd}(\beta)\right\} = &E\left\{\int\int_{(0,C^*]^2} \overline{Y}_{ijk}(s)\overline{Y}_{ijd}(t) \left[Z_{ij} - \mu_j(s)\right]\left[Z_{ij} - \mu_j(t)\right]dN_{ijk}(s)dN_{ijd}(t)\right\}\\
   &-E\left\{\int\int_{(0,C^*]^2} \overline{Y}_{ijk}(s)\overline{Y}_{ijd}(t) \left[Z_{ij} - \mu_j(s)\right]\left[Z_{ij} - \mu_j(t)\right]dN_{ijk}(s)d\Lambda_{ijd}(t)\right\}\\
   &-E\left\{\int\int_{(0,C^*]^2} \overline{Y}_{ijk}(s)\overline{Y}_{ijd}(t) \left[Z_{ij} - \mu_j(s)\right]\left[Z_{ij} - \mu_j(t)\right]d\Lambda_{ijk}(s)dN_{ijd}(t)\right\}\\
   &+E\left\{\int\int_{(0,C^*]^2} \overline{Y}_{ijk}(s)\overline{Y}_{ijd}(t) \left[Z_{ij} - \mu_j(s)\right]\left[Z_{ij} - \mu_j(t)\right]d\Lambda_{ijk}(s)d\Lambda_{ijd}(t)\right\}.
\end{split}
\end{align}

\noindent Similar to $E\left\{U_{ijk}(\beta)^2\right\}$, we must break each of the terms in \eqref{app_eq:cov_breakdown} into iterated expectations. For the first term, it may be computed as:
\begin{align*}
    E&\left\{\int\int_{(0,C^*]^2} \overline{Y}_{ijk}(s)\overline{Y}_{ijd}(t) \left[Z_{ij} - \mu_j(s)\right]\left[Z_{ij} - \mu_j(t)\right]dN_{ijk}(s)dN_{ijd}(t)\right\}\\
    &= E_{Z_{ij}}\left\{E_{Y_{ijk}(s), Y_{ijd}(t)| Z_{ij}}\left(E_{dN_{ijk}(s),dN_{ijd}(t)|Y_{ijk}(s), Y_{ijd}(t), Z_{ij}}\left[E_{Y^{\dagger}_{ijk}(s),Y^{\dagger}_{ijk}(t)| Z_{ij}, Y_{ijk}(s), Y_{ijd}(t), dN_{ijk}(s),dN_{ijd}(t)}\right.\right.\right.\\
    &\hspace{1cm}\left.\left.\left.\left\{\int\int_{(0,C^*]^2} \overline{Y}_{ijk}(s)\overline{Y}_{ijd}(t) \left[Z_{ij} - \mu_j(s)\right]\left[Z_{ij} - \mu_j(t)\right]dN_{ijk}(s)dN_{ijd}(t)\right\}\right]\right)\right\}\\
    &=E_{Z_{ij}}\left\{E_{Y_{ijk}(s), Y_{ijd}(t)| Z_{ij}}\left(E_{dN_{ijk}(s),dN_{ijd}(t)|Y_{ijk}(s), Y_{ijd}(t), Z_{ij}}\right.\right.\\
    &\hspace{1cm}\left.\left.\left[\int\int_{(0,C^*]^2} \mathscr{G}(s,t)Y_{ijk}(s)Y_{ijd}(t) \left[Z_{ij} - \mu_j(s)\right]\left[Z_{ij} - \mu_j(t)\right]dN_{ijk}(s)dN_{ijd}(t)\right]\right)\right\}\\
    &=E_{Z_{ij}}\left\{E_{Y_{ijk}(s), Y_{ijd}(t)| Z_{ij}}\left(\int\int_{(0,C^*]^2} \mathscr{G}(s,t)Y_{ijk}(s)Y_{ijd}(t) \left[Z_{ij} - \mu_j(s)\right]\left[Z_{ij} - \mu_j(t)\right]\right.\right.\\
    &\hspace{7cm}\left.\left.\times P(T_{ijk}=s, T_{ijd}=t|Y_{ijk}(s), Y_{ijd}(t), Z_{ij})dsdt\right)\right\}\\
    &=E_{Z_{ij}}\left\{\int\int_{(0,C^*]^2} \mathscr{G}(s,t)\left[Z_{ij} - \mu_j(s)\right]\left[Z_{ij} - \mu_j(t)\right]f(s,t|Z_{ij})dsdt\right\},
\end{align*}
where $f(s,t|Z_{ij})$ is the pairwise conditional density for $(T_{ijk}, T_{ijd})$.\\

\noindent For the second term in \eqref{app_eq:cov_breakdown}, it may be computed as:
\begin{align*}
    E&\left\{\int\int_{(0,C^*]^2} \overline{Y}_{ijk}(s)\overline{Y}_{ijd}(t) \left[Z_{ij} - \mu_j(s)\right]\left[Z_{ij} - \mu_j(t)\right]dN_{ijk}(s)d\Lambda_{ijd}(t)\right\}\\
    &= E_{Z_{ij}}\left\{E_{Y_{ijk}(s), Y_{ijd}(t)| Z_{ij}}\left(E_{dN_{ijk}(s)|Y_{ijk}(s), Y_{ijd}(t), Z_{ij}}\left[E_{Y^{\dagger}_{ijk}(s),Y^{\dagger}_{ijk}(t)| Z_{ij}, Y_{ijk}(s), Y_{ijd}(t), dN_{ijk}(s)}\right.\right.\right.\\
    &\hspace{1cm}\left.\left.\left.\left\{\int\int_{(0,C^*]^2} \overline{Y}_{ijk}(s)\overline{Y}_{ijd}(t) \left[Z_{ij} - \mu_j(s)\right]\left[Z_{ij} - \mu_j(t)\right]dN_{ijk}(s)d\Lambda_{ijd}(t)\right\}\right]\right)\right\}\\
    &= E_{Z_{ij}}\left\{E_{Y_{ijk}(s), Y_{ijd}(t)| Z_{ij}}\left(E_{dN_{ijk}(s)|Y_{ijk}(s), Y_{ijd}(t), Z_{ij}}\right.\right.\\
    &\hspace{1cm}\left.\left.\left[\int\int_{(0,C^*]^2} \mathscr{G}(s,t)Y_{ijk}(s)Y_{ijd}(t) \left[Z_{ij} - \mu_j(s)\right]\left[Z_{ij} - \mu_j(t)\right]dN_{ijk}(s)d\Lambda_{ijd}(t)\right]\right)\right\}\\
    % &= E_{Z_{ij}}\left\{E_{Y_{ijk}(s), Y_{ijd}(t)| Z_{ij}}\left(\int\int_{(0,C^*]^2} \mathscr{G}(s,t)Y_{ijk}(s)Y_{ijd}(t) \left\{Z_{ij} - \mu_j(s)\right\}\left\{Z_{ij} - \mu_j(t)\right\}\right.\right.\\
    % &\left.\left.\hspace{7cm}\times P\left(T_{ijk}=s, T_{ijd}=t|Z_{ij}\right)\lambda_{ijd}(t)dsdt\right)\right\}\\
    &= E_{Z_{ij}}\left\{\int\int_{(0,C^*]^2} \mathscr{G}(s,t)\left[Z_{ij} - \mu_j(s)\right]\left[Z_{ij} - \mu_j(t)\right] \frac{-\partial \mathscr{F}(s,t|Z_{ij})}{\partial s}\lambda_{ijd}(t)dsdt\right\},
\end{align*}
where $\mathscr{F}(s,t|Z_{ij})$ is the pairwise conditional survival function for $(T_{ijk}, T_{ijd})$, given the treatment status $Z_{ij}$.\\

\noindent Similarly, the third term of \eqref{app_eq:cov_breakdown} can be expressed as:
\begin{align*}
    E&\left\{\int\int_{(0,C^*]^2} \overline{Y}_{ijk}(s)\overline{Y}_{ijd}(t) \left[Z_{ij} - \mu_j(s)\right]\left[Z_{ij} - \mu_j(t)\right]d\Lambda_{ijk}(s)dN_{ijd}(t)\right\}\\
    &=E_{Z_{ij}}\left\{\int\int_{(0,C^*]^2} \mathscr{G}(s,t)\left[Z_{ij} - \mu_j(s)\right]\left[Z_{ij} - \mu_j(t)\right] \frac{-\partial \mathscr{F}(s,t|Z_{ij})}{\partial t}\lambda_{ijk}(s)dsdt\right\}.
\end{align*}

\noindent Finally, the last term of \eqref{app_eq:cov_breakdown} can be computed as:
\begin{align*}
    E&\left\{\int\int_{(0,C^*]^2} \overline{Y}_{ijk}(s)\overline{Y}_{ijd}(t) \left[Z_{ij} - \mu_j(s)\right]\left[Z_{ij} - \mu_j(t)\right]d\Lambda_{ijk}(s)d\Lambda_{ijd}(t)\right\}\\
    &=E_{Z_{ij}}\left\{E_{Y_{ijk}(s), Y_{ijd}(t)| Z_{ij}}\left(E_{Y^{\dagger}_{ijk}(s),Y^{\dagger}_{ijk}(t)| Z_{ij}, Y_{ijk}(s), Y_{ijd}(t)}\right.\right.\\
    &\hspace{1cm}\left.\left.\left[\int\int_{(0,C^*]^2} \overline{Y}_{ijk}(s)\overline{Y}_{ijd}(t) \left[Z_{ij} - \mu_j(s)\right]\left[Z_{ij} - \mu_j(t)\right]d\Lambda_{ijk}(s)d\Lambda_{ijd}(t)\right]\right)\right\}\\
    &=E_{Z_{ij}}\left\{E_{Y_{ijk}(s), Y_{ijd}(t)| Z_{ij}}\right.\\
    &\hspace{1cm}\left.\left(\int\int_{(0,C^*]^2} \mathscr{G}(s,t)Y_{ijk}(s)Y_{ijd}(t) \left[Z_{ij} - \mu_j(s)\right]\left[Z_{ij} - \mu_j(t)\right]\lambda_{ijk}(s)\lambda_{ijd}(t)\right)\right\}\\
    &=E_{Z_{ij}}\left\{\int\int_{(0,C^*]^2} \mathscr{G}(s,t) \left[Z_{ij} - \mu_j(s)\right]\left[Z_{ij} - \mu_j(t)\right]\mathscr{F}(s,t|Z_{ij})\lambda_{ijk}(s)\lambda_{ijd}(t)\right\}
\end{align*}

\noindent The derivation of the third component of $\Var\left\{U_{i++}(\beta)\right\}$,$\Cov\left\{U_{ijk}(\beta),U_{ild}(\beta)\right\}$ when $j \ne l$ and $k$ may be equal to $d$, is found in a similar manner and can be broken into four main terms:
\begin{align}
\begin{split}\label{app_eq:cov_betweePeriods_breakdown}
   E\left\{U_{ijk}(\beta)U_{ild}(\beta)\right\} = &E\left\{\int\int_{(0,C^*]^2} \overline{Y}_{ijk}(s)\overline{Y}_{ild}(t) \left[Z_{ij} - \mu_j(s)\right]\left[Z_{il} - \mu_l(t)\right]dN_{ijk}(s)dN_{ild}(t)\right\}\\
   &-E\left\{\int\int_{(0,C^*]^2} \overline{Y}_{ijk}(s)\overline{Y}_{ild}(t) \left[Z_{ij} - \mu_j(s)\right]\left[Z_{il} - \mu_l(t)\right]dN_{ijk}(s)d\Lambda_{ild}(t)\right\}\\
   &-E\left\{\int\int_{(0,C^*]^2} \overline{Y}_{ijk}(s)\overline{Y}_{ild}(t) \left[Z_{ij} - \mu_j(s)\right]\left[Z_{il} - \mu_l(t)\right]d\Lambda_{ijk}(s)dN_{ild}(t)\right\}\\
   &+E\left\{\int\int_{(0,C^*]^2} \overline{Y}_{ijk}(s)\overline{Y}_{ild}(t) \left[Z_{ij} - \mu_j(s)\right]\left[Z_{il} - \mu_l(t)\right]d\Lambda_{ijk}(s)d\Lambda_{ild}(t)\right\}.
\end{split}
\end{align}
The primary difference is that expectations must be taken with respect to study periods $j$ and $l$ instead of only period $j$. Thus, following the proof for \eqref{app_eq:cov_breakdown}, the four terms in \eqref{app_eq:cov_betweePeriods_breakdown} may be expressed as:
\begin{align*}
    E\left\{U_{ijk}(\beta)U_{ild}(\beta)\right\} = &E_{Z_{ij}, Z_{il}}\left\{\int\int_{(0,C^*]^2} \mathscr{G}(s,t)\left[Z_{ij} - \mu_j(s)\right]\left[Z_{il} - \mu_l(t)\right]f(s,t|Z_{ij}, Z_{il})dsdt\right\}\\
    &-E_{Z_{ij}, Z_{il}}\left\{\int\int_{(0,C^*]^2} \mathscr{G}(s,t)\left[Z_{ij} - \mu_j(s)\right]\left[Z_{il} - \mu_l(t)\right]\frac{-\partial \mathscr{F}(s,t|Z_{ij}, Z_{il})}{\partial s} \lambda_{ild}(t)dsdt\right\}\\
    &-E_{Z_{ij}, Z_{il}}\left\{\int\int_{(0,C^*]^2} \mathscr{G}(s,t)\left[Z_{ij} - \mu_j(s)\right]\left[Z_{il} - \mu_l(t)\right]\frac{-\partial \mathscr{F}(s,t|Z_{ij}, Z_{il})}{\partial t} \lambda_{ijk}(s)dsdt\right\}\\
    &+E_{Z_{ij}, Z_{il}}\left\{\int\int_{(0,C^*]^2} \mathscr{G}(s,t)\left[Z_{ij} - \mu_j(s)\right]\left[Z_{il} - \mu_l(t)\right]\mathscr{F}(s,t|Z_{ij}, Z_{il}) \lambda_{ijk}(s)\lambda_{ild}(t)dsdt\right\}.\\   
\end{align*}

\noindent Thus, because $E\left\{U_{i++}(\beta)\right\}=0$, we can write 
$\Var\{U_{ijk}(\beta)\}=E\{U_{ijk}(\beta)^2\}=E_{Z_{ij}}\{q_0(Z_{ij})\}$, where $q_0(Z_{ij}) = \int_0^{C^*}\mathscr{G}(s)\left\{Z_{ij} - \mu_j(s)\right\}^2 f(s|Z_{ij})ds$. 
%$\Var\{U_{ijk}({\beta})\}=E\{U_{ijk}({\beta})^2\}$, which can be expressed as $E_{Z_{ij}}\{q_0(j)\}$. 
%where again $E_{Z_{ij}}\{\cdotb\}$ is the expectation with respect to treatment at study period $j$, $W_j(s) = \frac{S^{(1)}_j(s;\hat{\beta})}{S^{(0)}_j(s;\hat{\beta})}$, $\mathscr{G}(s)$ is the survival function for the censoring time $C_{ijk}$, and $f(s|Z_{ij})$ is the conditional density of event time $T_{ijk}$.
Similarly, using the above derivations, the covariance term can be expanded as the sum of four expectations,
\begin{align*}
\Cov\{U_{ijk}(\beta), U_{ild}(\beta)\}%=
%E\{U_{ijk}(\beta)U_{ild}(\beta)\}\\
=E_{Z_{ij}Z_{il}}\left\{q_1(Z_{ij},Z_{il})+q_2(Z_{ij},Z_{il})+q_3(Z_{ij},Z_{il})+q_4(Z_{ij},Z_{il})\right\}  
\end{align*}
for any two study periods $\{j,l\}$ and any two individuals $\{k,d\}$ belonging to the same cluster $i$. These terms are defined as
%This covariance can further be expressed as the summation of four separate expectations:
% \begin{enumerate}[leftmargin=1cm,label=(\roman*)]
%     \item $E_{Z_{ij},Z_{il}}\left\{\int_0^{C_{ijk}}\int_0^{C_{ild}}\mathscr{G}(s,t) \left[Z_{ij} - W_j(s)\right]\left[Z_{il} - W_l(t)\right] f(s,t|Z_{ij},Z_{il}) dt ds\right\}$
%     \item $-E_{Z_{ij},Z_{il}}\left\{\int_0^{C_{ijk}}\int_0^{C_{ild}}\mathscr{G}(s,t) \left[Z_{ij} - W_j(s)\right]\left[Z_{il} - W_l(t)\right] \displaystyle\frac{-\partial \mathscr{F}(s,t|Z_{ij},Z_{il})}{\partial t}\lambda_{ijk}(s) dt ds\right\}$
%     \item $-E_{Z_{ij},Z_{il}}\left\{\int_0^{C_{ijk}}\int_0^{C_{ild}}\mathscr{G}(s,t) \left[Z_{ij} - W_j(s)\right]\left[Z_{il} - W_l(t)\right] \displaystyle\frac{-\partial \mathscr{F}(s,t|Z_{ij},Z_{il})}{\partial s}\lambda_{ild}(t) dt ds\right\}$
%     \item $E_{Z_{ij},Z_{il}}\left\{\int_0^{C_{ijk}}\int_0^{C_{ild}}\mathscr{G}(s,t) \left[Z_{ij} - W_j(s)\right]\left[Z_{il} - W_l(t)\right] \mathscr{F}(s,t|Z_{ij},Z_{il})\lambda_{ijk}(s)\lambda_{ild}(t) dt ds\right\}$,
% \end{enumerate}
\begin{align*}
    q_1(Z_{ij},Z_{il}) &= \int\int_{(0,C^*]^2}\mathscr{G}(s,t) \left\{Z_{ij} - \mu_j(s)\right\}\left\{Z_{il} - \mu_l(t)\right\} f(s,t|Z_{ij},Z_{il}) dt ds\\
    q_2(Z_{ij},Z_{il}) &= -\int\int_{(0,C^*]^2}\mathscr{G}(s,t) \left\{Z_{ij} - \mu_j(s)\right\}\left\{Z_{il} - \mu_l(t)\right\} \frac{-\partial \mathscr{F}(s,t|Z_{ij},Z_{il})}{\partial t}\lambda_{ijk}(s) dt ds\\
    q_3(Z_{ij},Z_{il}) &= -\int\int_{(0,C^*]^2}\mathscr{G}(s,t) \left\{Z_{ij} - \mu_j(s)\right\}\left\{Z_{il} - \mu_l(t)\right\} \frac{-\partial \mathscr{F}(s,t|Z_{ij},Z_{il})}{\partial s}\lambda_{ild}(t) dt ds\\
    q_4(Z_{ij},Z_{il}) &= \int\int_{(0,C^*]^2}\mathscr{G}(s,t) \left\{Z_{ij} - \mu_j(s)\right\}\left\{Z_{il} - \mu_l(t)\right\} \mathscr{F}(s,t|Z_{ij},Z_{il})\lambda_{ijk}(s)\lambda_{ild}(t) dt ds,
\end{align*}
where $E_{Z_{ij},Z_{il}}\{\cdotb\}$ is the expectation with respect to joint distribution of the treatment variables at study periods $j$ and $l$, $\mathscr{G}(s,t)$ is the bivariate survival function for the censoring times $(C_{ijk}, C_{ild})$, and $f(s,t|Z_{ij},Z_{il})$ and $\mathscr{F}(s,t|Z_{ij},Z_{il})$ are the bivariate conditional density and survival functions for $(T_{ijk},T_{ild})$ given levels of the treatment status, respectively.

%% ALLOCATION PROBABILITIES %%
\addcontentsline{toc}{subsection}{Sequence Allocation Probabilities}
\subsection*{Sequence Allocation Probabilities}
We note that $E_{Z_{ij}}\{\cdotb\}$ and $E_{Z_{ij},Z_{il}}\{\cdotb\}$ depend on the sequence allocation. With $J$ time periods, a cluster $i$ may be assigned to a treatment sequence with probability $\pi_b$, where $\sum_{b=1}^{(J-1)}\pi_b = 1$ and $\pi_0 = 0$. Thus, $\sum_{b=0}^{(j-1)}\pi_b$ is equal to the proportion of clusters on treatment at period $j$. From the law of total expectations, we can explicitly write $E_{Z_{ij}}\{q_0(z_{ij})\} = \sum_{b=0}^{(j-1)}\pi_b q_0(Z_{ij}=1)+ \left(1-\sum_{b=0}^{(j-1)}\pi_b\right)q_0(Z_{ij}=0)$.\\

\noindent For the components in the covariance expression that depend on the joint distribution of two treatment variables, there are four joint probabilities based on all combinations of $\{z_{ij},z_{il}\}$, given by:
\begin{enumerate}
    \item $P(Z_{ij}=1, Z_{il}=1)=\mathbb{I}\left(\min(j,l) >1\right)\sum_{b=0}^{\min{(j,l)}-1}\pi_b$
    \item $P(Z_{ij}=0, Z_{il}=1)=\mathbb{I}\left(\max{(j,l)} >1\right)\mathbb{I}\left(j>l\right)\sum_{b=l}^{j-1}\pi_b$
    \item $P(Z_{ij}=1, Z_{il}=0)=\mathbb{I}\left(\max{(j,l)} >1\right)\mathbb{I}\left(j<l\right)\sum_{b=j}^{l-1}\pi_b$
    \item $P(Z_{ij}=0, Z_{il}=0)=1-\sum_{b=0}^{\max{(j,l)}-1}\pi_b$
\end{enumerate}

%% EQUIVALNCE PROOF %%
\addcontentsline{toc}{subsection}{Equivalence of $\Upsilon_0(j)$ and $\sum_{a=0}^1P(Z_{ij}=a)\nu(Z_{ij}=a)$}
\subsection*{Equivalence of $\Upsilon_0(j)$ and $\sum_{a=0}^1P(Z_{ij}=a)\nu(Z_{ij}=a)$}
We will show that when the model is correctly specified, $\Upsilon_0(j)$ is equivalent to $E_{Z_{ij}}\left\{\nu(z_{ij})\right\} = \sum_{a=0}^1P(Z_{ij}=a)\nu(Z_{ij}=a)$ and thus, when there is no covariation between survival times (i.e., no within- or between-period correlation), that $\Var(\hat{\beta}) = \left\{nm\sum_{j=1}^J \Upsilon_0(j)\right\}^{-1}$. To do this we will first show that $\Upsilon_0(j) = E\left\{U^2_{ijk}(\beta)\right\} = \sum_{a=0}^1P(Z_{ij}=a)\nu(Z_{ij}=a)$ at a particular period $j$. Then will we show that $n^{-1}\sum_{i=1}^n\Var\{U_{i++}(\beta)\} = m\sum_{j=1}^JE\left\{U^2_{ijk}(\beta)\right\} = m\sum_{j=1}^J\Upsilon_0(j) = m\sum_{j=1}^J\sum_{a=0}^1P(Z_{ij}=a)\nu(Z_{ij}=a)$ when survival times within and between cluster-periods are independent.\\

\noindent First recall, given independent clusters, that
$$A(\beta) = E\left\{-{\partial U_{i++}(\beta)}/{\partial \beta}\right\}= \sum_{j=1}^J E_{Z_{ij}}\left\{\sum_{k=1}^m\nu(z_{ij})\right\} =  \sum_{j=1}^J \sum_{k=1}^mE_{Z_{ij}}\left\{\int_0^{C^*}\mathscr{G}(s)\mu_j(s)\left[1-\mu_j(s)\right]f(t|Z_{ij})ds\right\}$$
and
$$\Upsilon_0(j) = E_{Z_{ij}}\left\{\int_0^{C^*}\mathscr{G}(s)\left[Z_{ij}-\mu_j(s)\right]^2f(s|Z_{ij})ds\right\},$$
where 
$$\mu_j(s) = \frac{s_j^{(1)}(s;\beta)}{s_j^{(0)}(s;\beta)} = \frac{E\left\{\sum_{k=1}^m \overline{Y}_{ijk}(s)Z_{ij}\exp (\beta Z_{ij})\right\}}{E\left\{\sum_{k=1}^m \overline{Y}_{ijk}(s)\exp (\beta Z_{ij})\right\}}.$$
We also note that $s_j^{(0)}(s;\beta)$ and $s_j^{(1)}(s;\beta)$ are the almost sure limits of $S_j^{(0)}(s;\beta)$ and $S_j^{(1)}(s;\beta)$, respectively.\\

\noindent We can expand each component of $\mu_j(s)$:
\begin{align*}
    s_j^{(0)} &= E\left\{\sum_{k=1}^m \overline{Y}_{ijk}(s)\exp (\beta Z_{ij})\right\}\\
    &= E\left\{m Y_{ijk}(t)Y^\dagger_{ijk}(t)\exp (\beta Z_{ij})\right\}\\
    &= m\mathscr{G}(s)E_{Z_{ij}}\left\{\mathscr{F}_j(s|Z_{ij})\exp (\beta Z_{ij})\right\}\\
    &= m\mathscr{G}(s) \left\{\left(\sum_{b=0}^{j-1}\pi_b\right)\mathscr{F}_j(s|Z_{ij}=1)\exp (\beta) + \left(1-\sum_{b=0}^{j-1}\pi_b\right)\mathscr{F}_j(s|Z_{ij}=0)\right\},
\end{align*}
and
\begin{align*}
    s_j^{(1)} &= E\left\{\sum_{k=1}^m \overline{Y}_{ijk}(s)Z_{ij}\exp (\beta Z_{ij})\right\}\\
    &= E\left\{m Y_{ijk}(t)Y^\dagger_{ijk}(t)Z_{ij}\exp (\beta Z_{ij})\right\}\\
        &= m\mathscr{G}(s)E_{Z_{ij}}\left\{\mathscr{F}_j(s|Z_{ij})Z_{ij}\exp (\beta Z_{ij})\right\}\\
    &= m\mathscr{G}(s) \left\{\left(\sum_{b=0}^{j-1}\pi_b\right)\mathscr{F}_j(s|Z_{ij}=1)\exp (\beta)\right\}.
\end{align*}

\noindent Thus, $\mu_j(s)$ can be re-expressed as:
$$\mu_j(s) = \frac{\left\{\left(\sum_{b=0}^{j-1}\pi_b\right)\mathscr{F}_j(s|Z_{ij}=1)\exp (\beta)\right\}}{\left\{\left(\sum_{b=0}^{j-1}\pi_b\right)\mathscr{F}_j(s|Z_{ij}=1)\exp (\beta) + \left(1-\sum_{b=0}^{j-1}\pi_b\right)\mathscr{F}_j(s|Z_{ij}=0)\right\}},$$
and that
$$1-\mu_j(s) = \frac{\left\{\left(1-\sum_{b=0}^{j-1}\pi_b\right)\mathscr{F}_j(s|Z_{ij}=0)\right\}}{\left\{\left(\sum_{b=0}^{j-1}\pi_b\right)\mathscr{F}_j(s|Z_{ij}=1)\exp (\beta) + \left(1-\sum_{b=0}^{j-1}\pi_b\right)\mathscr{F}_j(s|Z_{ij}=0)\right\}}.$$

\noindent Substituting $\mu_j(s)$ and $1-\mu_j(s)$ into $\Upsilon_0(j)$, we get:
\begin{align*}
    \Upsilon_0(j) &= E_{Z_{ij}}\left\{\int_0^{C^*}\mathscr{G}(s)\left[Z_{ij}-\mu_j(s)\right]^2f(s|Z_{ij})ds\right\}\\
    &= \left(\sum_{b=0}^{j-1}\pi_b\right)\int_0^{C^*}\mathscr{G}(s)\left[1-\mu_j(s)\right]^2f(s|Z_{ij}=1)ds + \left(1-\sum_{b=0}^{j-1}\pi_b\right)\int_0^{C^*}\mathscr{G}(s)\left[-\mu_j(s)\right]^2f(s|Z_{ij}=0)ds\\
    &=\left(\sum_{b=0}^{j-1}\pi_b\right)\int_0^{C^*}\mathscr{G}(s)\left[1-\mu_j(s)\right]^2\mathscr{F}_j(s|Z_{ij}=1)\lambda_{0j}(s)\exp (\beta) ds\\
    &\hspace{1.5em}+ \left(1-\sum_{b=0}^{j-1}\pi_b\right)\int_0^{C^*}\mathscr{G}(s)\left[\mu_j(s)\right]^2\mathscr{F}_j(s|Z_{ij}=1)\lambda_{0j}(s)ds\\
    &= \int_0^{C^*}\mathscr{G}(s) \lambda_{0j}(s) \left\{\left(\sum_{b=0}^{j-1}\pi_b\right)\mathscr{F}_j(s|Z_{ij}=1)\exp (\beta) \times \frac{\left[\left(1-\sum_{b=0}^{j-1}\pi_b\right)\mathscr{F}_j(s|Z_{ij}=0)\right]^2}{\left[\left(\sum_{b=0}^{j-1}\pi_b\right)\mathscr{F}_j(s|Z_{ij}=1)\exp (\beta) + \left(1-\sum_{b=0}^{j-1}\pi_b\right)\mathscr{F}_j(s|Z_{ij}=0)\right]^2}\right.\\
    &\hspace{8.5em}\left. + \left(1-\sum_{b=0}^{j-1}\pi_b\right)\mathscr{F}_j(s|Z_{ij}=0)\times \frac{\left[\left(\sum_{b=0}^{j-1}\pi_b\right)\mathscr{F}_j(s|Z_{ij}=1)\exp (\beta)\right]^2}{\left[\left(\sum_{b=0}^{j-1}\pi_b\right)\mathscr{F}_j(s|Z_{ij}=1)\exp (\beta) + \left(1-\sum_{b=0}^{j-1}\pi_b\right)\mathscr{F}_j(s|Z_{ij}=0)\right]^2}\right\}ds\\
    &= \int_0^{C^*}\mathscr{G}(s) \lambda_{0j}(s) \left\{\frac{\left[\left(\sum_{b=0}^{j-1}\pi_b\right)\mathscr{F}_j(s|Z_{ij}=1)\exp (\beta)\right] \times \left[\left(1-\sum_{b=0}^{j-1}\pi_b\right)\mathscr{F}_j(s|Z_{ij}=0)\right]}{\left[\left(\sum_{b=0}^{j-1}\pi_b\right)\mathscr{F}_j(s|Z_{ij}=1)\exp (\beta) + \left(1-\sum_{b=0}^{j-1}\pi_b\right)\mathscr{F}_j(s|Z_{ij}=0)\right]}\right\}ds.
\end{align*}

\noindent Performing the same substitution for $E_{Z{ij}}\left\{\nu(Z_{ij})\right\}$ gives us:
\begin{align*}
     E_{Z{ij}}\left\{\nu(Z_{ij})\right\} &= E_{Z_{ij}}\left\{\int_0^{C^*}\mathscr{G}(s)\mu_j(s)\left[1-\mu_j(s)\right]f(t|Z_{ij})ds\right\}\\
     &= \left(\sum_{b=0}^{j-1}\pi_b\right)\int_0^{C^*}\mathscr{G}(s)\mu_j(s)\left[1-\mu_j(s)\right]f(t|Z_{ij}=1)ds + \left(1-\sum_{b=0}^{j-1}\pi_b\right)\int_0^{C^*}\mathscr{G}(s)\mu_j(s)\left[1-\mu_j(s)\right]f(t|Z_{ij}=0)ds\\
     &=\left(\sum_{b=0}^{j-1}\pi_b\right)\int_0^{C^*}\mathscr{G}(s)\mu_j(s)\left[1-\mu_j(s)\right]\mathscr{F}_j(s|Z_{ij}=1)\lambda_{0j}(s)\exp (\beta)ds\\
     &\hspace{1.5em}+ \left(1-\sum_{b=0}^{j-1}\pi_b\right)\int_0^{C^*}\mathscr{G}(s)\mu_j(s)\left[1-\mu_j(s)\right]\mathscr{F}_j(s|Z_{ij}=0)\lambda_{0j}(s)ds\\
     &=\int_0^{C^*}\mathscr{G}(s)\lambda_{0j}(s) \left\{\mu_j(s)[1-\mu_j(s)]\times \left[\left(\sum_{b=0}^{j-1}\pi_b\right) \mathscr{F}_j(s|Z_{ij}=1)\exp (\beta) + \left(1-\sum_{b=0}^{j-1}\pi_b\right)\mathscr{F}_j(s|Z_{ij}=0)\right]\right\}ds\\
     &= \int_0^{C^*}\mathscr{G}(s)\lambda_{0j}(s)\left\{\frac{\left[\left(\sum_{b=0}^{j-1}\pi_b\right)\mathscr{F}_j(s|Z_{ij}=1)\exp (\beta)\right] \times\left[\left(1-\sum_{b=0}^{j-1}\pi_b\right)\mathscr{F}_j(s|Z_{ij}=0)\right]}{\left[\left(\sum_{b=0}^{j-1}\pi_b\right)\mathscr{F}_j(s|Z_{ij}=1)\exp (\beta) + \left(1-\sum_{b=0}^{j-1}\pi_b\right)\mathscr{F}_j(s|Z_{ij}=0)\right]^2}\right.\\
     &\hspace{9em}\left.\times\left[\left(\sum_{b=0}^{j-1}\pi_b\right) \mathscr{F}_j(s|Z_{ij}=1)\exp (\beta) + \left(1-\sum_{b=0}^{j-1}\pi_b\right)\mathscr{F}_j(s|Z_{ij}=0)\right]\right\}ds\\
     &=\int_0^{C^*}\mathscr{G}(s)\lambda_{0j}(s)\left\{\frac{\left[\left(\sum_{b=0}^{j-1}\pi_b\right)\mathscr{F}_j(s|Z_{ij}=1)\exp (\beta)\right]\times\left[\left(1-\sum_{b=0}^{j-1}\pi_b\right)\mathscr{F}_j(s|Z_{ij}=0)\right]}{\left[\left(\sum_{b=0}^{j-1}\pi_b\right)\mathscr{F}_j(s|Z_{ij}=1)\exp (\beta) + \left(1-\sum_{b=0}^{j-1}\pi_b\right)\mathscr{F}_j(s|Z_{ij}=0)\right]}\right\}ds
\end{align*}

\noindent After these substitutions, it is clear that $\Upsilon_0(j) = E_{Z{ij}}\left\{\nu(Z_{ij})\right\}$.\\

\noindent Now we will show that $n^{-1}\sum_{i=1}^n\Var\{U_{i++}(\beta)\} = m\sum_{j=1}^JE\left\{U^2_{ijk}(\beta)\right\} = m\sum_{j=1}^J\Upsilon_0(j)$ when survival times within and between cluster-periods are independent. First recall that it is given $E\{U_{i++}(\beta)\} = 0$. Thus $\Var\{U_{i++}(\beta)\} = \sum_{j=1}^J\sum_{l=1}^J\sum_{k=1}^m\sum_{d=1}^mE\{U_{ijk}(\beta)U_{ild}(\beta)\}$. This expectation can be broken down into three cases: $j=l$, $k=d$; $j\ne l$, $k$ may be equal to $d$; $j=l$, $k \ne d$. As we have shown the first case above, we will address the two remaining cases separately.\\

\noindent When $j \ne l$, recall that we assume there is no covariation between survival times such that $T_{ijk}\perp T_{ild}$. Thus,
$E\{U_{ijk}(\beta)U_{ild}(\beta)\} = E\{U_{ijk}(\beta)\}E\{U_{ild}(\beta)\}=0$. %\approx E\{U^2_{ijk}(\beta)\}\\%&= \Upsilon_0(j).
Similarly for the third case when $j=l$ but $k\ne d$, we again invoke the assumption that survival times within a cluster-period are independent such that $T_{ijk}\perp T_{ijd}$. Thus, $E\{U_{ijk}(\beta)U_{ijd}(\beta)\} = E\{U_{ijk}(\beta)\}E\{U_{ijd}(\beta)\}=0$. %\approx E\{U^2_{ijk}(\beta)\}\\%&= \Upsilon_0(j).
Therefore, $n^{-1}\sum_{i=1}^n\Var\{U_{i++}(\beta)\} = m\sum_{j=1}^J\sum_{j=1}^JE\left\{U^2_{ijk}(\beta)\right\} = m\sum_{j=1}^J\Upsilon_0(j)$.\\

%% Then show that when there's no covariation, var(beta) = (nm Upsilon_0(beta))^-1 %%
\noindent Similar to previously, we note that $A(\beta) = \sum_{j=1}^J E\left\{\sum_{k=1}^m\nu(Z_{ij})\right\} = n^{-1}\sum_{i=1}^n\sum_{j=1}^J\sum_{k=1}^m \Upsilon_0(j)$. %Thus, 
% \begin{align*}
%     A(\beta) &= m\sum_{j=1}^J\sum_{a=0}^1P(Z_{ij}=a)\nu(Z_{ij}=a)\\
%     &= n^{-1}\sum_{i=1}^n E\left\{-\frac{\partial^2 l(t;\beta)}{\partial \beta^2}\right\}\\
%     &= n^{-1}\sum_{i=1}^n E\left\{\left(\frac{\partial l(t; \beta)}{\partial \beta}\right)\right\}\\
%     &= m\sum_{j=1}^J\sum_{j=1}^JE\left\{U^2_{ijk}(\beta)\right\}\\
%     &= m\sum_{j=1}^J\Upsilon_0(j)
% \end{align*}
Therefore, $\Var(\hat{\beta})=A^{-1}(\beta)B(\beta)A^{-1}(\beta) = \left\{nm\sum_{j=1}^J \Upsilon_0(j)\right\}^{-1}$.%$\Var(\hat{\beta})= n^{-1}\sum_{i=1}^n\Var\{U_{i++}(\beta)\}=\left\{nm\sum_{j=1}^J \Upsilon_0(j)\right\}^{-1}$.

%% THEOREM 1 %%
\addcontentsline{toc}{subsection}{THEOREM 1: Derivation of $\Var({\hat{\beta}})$}
\subsection*{THEOREM 1: Derivation of $\Var({\hat{\beta}})$}
\emph{Theorem 1}: Assuming known survival and censoring distributions and correct model specification, the variance of the treatment effect estimator based on a period-stratified Cox proportional hazards model is
\begin{equation}\label{eq:var-robust2}
    \Var(\hat{\beta})=\left\{nm\sum_{j=1}^J \Upsilon_0(j)\right\}^{-1}\times \left\{1 + (m-1)\rho_w + m(J-1)\rho_b\right\},
\end{equation}
where $\rho_w = \displaystyle \frac{\sum_{j=1}^J \Upsilon_1(j,j)}{\sum_{j=1}^J \Upsilon_0(j)}$ and  $\rho_b = \displaystyle \frac{\mathop{\sum_{j=1}^J \sum_{l=1}^J}_{j\ne l} \Upsilon_1(j,l)}{(J-1)\sum_{j=1}^J \Upsilon_0(j)}$.\\
% Assuming known survival and censoring distributions, the variance of the treatment effect estimator based on a period-stratified Cox proportional hazards model is
% \begin{equation}\label{eq:var-robust}
%     \Var(\hat{\beta})=\left\{nm\sum_{j=1}^J \Upsilon_0(j)\right\}^{-1}\times \left\{1 + (m-1)\rho_w + m(J-1)\rho_b\right\},
%     %\Var(\hat{\beta})=\frac{\sum_{j=1}^J \Upsilon_0(j)}{nm\left\{\sum_{j=1}^J \Upsilon_2(j)\right\}^2}\times \left\{1 + (m-1)\rho_w + m(J-1)\rho_b\right\},
% \end{equation}
% where $\rho_w = \frac{\sum_{j=1}^J \Upsilon_1(j,j)}{\sum_{j=1}^J \Upsilon_0(j)}$ and  $\rho_b = \frac{\mathop{\sum_{j=1}^J \sum_{l=1}^J}_{j\ne l} \Upsilon_1(j,l)}{(J-1)\sum_{j=1}^J \Upsilon_0(j)}$.\\

\noindent\textbf{Proof:} Recall that sandwich variance of \citet{lin_cox_1994} takes the form $A^{-1}(\beta)B(\beta)A^{-1}(\beta)$, where $A^{-1}(\beta) = E\left\{-\partial U_{i++}(\beta)/\partial \beta\right\}^{-1}$ and $B(\beta) = n^{-1}\sum_{i=1}^n E\left\{U_{i++}(\hat{\beta})^2\right\}$.\\

\noindent Recall that $A^{-1}(\beta)$ is of the form,
$$A^{-1}({\beta})=\left\{m\sum_{j=1}^J \sum_{a=0}^1P(Z_{ij}=a)\nu(Z_{ij}=a)\right\}^{-1},$$
where $\nu(Z_{ij})=\int_0^{C^*} \mathscr{G}(t) \mu_j(t)\left\{1 - \mu_j(t)\right\} f(t|Z_{ij})dt$. Also recall that $B(\beta)$ is of the form,
$$B({\beta}) = m\sum_{j=1}^J\Upsilon_0(j) + m(m-1)\sum_{j=1}^J \Upsilon_1(j,j) + m^2\mathop{\sum_{j=1}^J \sum_{l=1}^J}_{j \ne l}\Upsilon_1(j,l),$$
where $\Upsilon_0(j) = \sum_{a=0}^1P(Z_{ij}=a)q_0(Z_{ij}=a)$ and $\Upsilon_1(j,l) = \sum_{a=0}^1\sum_{a'=0}^1P(Z_{ij}=a, Z_{il}=a')\sum_{r=1}^4 q_r(Z_{ij}=a, Z_{il}=a')$.\\

\noindent It was previously shown that $\sum_{a=0}^1P(Z_{ij}=a)\nu(Z_{ij}=a) = \Upsilon_0(j)$, such that $A^{-1}({\beta})=\left\{m\sum_{j=1}^J \Upsilon_0(j)\right\}^{-1}$.\\

\noindent Combining these gives us,
\begin{equation}\label{eq:var_b_proof}
    \Var(\hat{\beta}) = \frac{\sum_{j=1}^J \Upsilon_0(j) + (m-1)\sum_{j=1}^J \Upsilon_1(j,j) + m\mathop{\sum_{j=1}^J \sum_{l=1}^J}_{j\ne l} \Upsilon_1(j,l)}{nm\left\{\sum_{j=1}^J \Upsilon_0(j)\right\}^2}.
\end{equation}

\noindent Noting that $\rho_w = \rho_w = \frac{\sum_{j=1}^J \Upsilon_1(j,j)}{\sum_{j=1}^J \Upsilon_0(j)}$ and $\rho_b = \frac{\mathop{\sum_{j=1}^J \sum_{l=1}^J}_{j\ne l} \Upsilon_1(j,l)}{(J-1)\sum_{j=1}^J \Upsilon_0(j)}$, the variance can be rewritten as,
$$\Var(\hat{\beta})=\left\{nm\sum_{j=1}^J \Upsilon_0(j)\right\}^{-1}\times \left\{1 + (m-1)\rho_w + m(J-1)\rho_b\right\}$$

%% REMARK 2 %%
\addcontentsline{toc}{subsection}{REMARK 2: Connection to DE of independence GEE for continuous outcomes}
\subsection*{REMARK 2: Connection to DE of independence GEE for continuous outcomes}
%% (1) show equivalence of Tian and Wang %%
The design effect defined in Remark 2 is of a similar form to the design effect for SW-CRT independence GEEs with continuous outcomes. To explicitly connect these two, first us define $\boldsymbol{Z}_i = (Z_{i1}, \dots, Z_{iJ})^T$, $\overline{Z}_j = n^{-1}\sum_{i=1}^n Z_{ij}$, $\boldsymbol{M}_i = \left((Z_{i1} - \overline{Z}_1), \dots, (Z_{iJ} - \overline{Z}_J) \right)^T$, $U=\sum_{i=1}^n\sum_{j=1}^JZ_{ij}$, $W =\sum_{j=1}^J\left(\sum_{i=1}^nZ_{ij}\right)^2$, and $V =\sum_{i=1}^n\left(\sum_{j=1}^JZ_{ij}\right)^2$. Let $Y_{ijk}$ be the continuous outcome measure for individual $k$ in cluster $i$ at period $j$. Also let $\boldsymbol{\Omega}$ and $\boldsymbol{\Phi}$ be $J\times J$ basis matrices such that 
\begin{equation*}
\boldsymbol{\Omega} = \begin{pmatrix}
1 & r_{12}^* & \cdots & r_{1J}^* \\
r_{12^*} & 1 & \cdots & r_{2J}^* \\
\vdots  & \vdots  & \ddots & \vdots  \\
r_{1J}^* & r_{2J}^* & \cdots & 1 
\end{pmatrix},\hspace{0.5cm}
\boldsymbol{\Phi} = \begin{pmatrix}
r_{11} & r_{12} & \cdots & r_{1J} \\
r_{12} & r_{22} & \cdots & r_{2J} \\
\vdots  & \vdots  & \ddots & \vdots  \\
r_{1J} & r_{2J} & \cdots & r_{JJ} 
\end{pmatrix}    
\end{equation*}
and $Corr(\boldsymbol{Y_i}) = \boldsymbol{I}_m\otimes (\boldsymbol{\Omega} - \boldsymbol{\Phi}) + (\boldsymbol{\mathbbm{1}}_m\boldsymbol{\mathbbm{1}}_m^T)\otimes \boldsymbol{\Phi}$. For cross-sectional studies, let $r_{jj} = \alpha_0$ be the within-period correlation for two subjects in period $j$ and $r_{jj'}=r_{jj'}^*=\alpha_1$ be the between-period correlation for two subjects in periods $j$ and $j'$, respectively; this creates a nested-exchangeable correlation structure \citep{hooper_sample_2016,li_marginal_2022}.\\

\noindent \citet{wang_sample_2021} derived the treatment effect variance of a continuous outcome SW-CRT as
\begin{equation}\label{var-wang}
\Var(\hat{\beta}) = \frac{\sigma^2\sum_{s=1}^S p_s(\boldsymbol{v}_s - \overline{\boldsymbol{u}})^T [\boldsymbol{\Omega} + (m-1)\boldsymbol{\Phi}](\boldsymbol{v}_s - \overline{\boldsymbol{u}})}{m[\sum_{j=1}^J \overline{u}_j(1-\overline{u}_j)]^2},
\end{equation}
where $\sigma^2$ is the marginal variance of the outcome, $\boldsymbol{p}_s$ is the probability of a cluster having a particular treatment sequence $s$, $\boldsymbol{v}_s$ represents a treatment sequence $s$, and $\overline{\boldsymbol{u}}$ is a $J$-length vector of the proportion of subjects receiving intervention at period $j$.\\

\noindent Note that $\lim_{n\rightarrow \infty} n^{-1}\sum_{i=1}^n \boldsymbol{Z}_i^T\boldsymbol{Z}_i = \sum_{s=1}^S p_s\boldsymbol{v}_s^T\boldsymbol{v}_s$. Therefore we may write the $\Var(\hat{\beta})$ as
$$\Var(\hat{\beta}) = \frac{\sigma^2 n^{-1}\sum_{i=1}^n(\boldsymbol{Z}_i - \overline{\boldsymbol{Z}})^T [\boldsymbol{\Omega} + (m-1)\boldsymbol{\Phi}](\boldsymbol{Z}_i - \overline{\boldsymbol{Z}})}{m[\sum_{j=1}^J \overline{Z}_j(1-\overline{Z}_j)]^2}.$$

\noindent This is identical to the treatment effect variance derived by \citet{tian2024information}
\begin{equation}\label{var-tian}
    \Var(\hat{\beta}) = \frac{\sigma^2}{(U-n^{-1}W)^2}\sum_{i=1}^n\boldsymbol{Z}^T_i[\boldsymbol{\Omega} + (m-1)\boldsymbol{\Phi}]\boldsymbol{Z}^T_i - \frac{\sigma^2}{n(U-n^{-1}W)^2}\left(\sum_{i=1}^n\boldsymbol{Z}_i^T\right)[\boldsymbol{\Omega} + (m-1)\boldsymbol{\Phi}]\left(\sum_{i=1}^n\boldsymbol{Z}_i\right).
\end{equation}

\noindent Expanding this to the cluster-period level, we may rewrite the variance expression (assuming the nested exchangeable correlation structure) as
\begin{align*}
\Var(\hat{\beta}) &= \frac{\sigma^2}{m(U-n^{-1}W)^2}\left\{\sum_{i=1}^n\sum_{j=1}^J[1 + (m-1)\alpha_0]Z_{ij} + \sum_{i=1}^n\mathop{\sum_{j=1}^J\sum_{j'=1}^J}_{j\ne j'}[\alpha_1 + (m-1)\alpha_1]Z_{ij}Z_{ij'}\right\}\\
&~~- \frac{\sigma^2}{nm(U-n^{-1}W)^2}\left\{\sum_{j=1}^J[1 + (m-1)\alpha_0]\left(\sum_{i=1}^nZ_{ij}\right)^2 + \mathop{\sum_{j=1}^J\sum_{j'=1}^J}_{j\ne j'}[\alpha_1 + (m-1)\alpha_1]\left(\sum_{i=1}^nZ_{ij}\right)\left(\sum_{i=1}^nZ_{ij'}\right)\right\}\\
&=\frac{\sigma^2}{m(U-n^{-1}W)^2}\left\{\sum_{i=1}^n\sum_{j=1}^J(1 + (m-1)\alpha_0)Z_{ij} + \sum_{i=1}^n\mathop{\sum_{j=1}^J\sum_{j'=1}^J}_{j\ne j'}m\alpha_1Z_{ij}Z_{ij'}\right\}\\
&~~- \frac{\sigma^2}{nm(U-n^{-1}W)^2}\left\{\sum_{j=1}^J[1 + (m-1)\alpha_0]\left(\sum_{i=1}^nZ_{ij}\right)^2 + \mathop{\sum_{j=1}^J\sum_{j'=1}^J}_{j\ne j'}m\alpha_1\left(\sum_{i=1}^nZ_{ij}\right)\left(\sum_{i=1}^nZ_{ij'}\right)\right\}.
\end{align*}

\noindent Rearranging, we have
\begin{align*}
\Var(\hat{\beta}) = \frac{\sigma^2}{m(U-n^{-1}W)^2}&{\Bigg\{}(U-n^{-1}W) + (m-1)\alpha_0(U-n^{-1}W)\\ 
&\left.+ m\alpha_1 \left[\sum_{i=1}^n\mathop{\sum_{j=1}^J\sum_{j'=1}^J}_{j\ne j'} Z_{ij}Z_{ij'} - n^{-1}\mathop{\sum_{j=1}^J\sum_{j'=1}^J}_{j\ne j'}\left(\sum_{i=1}^nZ_{ij}\right)\left(\sum_{i=1}^nZ_{ij'}\right)\right]\right\}.
\end{align*}

%% (2) Show Tian and Wang equivalent to tr(Sigma) form %%
\noindent Note that $n^{-1}\sum_{i=1}^n \boldsymbol{M}_i\boldsymbol{M}_i^T=\boldsymbol{\Sigma}$ is the covariance matrix of the intervention vector under a specific design and $(U-n^{-1}W)=\tr(\boldsymbol{\Sigma)}$. Also note that $\sum_{i=1}^n\mathop{\sum_{j=1}^J\sum_{j'=1}^J}_{j\ne j'} Z_{ij}Z_{ij'} - n^{-1}\mathop{\sum_{j=1}^J\sum_{j'=1}^J}_{j\ne j'}\left(\sum_{i=1}^nZ_{ij}\right)\left(\sum_{i=1}^nZ_{ij'}\right) = \boldsymbol{\mathbbm{1}}^T\boldsymbol{\Sigma}\boldsymbol{\mathbbm{1}} - \tr(\boldsymbol{\Sigma})$, where $\boldsymbol{\mathbbm{1}}$ is a vector of 1s. Thus we can express the general GEE variance under a working independence assumption for a cross-sectional SW-CRT with continuous outcomes as
\begin{equation}\label{var:gee-cont}
    \Var(\hat{\beta}) = \frac{\sigma^2}{nm\tr(\boldsymbol{\Sigma})} \times \left\{1 + (m-1)\alpha_0 + m(J-1)\alpha_1 \frac{\boldsymbol{\mathbbm{1}}^T\Sigma\boldsymbol{\mathbbm{1}} - \tr(\boldsymbol{\Sigma})}{(J-1)\tr(\boldsymbol{\Sigma})}\right\}.
\end{equation}
%$$\frac{\sigma^2}{nmA^2}\{\tr(\Sigma) + (m-1)\alpha_0 \tr(\Sigma) + m\alpha_1(\boldsymbol{\mathbbm{1}}^T\Sigma \boldsymbol{\mathbbm{1}} - \tr(\Sigma))\},$$

%Let $\Omega = \boldsymbol{\mathbbm{1}}_J^T\boldsymbol{\mathbbm{1}}_J (J-1)\rho_b + (1-(J-1)\rho_b)\mathbf{I}_J$ and $\Phi = \boldsymbol{\mathbbm{1}}_J^T\boldsymbol{\mathbbm{1}}_J (J-1)\rho_b + (\rho_w - (J-1)\rho_b)\mathbf{I}_J$.

%% (3) Map to rho_w and rho_b %%
\noindent Recall variance (9) from Theorem 1 can be expressed as
$$\Var(\hat{\beta})=\left\{nm\sum_{j=1}^J \Upsilon_0(j)\right\}^{-1}\times \left\{1 + (m-1)\rho_w + m(J-1)\rho_b\right\}.$$
There are obviously clear connections between this variance and the variance \eqref{var:gee-cont} for continuous outcome SW-CRTs. First, $\rho_w$ can be thought similar to the within-period correlation $\alpha_0$ of a continuous outcome cross-sectional SW-CRT, but defined on the martingale scale. In addition, $\rho_b$ can be thought of as similar to the between-period correlation $\alpha_1$ of a continuous outcome cross-sectional SW-CRT multiplied by the generalized ICC of the intervention $\frac{\boldsymbol{\mathbbm{1}}^T\boldsymbol{\Sigma}\boldsymbol{\mathbbm{1}} - \tr(\boldsymbol{\Sigma})}{(J-1)\tr(\boldsymbol{\Sigma})}$ as defined generally by \citet{kistner_exact_2004} and in SW-CRTs with subclusters by \citet{davis-plourde_sample_2021}.
\addcontentsline{toc}{section}{Web Appendix D}
\section*{Web Appendix D}\label{D}

\addcontentsline{toc}{subsection}{Nested Archimedean Copulas in Power Calculation}
\subsection*{Nested Archimedean Copulas in Power Calculation}
%Nested Archimedean copulas are an attractive approach for inducing multiple dependencies on survival data as they only require specifying the form of the marginal univariate distribution of event times, determining which transformations will be used, and the values of dependency parameters .
To conduct power calculations, one can directly specify the survival distributions for the censoring and event times to calculate $\rho_w$ and $\rho_b$ for main-text equation (9) in the Wald testing paradigm, or to directly calculate $\kappa^{H_c}_w$ and $\kappa^{H_c}_b$ via main-text equation (10) for the robust score paradigm. In formulating these bivariate distributions, it is critical to incorporate a dependency structure with separate within-period and between-period components. While there are several potential choices for this specification, we consider the nested Archimedean copula approach \citep{mcneil_sampling_2008} with Gumbel transformations \citep{gumbel_bivariate_1960}, which we outline below.\\

\noindent To begin, assume event times follow an exponential distribution, $T_{ijk} \sim \text{Exp}\left(\lambda_{ij}\right)$, such that the marginal survival function takes the form: 
$$\mathscr{F}(t_{ijk}) = \exp\left(-\lambda_{ij}t\right)=\exp\left(-\lambda_{0j}te^{\beta Z_{ij}}\right).$$ 
To approximate the bivariate distribution for two event times $T_{ijk}$ and $T_{ild}$, we can apply a nested Gumbel copula transformation, $\psi_0^{-1}(x; \theta_0) = \left\{-\ln(x)\right\}^{\theta_0}$, to map their marginal survival functions from $[0,1] \rightarrow [0, \infty)$, add them together, and then map them back to the $[0,1]$ space with $\psi_0(x; \theta_0) = \exp\left(-x^{1/\theta_0}\right)$, where $\theta_0$ would be a dependency parameter \citep{gumbel_bivariate_1960}:
$$\mathscr{F}(t_{ijk}, t_{ild}) = \psi_0\left(\psi_0^{-1}\left\{\mathscr{F}(t_{ijk})\right\} + \psi_{0}^{-1}\left\{\mathscr{F}(t_{ild})\right\}\right).$$ 
This parameter induces one level of dependency or correlation on the event times, such as being in the same cluster.\\

\noindent To induce a second level of dependency, such as two individuals being within the same period, one can perform a second set of transformations --- $\psi_{01}(x; \theta_{01})$, $\psi_{01}^{-1}(x; \theta_{01})$ --- within the original copula: 
$$\mathscr{F}(t_{ijk}, t_{ijk'}, t_{ild}, t_{ild'}) = \psi_0\left(\psi_0^{-1}\left(\psi_{01}\left\{\psi_{01}^{-1}\{\mathscr{F}(t_{ijk})\} + \psi_{01}^{-1}\{\mathscr{F}(t_{ijk'})\}\right\}+ \psi_{01}\left\{\psi_{01}^{-1}\{\mathscr{F}(t_{ild})\} + \psi_{01}^{-1}\{\mathscr{F}(t_{ild'})\}\right\}\right)\right),$$ 
% \begin{align*}
%     \mathscr{F}(t_{ijk}, t_{ijk'}, t_{ild}, t_{ild'}) &= \psi_0\left(\psi_0^{-1}\left(\psi_{01}\left\{\psi_{01}^{-1}\{\mathscr{F}(t_{ijk})\} + \psi_{01}^{-1}\{\mathscr{F}(t_{ijk'})\}\right\}\right.\right.\\
%     &\left.\left.~~+ \psi_{01}\left\{\psi_{01}^{-1}\{\mathscr{F}(t_{ild})\} + \psi_{01}^{-1}\{\mathscr{F}(t_{ild'})\}\right\}\right)\right),
% \end{align*}
with the condition that $\theta_{01} > \theta_0$.\\

\noindent If we are comparing two individuals who share at least one level of dependency, one set of these transformations will negate the other. For example, for two event times in different periods $j,l$ but the same cluster $i$, the bivariate conditional survival function would simply be expressed as: 
$$\mathscr{F}(t_{ijk}, t_{ild}) = \exp\left\{-\left[(\lambda_{ij}t_{ijk})^{\theta_0} + (\lambda_{il}t_{ild})^{\theta_0} \right]^{1/\theta_0}\right\}.$$ 
% \begin{align*}
% \mathscr{F}(t_{ijk}, t_{ild}) &= \psi_0\left(\psi_0^{-1}\left(\psi_{01}\left\{\psi_{01}^{-1}\{\mathscr{F}(t_{ijk})\}\right\} + \psi_{01}\left\{\psi_{01}^{-1}\{\mathscr{F}(t_{ild})\}\right\}\right)\right)\\
% %&= \psi_0\left(\psi_0^{-1}\left(\mathscr{F}(t_{ijk}) + \mathscr{F}(t_{ild})\right)\right)\\
% &= \exp\left\{-\left[(\lambda_{ij}t_{ijk})^{1/\theta_0} + (\lambda_{il}t_{ild})^{1/\theta_0} \right]^{\theta_0}\right\},  
% \end{align*}
%where the inner $\psi_{01}(x)$, $\psi_{01}^{-1}(x)$ functions negate each other. 
\noindent On the other hand, for two event times in the same period $j$ and same cluster $i$, the bivariate conditional survival function would be expressed as: 
$$\mathscr{F}(t_{ijk}, t_{ijd})=\exp\left\{-\left[(\lambda_{ij}t_{ijk})^{\theta_{01}} + (\lambda_{il}t_{ijd})^{\theta_{01}} \right]^{1/\theta_{01}}\right\}.$$ 
% \begin{align*}
% \mathscr{F}(t_{ijk}, t_{ijd}) &=\psi_0\left(\psi_0^{-1}\left(\psi_{01}\left\{\psi_{01}^{-1}\{\mathscr{F}(t_{ijk})\} + \psi_{01}^{-1}\{\mathscr{F}(t_{ijd})\}\right\}\right)\right)\\
% %&= \psi_{01}\left(\psi_{01}^{-1}\left(\mathscr{F}(t_{ijk}) + \mathscr{F}(t_{ijd})\right)\right)\\
% &= \exp\left\{-\left[(\lambda_{ij}t_{ijk})^{1/\theta_{01}} + (\lambda_{il}t_{ijd})^{1/\theta_{01}} \right]^{\theta_{01}}\right\},
% \end{align*}
%where the outer $\psi_{0}(x)$, $\psi_{0}^{-1}(x)$ functions negate each other. 
A similar approach was taken by \citet{li_sample_2022} to generate clustered survival times with a three-level data structure.\\

%It then becomes a question as to what values to assume for dependency parameters $\theta_0$ and $\theta_{01}$. 
\noindent The dependency parameter for a nested Gumbel copula can be interpreted as a transformation of the rank-based correlation measure Kendall's tau ($\tau$): $\theta = 1/(1-\tau)$. Therefore, when integrating nested Gumbel copulas into our power calculation approach,
%if we assume there is $\tau_w$ correlation between individuals in the same cluster and period (within-period correlation) and $\tau_b$ correlation between individuals in the same cluster and different periods (between-period correlation), 
we can set the dependency parameters for the copula to be $\theta_0=1/(1-\tau_b)$ and $\theta_{01}=1/(1-\tau_w)$%$\theta_0=1-\tau_b$ and $\theta_{01}=1-\tau_w$,
where $\tau_b$ and $\tau_w$ refer to the between-period and within-period correlation on the scale of Kendall's tau. 
%$\tau = (\theta - 1)/\theta$, $\theta\in [1,\infty)$.
In our experiences with a balanced design (where an equal number of clusters are assigned to each sequence), if one assumes the bivariate conditional survival functions follow a nested Gumbel copula structure with $\theta_0=1/(1-\tau_b)$ and $\theta_{01}=1/(1-\tau_w)$, the resulting within-period generalized ICC, $\rho_w$, matches closely to the value for $\tau_w$, whereas the resulting between-period generalized ICC, $\rho_b$, tends to be smaller than $\tau_b$. Further exploration of this relationship can be found in Web Figure 1 in Web Appendix F.
\newpage
\addcontentsline{toc}{section}{Web Appendix E: Relationship between g-ICC and Kendall's tau}
\section*{Web Appendix E: Relationship between g-ICC and Kendall's tau}\label{E}
As discussed in the main text, there are two options for which to use main text variance equations (9) and (10) for power calculations. First, one can directly assume specific values for the within-period and between-period g-ICCs and then use equation (9). While this is computationally simple, it is not immediately obvious how specific g-ICC values map to features of the within-cluster censoring and event outcome distributions, such as within-period and between-period Kendall's tau.\\

\noindent Recall from Appendix D, the dependency parameters in a nested Archimedean copula can be interpreted as transformations of the rank-based correlation measure Kendall's tau ($\tau$). Under Gumbel copulas, we can set the dependency parameters for the copula to be $\theta_0 = 1/(1-\tau_b)$ and $\theta_{01} = 1/(1-\tau_w)$ where $\tau_b$ and $\tau_w$ refer to the between-period and within-period correlation on the scale of Kendall's tau.\\

\noindent To better understand how g-ICC values map to Kendall's tau across multiple design variations, we provide some initial exploratory results under specific examples. We assume survival times had a bivariate distribution that followed a nested Archimedean gumbel copula with a within-period Kendall's tau of $\tau_w$ and a between-period Kendall's tau of $\tau_b$ and censoring times followed an independent Uniform distribution. We also assume a varying number of study periods $J\in\{3,6,11\}$ and a baseline hazard that progressively increases with time, $\lambda_{0j}(t|Z_{ij}) = \lambda_0 + 0.05(j-1)$, to induce a non-zero period effect. Following \citet{zhong_sample_2015} and \citet{wang_improving_2023}, we set $\lambda_0$ as the solution to $P(T_{i1k} > C^*|Z_{i1}=0)=p_a$ in the first study period given a reference administrative censoring rate $p_a$; here we consider $p_a=20$\%. Under these assumptions, we directly calculate the marginal variance, within-period, and between-period covariances of the score, as well as the within-period and between-period g-ICCs.\\

\noindent Figure \ref{fig:gicc-tau} depicts two relationships: that of within-period g-ICC and within-period Kendall's tau for fixed values of between-period Kendall's tau $\tau_b$ (top panel), and that of between-period g-ICC and between-period Kendall's tau for fixed values of within-period Kendall's tau $\tau_w$ (bottom panel). We see that the within-period g-ICC matches closely to the value for the within-period Kendall's tau $\tau_w$, and that those does not change as the number of study periods increases. On the other hand, we observe that the between-period g-ICC tends to be smaller than a given between-period Kendall's tau $\tau_b$, and that the strength of this relationship does change with number of study periods; in other words, this relationship tends to be more sensitive to values of the remaining design parameters. For example we see that, under $J=3$ periods, the between-period g-ICC remains near $0$ across $\tau_b$ and $\tau_w$, while under $J=11$ periods the value for the between-period g-ICC increases to approximately half of the between-period Kendall's tau $\tau_b$. Interestingly, this is very similar to observations found by \citet{meng2023simulating} in parallel-arm CRTs.\\

\renewcommand\thefigure{E.\arabic{figure}}
\setcounter{figure}{0}
\begin{figure}[h]
    \centering
    \includegraphics[width=\textwidth]{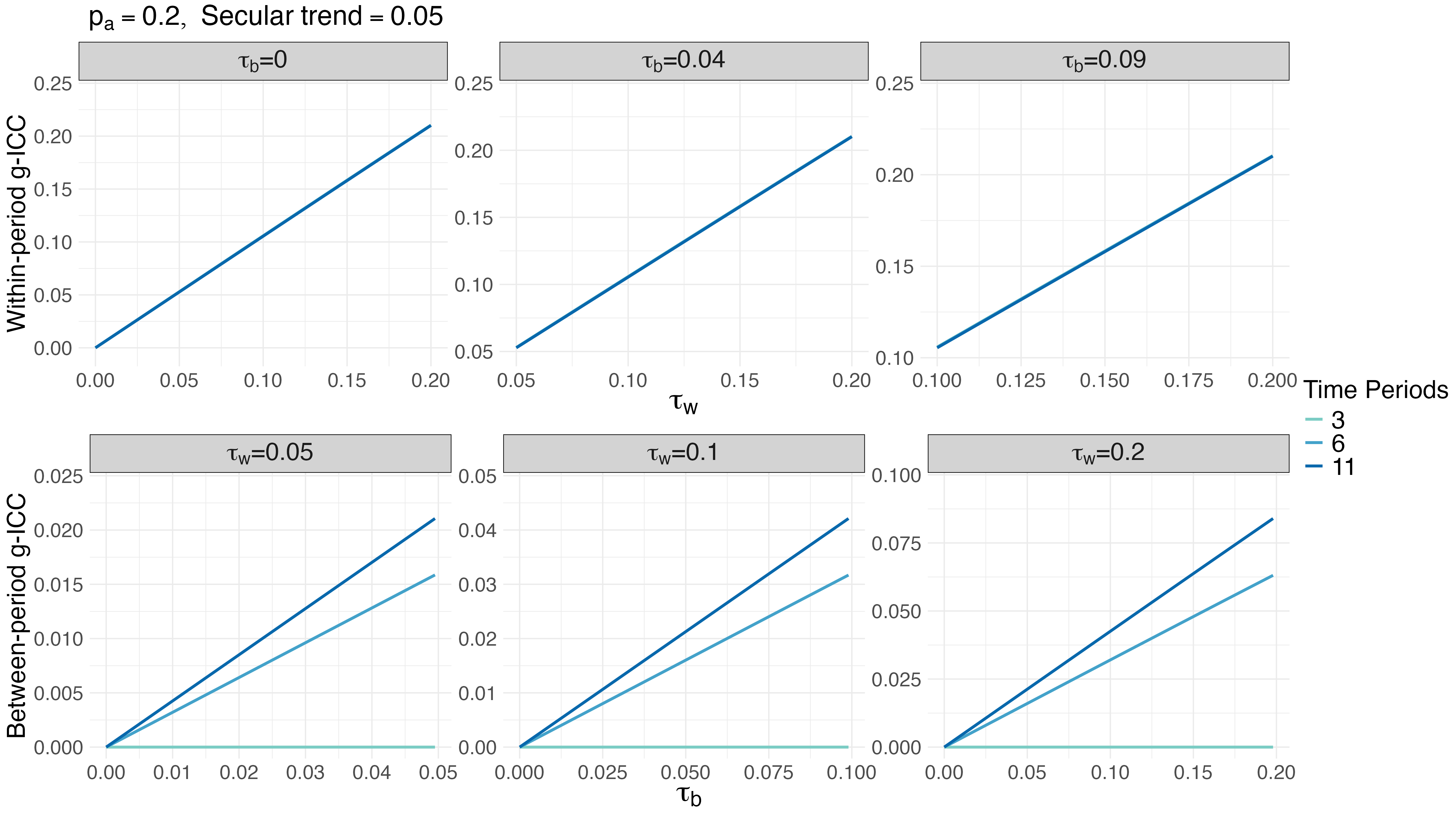}
    \caption{\label{fig:gicc-tau}Relationship between within-period (top panel) and between-period (bottom panel) Kendall's tau and generalized ICC for different numbers of periods, assuming $n=30$ clusters, a cluster-period size of $m=50$, $20$\% reference administrative censoring, uniform loss to follow-up censoring, and a baseline hazard that increases by $5$\% with each period.}
\end{figure}

In calculations not shown, we also examined the effect of different administrative censoring rates ($p_a=\{5\%, 20\%\}$) and secular trends ($\{\lambda_{0j}(t|Z_{ij}) = \lambda_0 - 0.05(j-1),\lambda_{0j}(t|Z_{ij}) = \lambda_0,\lambda_{0j}(t|Z_{ij}) = \lambda_0 + 0.05(j-1)\}$), 
%($\{-0.05, 0, 0.05\}$), 
finding that varying these factors did not significantly change the relationships observed above.
\addcontentsline{toc}{section}{Web Appendix F: Data Example Sensitivity Analyses}
\section*{Web Appendix F: Data Example Sensitivity Analyses}\label{F}
\renewcommand\thefigure{F.\arabic{figure}}
\setcounter{figure}{0}

In the data application in Section 5, we investigated how sensitive power calculations were to choice of within-period and between-period Kendall's tau. We saw that power decreased with increasing within- or between-period correlation, decreasing more quickly if both correlations increase simultaneously. Below in Figure \ref{fig:sensitivity-gicc} we show how power shifts with changes to within-period and between-period g-ICC. Similar to Kendall's tau, larger within-period and between-period g-ICCs result in smaller predicted power, and power is more robust to changes in between-period g-ICC when the within-period g-ICC is small. The major difference is that while the between-period Kendall's tau could be as large as the within-period Kendall's tau, we see that the between-period g-ICC ranges from $0-30$\% of the within-period g-ICC - this is due to differences in the definition of the correlation parameters, and we refer to Web Appendix E for more empirical exploratory results on this point.\\

\begin{figure}
    \centering
    \includegraphics[width=\textwidth]{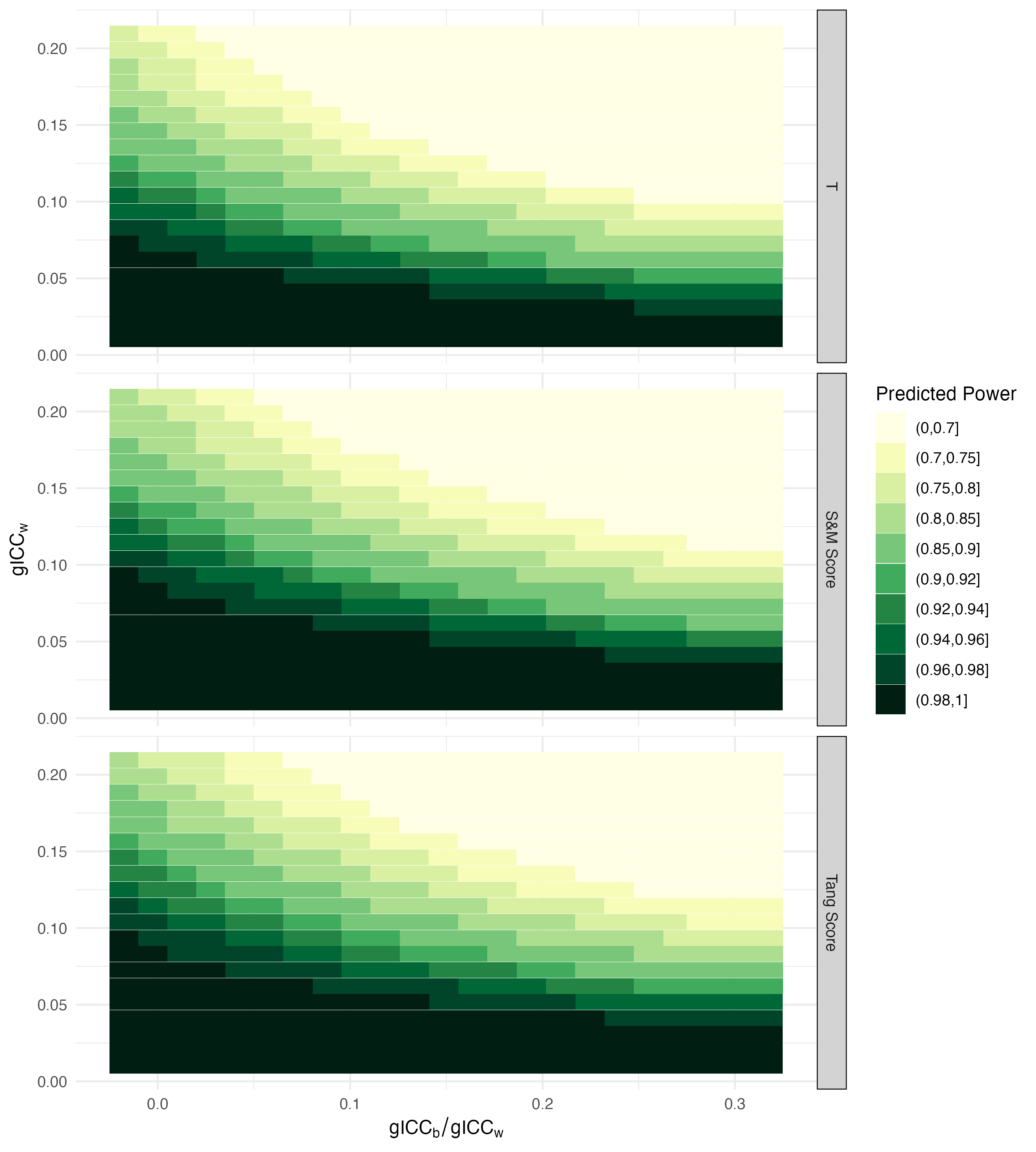}%{figures/mitchell-power-sensitivity-tau-02.png}
    \caption{\label{fig:sensitivity-gicc} Contour plots of predicted power trends across within-period g-ICC and the ratio of between- and within-period g-ICC ($\rho_b/\rho_w$) within our application study of the CATH TAG trial, assuming $n=20$ clusters and a baseline hazard that increases by 5\% at each subsequent time period. The top row represents trends when power is predicted using the Wald $t$-test formula, the middle row when using the \citep{self_powersample_1988} robust score test formula, and the bottom row when using the \citep{tang_improved_2021} robust score test formula. Darker colors correspond to greater predicted power.}
\end{figure}

Understanding how within-period and between-period correlations may affect power in the specific case of an increasing baseline hazard function,  we will now examine how sensitive our power calculation is to choice of baseline hazard -- a design parameter, much like within-period and between-period ICCs, that investigators are likely to have little information on at the design stage.\\

\noindent Assuming a baseline hazard that decreases by $5$\% with each additional period and minimal administrative censoring ($p_a=5\%$), such that $\lambda_{0j}(t) = \lambda_0 - 0.05(j-1)$, our Wald-based formula estimates we would have $79.7$\% power to detect a treatment effect of $\beta=0.4$ (HR=$1.5$) with $n=20$ total clusters, similar to predictions made assuming a $5$\% increasing baseline hazard. On the other hand, our robust score-based formulas using the \citet{self_powersample_1988} and \citet{tang_improved_2021} methods estimate $84.1$\% and $85.0$\% power under $20$ clusters, respectively -- largely the same as was predicted under increasing baseline hazards.\\

\noindent If we instead assume that the baseline hazard does not change with time, such at $\lambda_{0j}(t) = \lambda_0 - 0(j-1) = 1$, our Wald-based formula estimates $80.3$\% power under the same sample size, while our robust score-based formulas using the \citet{self_powersample_1988} and \citet{tang_improved_2021} methods predict $84.9$\% and $85.8$\% power, respectively.\\

\begin{figure}
    \centering
    \includegraphics[width=\textwidth]{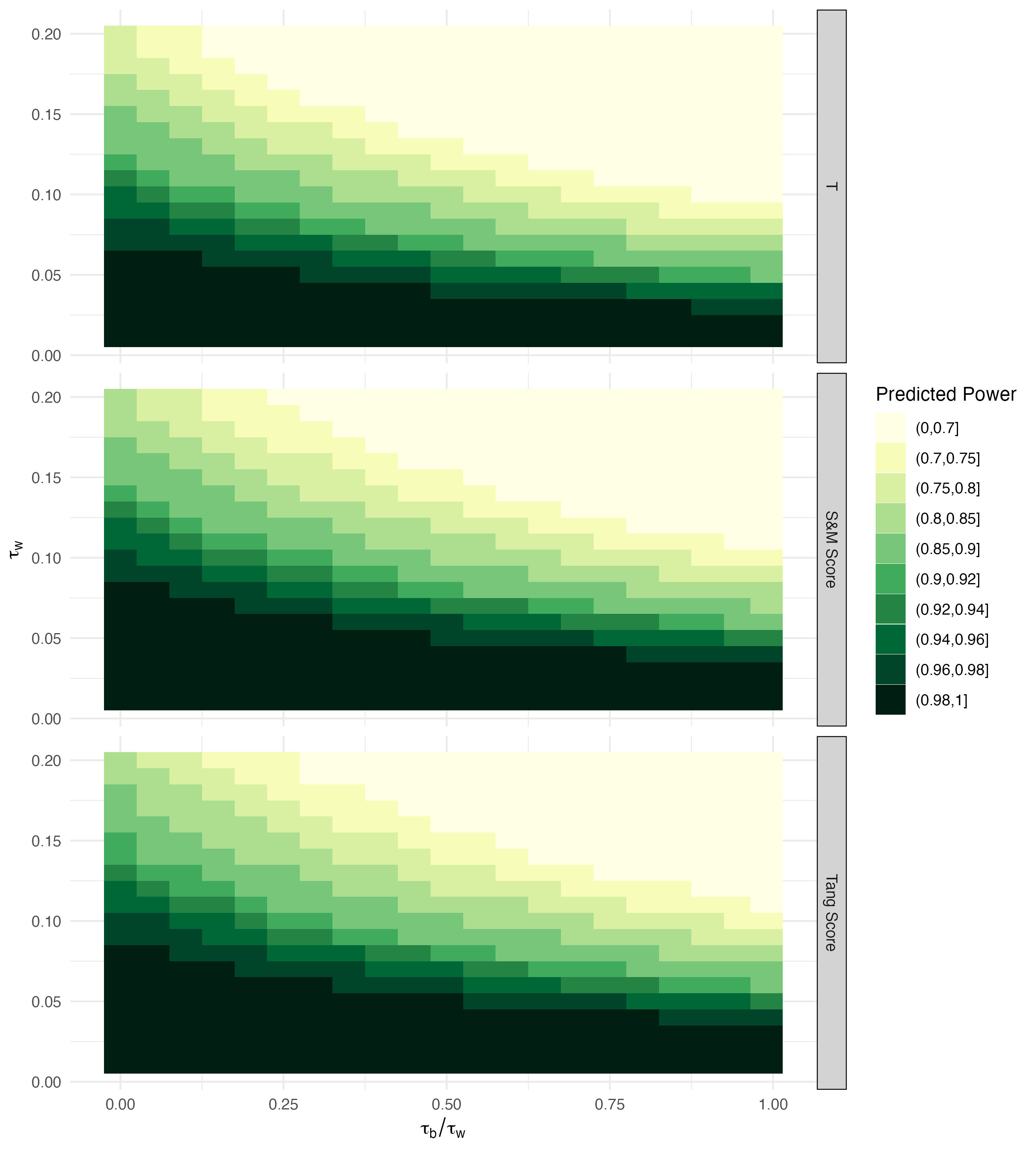}%{figures/mitchell-power-sensitivity-tau-02.png}
    \caption{\label{fig:sensitivity-decrease} Contour plots of predicted power trends across within-period Kendall's tau ($\tau_w$) and the ratio of between- and within-period Kendall's tau ($\tau_b/\tau_w$) within our application study of the CATH TAG trial, assuming a constant baseline hazard across time and $n=20$ clusters. The top row represents trends when power is predicted using the Wald $t$-test formula, the middle row when using the \citep{self_powersample_1988} robust score test formula, and the bottom row when using the \citep{tang_improved_2021} robust score test formula. Darker colors correspond to greater predicted power.}
\end{figure}

\begin{figure}
    \centering
    \includegraphics[width=\textwidth]{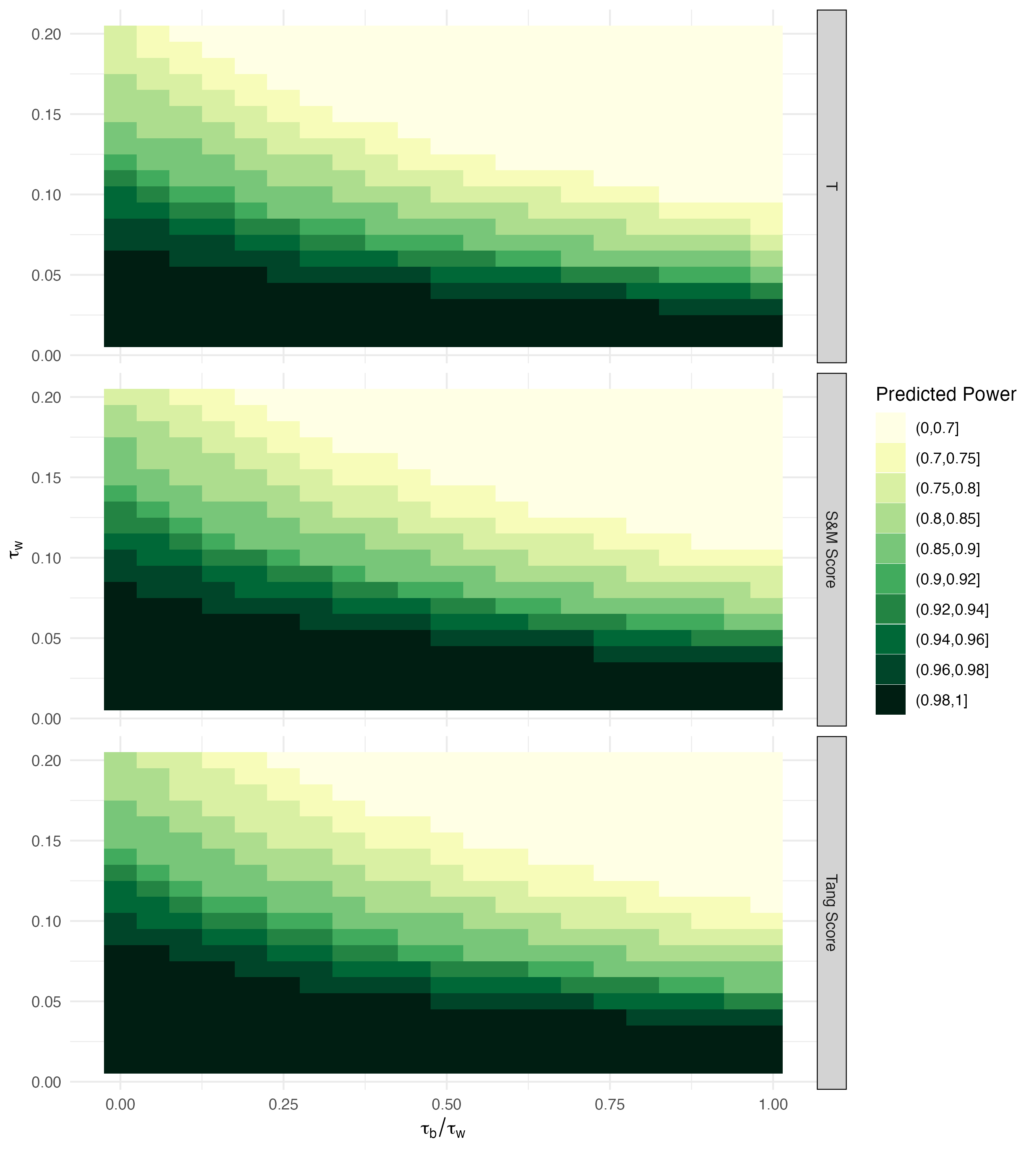}%{figures/mitchell-power-sensitivity-tau-02.png}
    \caption{\label{fig:sensitivity-constant} Contour plots of predicted power trends across within-period Kendall's tau ($\tau_w$) and the ratio of between- and within-period Kendall's tau ($\tau_b/\tau_w$) within our application study of the CATH TAG trial, assuming $n=20$ clusters and a baseline hazard that decreases by 5\% at each subsequent time period. The top row represents trends when power is predicted using the Wald $t$-test formula, the middle row when using the \citep{self_powersample_1988} robust score test formula, and the bottom row when using the \citep{tang_improved_2021} robust score test formula. Darker colors correspond to greater predicted power.}
\end{figure}

\noindent In Figures \ref{fig:sensitivity-decrease} and \ref{fig:sensitivity-constant}, we see how predicted power for such trials changed over varying $\tau_w$ and $\tau_b$; for each baseline hazard scenario, we assume $n=20$ clusters. We see that within each baseline hazard, the effect of Kendall's tau is the same as was observed in the increasing baseline hazard scenario.

\pagebreak
\newpage
\addcontentsline{toc}{section}{Web Appendix G: Web Figures \& Tables}
\section*{Web Appendix G: Web Figures \& Tables}\label{G}
\renewcommand\thefigure{\arabic{figure}}
\renewcommand\thetable{\arabic{table}}
\setcounter{figure}{0}
\setcounter{table}{0}

\renewcommand{\figurename}{Web Figure}

\renewcommand{\tablename}{Web Table}

\begin{figure}[h]
    \centering
    \includegraphics[width=\textwidth]{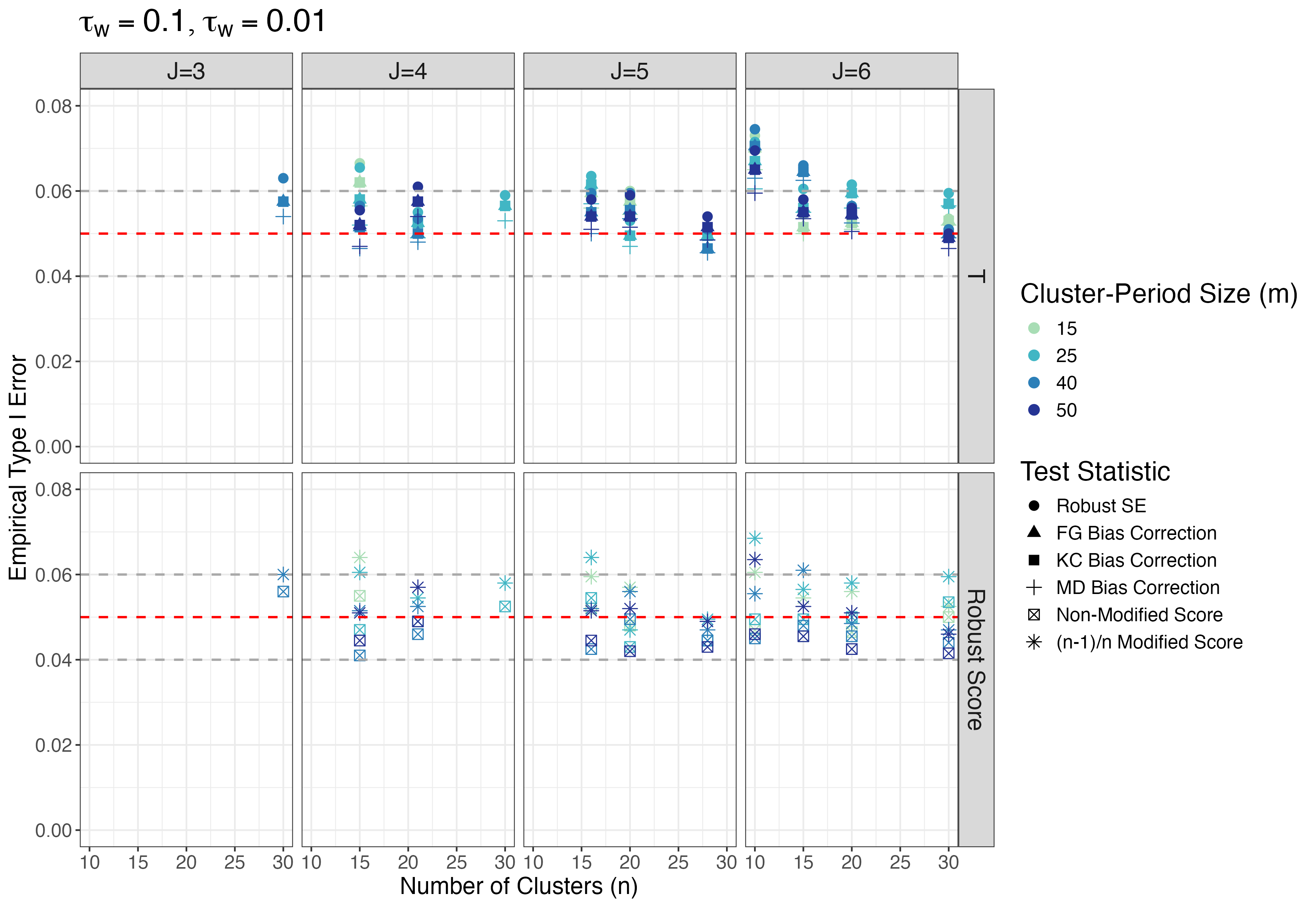}
    \caption{\label{fig:typeI101}Empirical type I error rates for hypotheses testing paradigms when within-period Kendall's $\tau_w=0.1$ and between-period Kendall's $\tau_b=0.01$, given $n$ clusters of cluster-period size $m$ are transitioned onto intervention over $J$ periods (columns). The top row displays empirical type I error results for Wald $t$-tests using a robust sandwich variance (Robust SE) as well as \citep{fay_small-sample_2001} (FG), \citep{kauermann_note_2001} (KC), and \citep{mancl_covariance_2001} (MD) finite-sample adjusted variances (top row). The bottom row displays empirical type I error results for robust (Non-Modified Score) and modified robust score tests ($(n-1)/n$ Modified Score). The red dotted line represents the nominal 5\% error rate and gray dotted lines represent simulation 95\% confidence intervals.}
\end{figure}

\begin{figure}
    \centering
    \includegraphics[width=\textwidth]{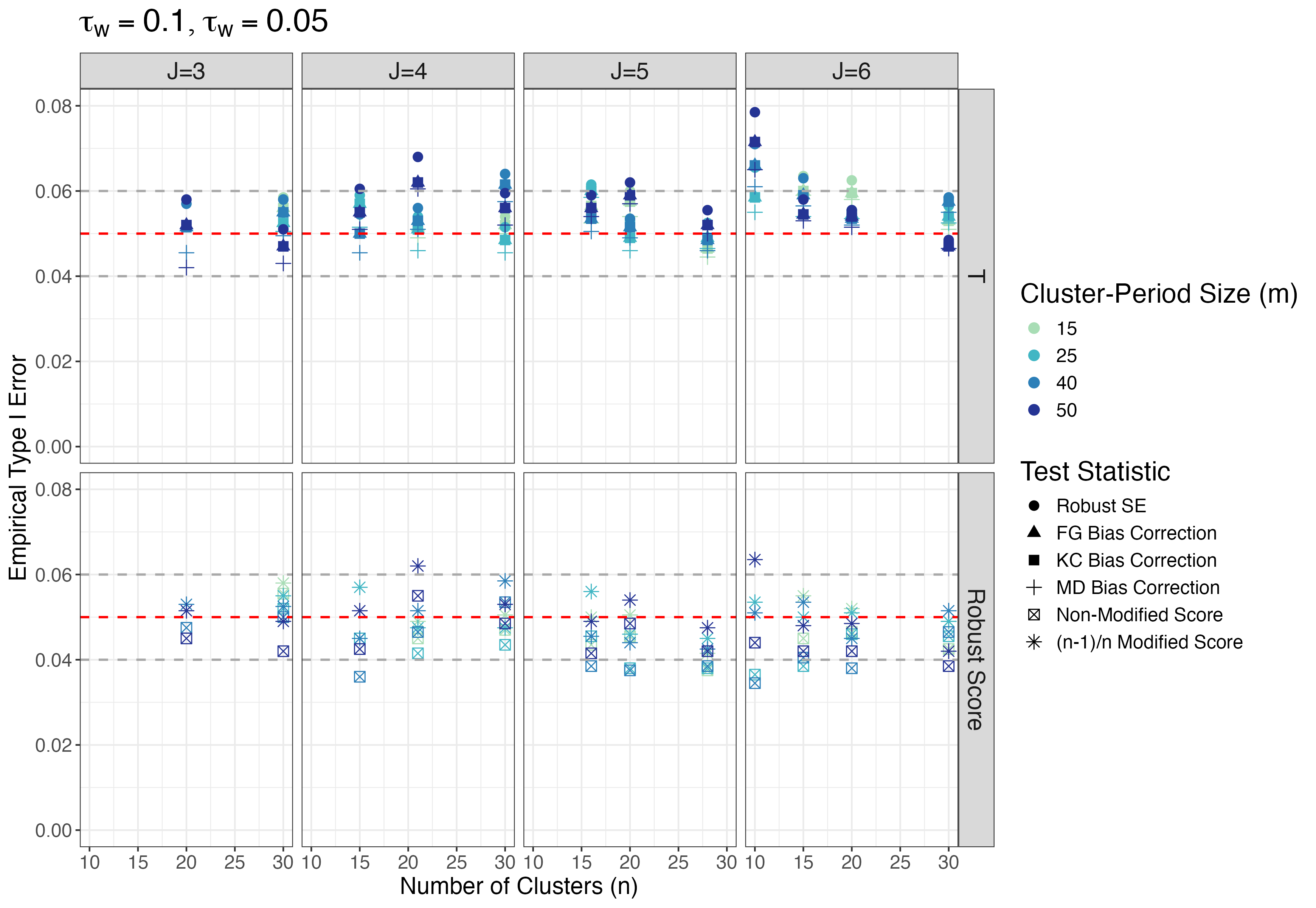}%{figures/typeIError-tau105-092723.png}
    \caption{\label{fig:typeI105}Empirical type I error rates for hypotheses testing paradigms when within-period Kendall's $\tau_w=0.1$ and between-period Kendall's $\tau_b=0.05$, given $n$ clusters of cluster-period size $m$ are transitioned onto intervention over $J$ periods (columns). The top row displays empirical type I error results for Wald $t$-tests using a robust sandwich variance (Robust SE) as well as \citep{fay_small-sample_2001} (FG), \citep{kauermann_note_2001} (KC), and \citep{mancl_covariance_2001} (MD) finite-sample adjusted variances (top row). The bottom row displays empirical type I error results for robust (Non-Modified Score) and modified robust score tests ($(n-1)/n$ Modified Score). The red dotted line represents the nominal 5\% error rate and gray dotted lines represent simulation 95\% confidence intervals.}
\end{figure}

\begin{figure}
    \centering
    \includegraphics[width=\textwidth]{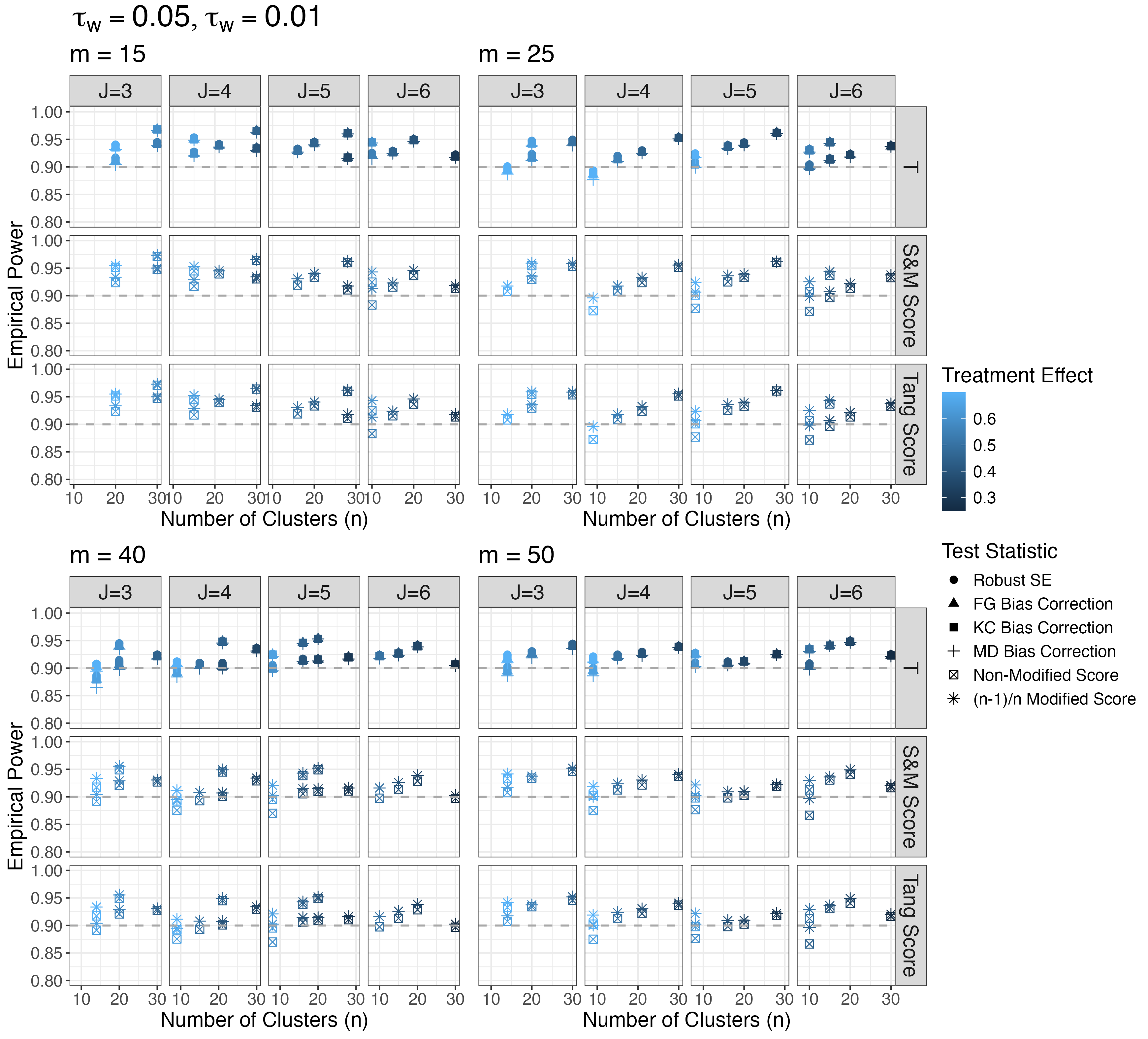}%{figures/empPower-032723.png}
    \caption{\label{fig:empPower0105}Empirical power of hypothesis testing paradigms when within-period Kendall's $\tau_w=0.05$ and between-period Kendall's $\tau_b=0.01$, given $n$ clusters of cluster-period size $m$ are transitioned onto intervention over $J$ periods (columns) under a given treatment effect magnitude (color scale; lighter colors represent larger magnitude). The top row displays empirical power results for Wald $t$-tests using a robust sandwich variance (Robust SE) as well as \citep{fay_small-sample_2001} (FG), \citep{kauermann_note_2001} (KC), and \citep{mancl_covariance_2001} (MD) finite-sample adjusted variances. The bottom row displays empirical power results for robust (Non-Modified Score) and modified robust score tests ($(n-1)/n$ Modified Score). The gray dotted line represents 90\% power for reference.}
\end{figure}

\begin{figure}
    \centering
    \includegraphics[width=\textwidth]{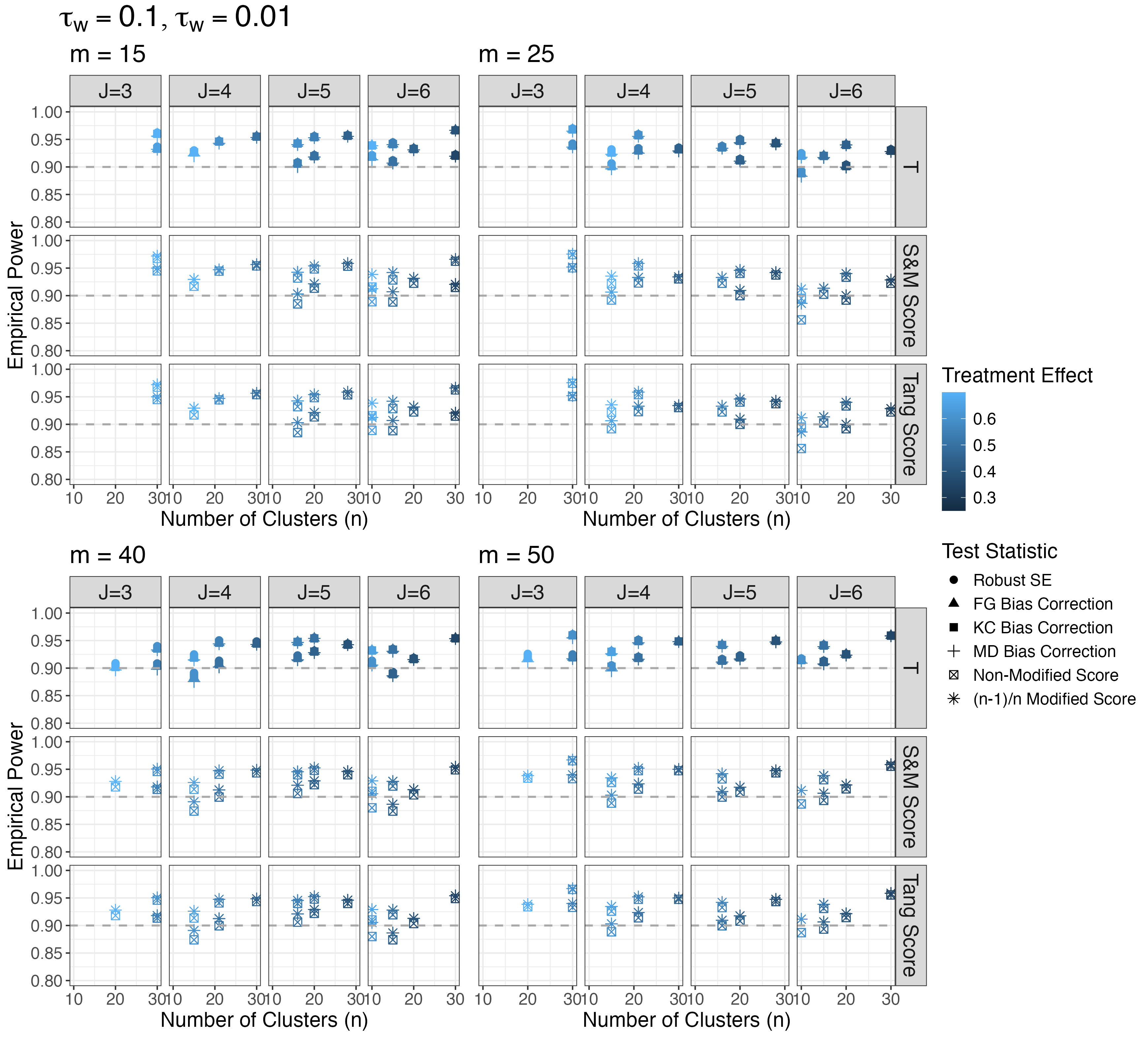}%{figures/empPower-032723.png}
    \caption{\label{fig:empPower105}Empirical power of hypothesis testing paradigms when within-period Kendall's $\tau_w=0.1$ and between-period Kendall's $\tau_b=0.01$, given $n$ clusters of cluster-period size $m$ are transitioned onto intervention over $J$ periods (columns) under a given treatment effect magnitude (color scale; lighter colors represent larger magnitude). The top row displays empirical power results for Wald $t$-tests using a robust sandwich variance (Robust SE) as well as \citep{fay_small-sample_2001} (FG), \citep{kauermann_note_2001} (KC), and \citep{mancl_covariance_2001} (MD) finite-sample adjusted variances. The bottom row displays empirical power results for robust (Non-Modified Score) and modified robust score tests ($(n-1)/n$ Modified Score). The gray dotted line represents 90\% power for reference.}
\end{figure}

\begin{figure}
    \centering
    \includegraphics[width=\textwidth]{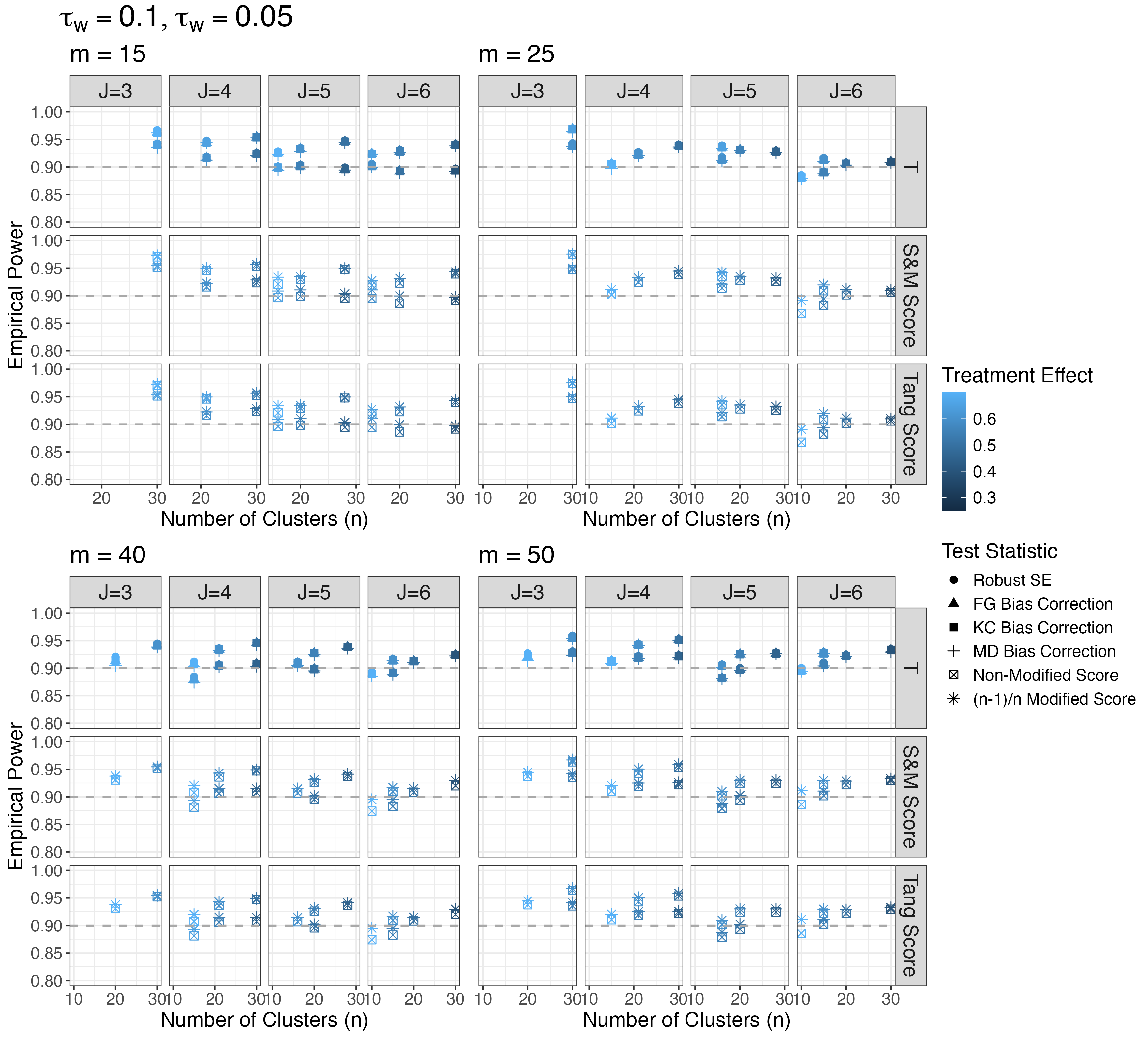}%{figures/empPower-032723.png}
    \caption{\label{fig:empPower101}Empirical power of hypothesis testing paradigms when within-period Kendall's $\tau_w=0.1$ and between-period Kendall's $\tau_b=0.05$, given $n$ clusters of cluster-period size $m$ are transitioned onto intervention over $J$ periods (columns) under a given treatment effect magnitude (color scale; lighter colors represent larger magnitude). The top row displays empirical power results for Wald $t$-tests using a robust sandwich variance (Robust SE) as well as \citep{fay_small-sample_2001} (FG), \citep{kauermann_note_2001} (KC), and \citep{mancl_covariance_2001} (MD) finite-sample adjusted variances. The bottom row displays empirical power results for robust (Non-Modified Score) and modified robust score tests ($(n-1)/n$ Modified Score). The gray dotted line represents 90\% power for reference.}
\end{figure}

\begin{figure}
    \centering
    \includegraphics[width=\textwidth]{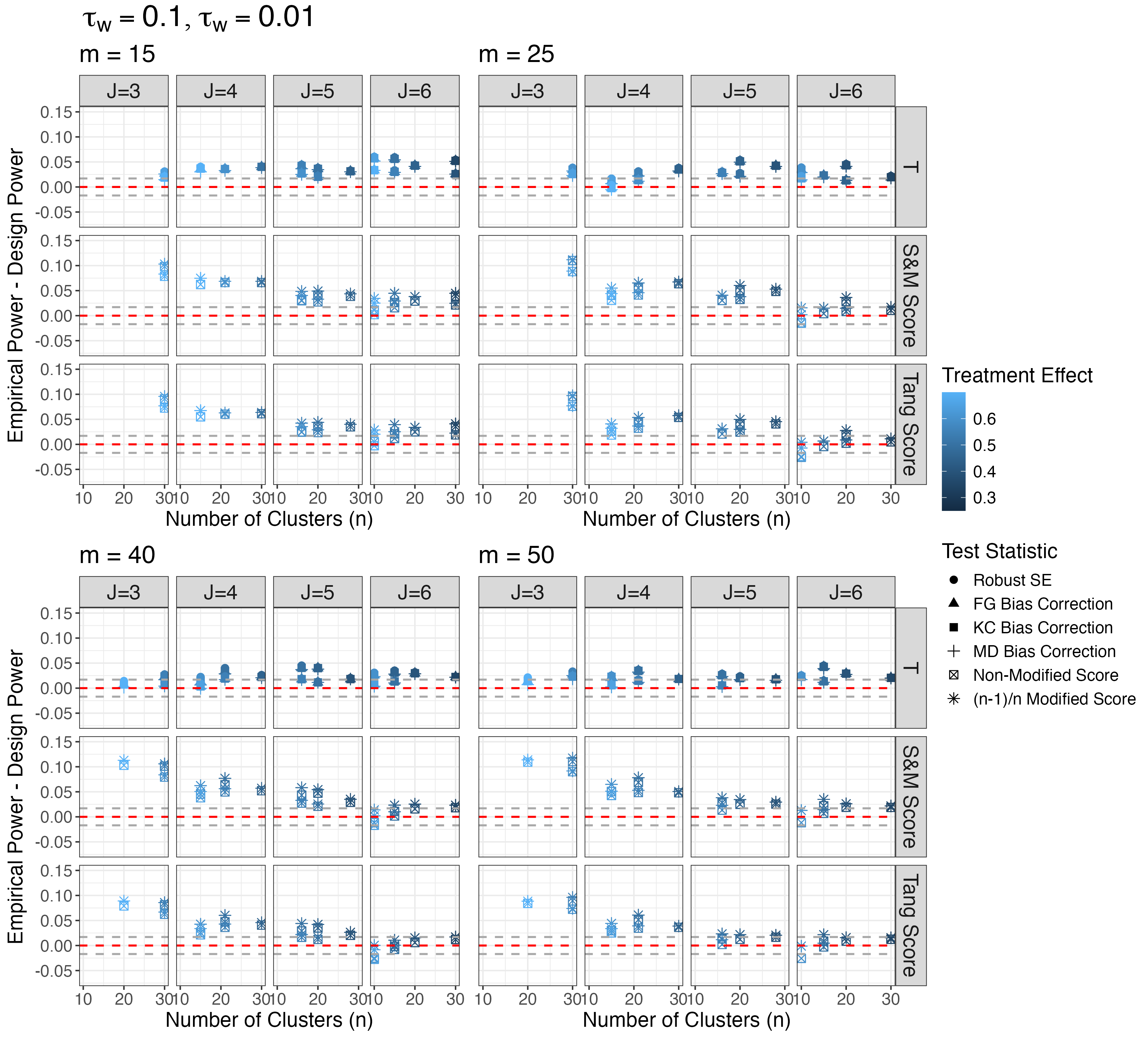}
    \caption{\label{fig:powerDiff101}Difference between empirical and predicted power of hypothesis testing paradigms when within-period Kendall's $\tau_w=0.1$ and between-period Kendall's $\tau_b=0.01$, given $n$ clusters of cluster-period size $m$ are transitioned onto intervention over $J$ periods (columns) under a given treatment effect magnitude (color scale; lighter colors represent larger magnitude). The top row displays difference in power for Wald $t$-tests using a robust sandwich variance (Robust SE) as well as \citep{fay_small-sample_2001} (FG), \citep{kauermann_note_2001} (KC), and \citep{mancl_covariance_2001} (MD) finite-sample adjusted variances. The middle and bottom rows displays difference in power for robust (Non-Modified Score) and modified robust score tests ($(n-1)/n$ Modified Score) when power is predicted using the \citep{self_powersample_1988} methods (middle row) and the \citep{tang_improved_2021} methods (bottom row). The red dotted line represents a difference of $0$ and the gray dotted lines represent simulation 95\% confidence intervals.}
\end{figure}

\begin{figure}
    \centering
    \includegraphics[width=\textwidth]{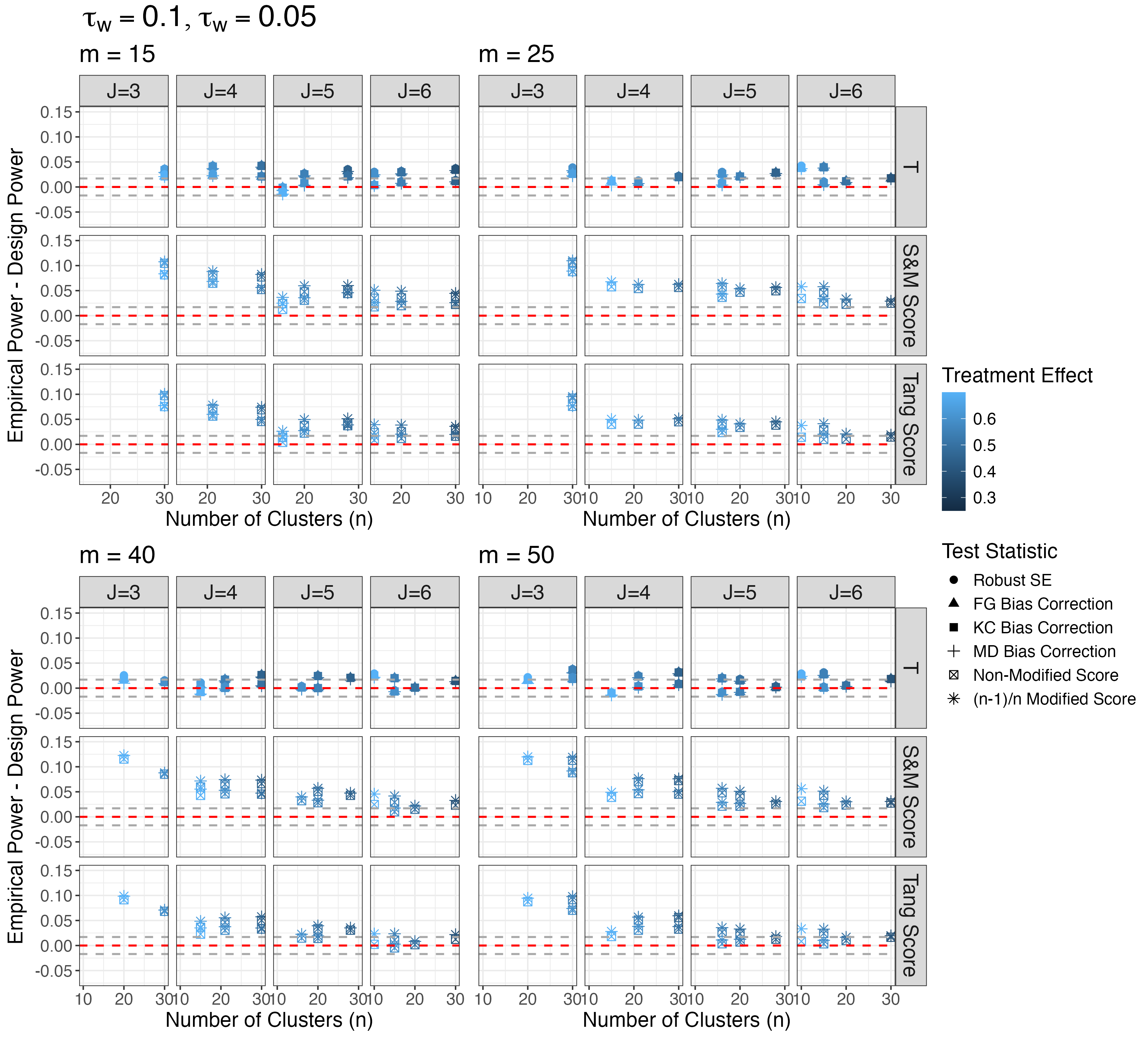}
    \caption{\label{fig:powerDiff105}Difference between empirical and predicted power of hypothesis testing paradigms when within-period Kendall's $\tau_w=0.1$ and between-period Kendall's $\tau_b=0.05$, given $n$ clusters of cluster-period size $m$ are transitioned onto intervention over $J$ periods (columns) under a given treatment effect magnitude (color scale; lighter colors represent larger magnitude). The top row displays difference in power for Wald $t$-tests using a robust sandwich variance (Robust SE) as well as \citep{fay_small-sample_2001} (FG), \citep{kauermann_note_2001} (KC), and \citep{mancl_covariance_2001} (MD) finite-sample adjusted variances. The middle and bottom rows displays difference in power for robust (Non-Modified Score) and modified robust score tests ($(n-1)/n$ Modified Score) when power is predicted using the \citep{self_powersample_1988} methods (middle row) and the \citep{tang_improved_2021} methods (bottom row). The red dotted line represents a difference of $0$ and the gray dotted lines represent simulation 95\% confidence intervals.}
\end{figure}

%% moved this back to main text %%
% \begin{figure}
%     \centering
%     \includegraphics[width=\textwidth]{figures/mitchell-power-sensitivity-tau-02-green-increasing-b18.png}%{figures/mitchell-power-sensitivity-tau-02.png}
%     \caption{\label{fig:sensitivity} Contour plots of predicted power trends across within-period Kendall's tau ($\tau_w$) and the ratio of between- and within-period Kendall's tau ($\tau_b/\tau_w$) within our application study of the CATH TAG trial, assuming a baseline hazard that increases by 5\% at each subsequent time period. The top row represents trends when power is predicted using the Wald $t$-test formula, the middle row when using the \citep{self_powersample_1988} robust score test formula, and the bottom row when using the \citep{tang_improved_2021} robust score test formula. Darker colors correspond to greater predicted power.}
% \end{figure}

\begin{table}\label{tab:simsettings2}
    \caption{\label{tab:simsettings} Simulation scenarios considered in Section 4. Checkmarks ($\checkmark$) indicate that simulations were run under a particular effect size value $\beta$, number of clusters $n$, cluster-period size $m$, and number of time periods $J$.}
    \vspace{0.25cm}
    \centering
    \begin{tabular}{l|llll|llll|llll|llll}
    \hline
    \multicolumn{17}{c}{$\boldsymbol{J=3}$}\\
    \hline
    $n$&\multicolumn{4}{c|}{}&\multicolumn{4}{c}{$14$} & \multicolumn{4}{c}{20} & \multicolumn{4}{c}{30}\\
    \hline
    $m$& \multicolumn{4}{c|}{}& $15$ &$25$ & $40$ & $50$ & $15$ & $25$ & $40$ & $50$& $15$ & $25$ & $40$ & $50$\\
    \hline
    $\beta=0.7$ &\multicolumn{4}{c|}{}& & $\checkmark$ & $\checkmark$ &$\checkmark$ &&&& &&&&\\
    $0.65$ & \multicolumn{4}{c|}{}&& & $\checkmark$ &$\checkmark$ & $\checkmark$ & $\checkmark$ &&& &&&\\
    $0.6$ & \multicolumn{4}{c|}{}&&& &&& $\checkmark$ &  $\checkmark$&$\checkmark$ & &&&\\
    $0.55$ &\multicolumn{4}{c|}{}&&& &&& &  $\checkmark$& $\checkmark$ & $\checkmark$ &&&\\
    $0.5$ & \multicolumn{4}{c|}{}&&& &&& &&& & $\checkmark$& &\\
    $0.45$ & \multicolumn{4}{c|}{}&&& &&& &&& & &$\checkmark$ &$\checkmark$\\
    $0.4$&\multicolumn{4}{c|}{}&\multicolumn{4}{c|}{}&\multicolumn{4}{c|}{}&\\
    $0.35$&\multicolumn{4}{c|}{}&\multicolumn{4}{c|}{}&\multicolumn{4}{c|}{}&\\
    $0.3$&\multicolumn{4}{c|}{}&\multicolumn{4}{c|}{}&\multicolumn{4}{c|}{}&\\
    $0.25$ &\multicolumn{4}{c|}{}&\multicolumn{4}{c|}{}&\multicolumn{4}{c|}{}&\\
    \hline
    \hline
    \multicolumn{17}{c}{$\boldsymbol{J=4}$}\\
    \hline
    $n$&\multicolumn{4}{c}{$9$}&\multicolumn{4}{c}{$15$} & \multicolumn{4}{c}{21} & \multicolumn{4}{c}{30}\\
    \hline
    $m$& $15$ &$25$ & $40$ & $50$& $15$ &$25$ & $40$ & $50$ & $15$ & $25$ & $40$ & $50$& $15$ & $25$ & $40$ & $50$\\
    \hline
    $\beta=0.7$ & & $\checkmark$ & $\checkmark$ & $\checkmark$ & \multicolumn{4}{c|}{}& \multicolumn{4}{c|}{}& \\
    $0.65$ & && $\checkmark$ & $\checkmark$ & $\checkmark$ &&&& \multicolumn{4}{c|}{}& \\
    $0.6$ & \multicolumn{4}{c|}{}&$\checkmark$  &&& &\multicolumn{4}{c|}{}\\
    $0.55$ &\multicolumn{4}{c|}{}& &$\checkmark$  && &\multicolumn{4}{c|}{}& \\
    $0.5$ & \multicolumn{4}{c|}{}&&&$\checkmark$&$\checkmark$ & $\checkmark$  &&&&\\
    $0.45$ & \multicolumn{4}{c|}{}&\multicolumn{4}{c|}{}& &$\checkmark$&$\checkmark$& & $\checkmark$\\
    $0.4$ & \multicolumn{4}{c|}{}&\multicolumn{4}{c|}{}&&&$\checkmark$& $\checkmark$& $\checkmark$ & $\checkmark$\\
    $0.35$ &\multicolumn{4}{c|}{}&\multicolumn{4}{c|}{}&\multicolumn{4}{c|}{}&  & &$\checkmark$&$\checkmark$\\
    $0.3$ &\multicolumn{4}{c|}{}&\multicolumn{4}{c|}{}&\multicolumn{4}{c|}{}&\\
    $0.25$ &\multicolumn{4}{c|}{}&\multicolumn{4}{c|}{}&\multicolumn{4}{c|}{}&\\
    \hline
    \hline
    \multicolumn{17}{c}{$\boldsymbol{J=5}$}\\
    \hline
    $n$&\multicolumn{4}{c}{$8$}&\multicolumn{4}{c}{$16$} & \multicolumn{4}{c}{20} & \multicolumn{4}{c}{28}\\
    \hline
    $m$& $15$ &$25$ & $40$ & $50$& $15$ &$25$ & $40$ & $50$ & $15$ & $25$ & $40$ & $50$& $15$ & $25$ & $40$ & $50$\\
    \hline
    $\beta=0.7$ && $\checkmark$ &&& \multicolumn{4}{c|}{}& \multicolumn{4}{c|}{}& \\
    $0.65$ & &$\checkmark$& $\checkmark$ & $\checkmark$ & \multicolumn{4}{c|}{} & \multicolumn{4}{c|}{}& \\
    $0.6$ & & &$\checkmark$  & $\checkmark$ &\multicolumn{4}{c|}{} & \multicolumn{4}{c|}{}\\
    $0.55$ &\multicolumn{4}{c|}{}& \multicolumn{4}{c|}{}&\multicolumn{4}{c|}{}& \\
    $0.5$ & \multicolumn{4}{c|}{}& $\checkmark$ &&&&\multicolumn{4}{c|}{}\\
    $0.45$ & \multicolumn{4}{c|}{}&&$\checkmark$ & $\checkmark$&&$\checkmark$ & &&&\\
    $0.4$ & \multicolumn{4}{c|}{}& &&$\checkmark$& $\checkmark$& & $\checkmark$ &$\checkmark$&&$\checkmark$\\
    $0.35$ &\multicolumn{4}{c|}{}&\multicolumn{4}{c|}{}& && $\checkmark$ & $\checkmark$& $\checkmark$&$\checkmark$\\
    $0.3$ &\multicolumn{4}{c|}{}&\multicolumn{4}{c|}{}&\multicolumn{4}{c|}{}& && $\checkmark$&$\checkmark$\\
    $0.25$ &\multicolumn{4}{c|}{}&\multicolumn{4}{c|}{}&\multicolumn{4}{c|}{}&\\
        \hline
    \hline
    \multicolumn{17}{c}{$\boldsymbol{J=6}$}\\
    \hline
    $n$&\multicolumn{4}{c}{$10$}&\multicolumn{4}{c}{$15$} & \multicolumn{4}{c}{20} & \multicolumn{4}{c}{30}\\
    \hline
    $m$& $15$ &$25$ & $40$ & $50$& $15$ &$25$ & $40$ & $50$ & $15$ & $25$ & $40$ & $50$& $15$ & $25$ & $40$ & $50$\\
    \hline
    $\beta=0.7$ &\multicolumn{4}{c|}{}&\multicolumn{4}{c|}{}&\multicolumn{4}{c|}{}&\\
    $0.65$ &\multicolumn{4}{c|}{}&\multicolumn{4}{c|}{}&\multicolumn{4}{c|}{}&\\
    $0.6$ & $\checkmark$ &&& &\multicolumn{4}{c|}{}&\multicolumn{4}{c|}{}&\\
    $0.55$ & $\checkmark$ & $\checkmark$ &&& \multicolumn{4}{c|}{}& \multicolumn{4}{c|}{}&\\
    $0.5$ & &$\checkmark$ & $\checkmark$ & $\checkmark$ & \multicolumn{4}{c|}{}& \multicolumn{4}{c|}{}&\\
    $0.45$ & &&& $\checkmark$ & $\checkmark$ & $\checkmark$ & &&\multicolumn{4}{c|}{}&\\
    $0.4$ & \multicolumn{4}{c|}{}& & $\checkmark$ & $\checkmark$ & $\checkmark$ & $\checkmark$ & &&&\\
    $0.35$ & \multicolumn{4}{c|}{}&\multicolumn{4}{c|}{}& & $\checkmark$ & $\checkmark$ & $\checkmark$ & \\
    $0.3$ &\multicolumn{4}{c|}{}&\multicolumn{4}{c|}{}&\multicolumn{4}{c|}{}& $\checkmark$ & $\checkmark$&&\\
    $0.25$ &\multicolumn{4}{c|}{}&\multicolumn{4}{c|}{}&\multicolumn{4}{c|}{}&&& $\checkmark$ & $\checkmark$
    \end{tabular}
    
\end{table}

\clearpage 
\addcontentsline{toc}{section}{Web Appendix H: Tutorial of Shiny Web Application}
\section*{Web Appendix H: Tutorial of Shiny Web Application}\label{H}
\renewcommand\thefigure{H.\arabic{figure}}
\setcounter{figure}{0}

We have created an online R Shiny application that allows users to input study design parameters to estimate the power such a SW-CRT would have or the number of clusters required to achieve a particular power threshold using the methods developed in this article. The application can be accessed at: \url{https://mary-ryan.shinyapps.io/survival-SWD-app/}; source code can be found at: \url{https://github.com/maryryan/survivalSWCRT}.\\

\noindent The application is comprised of two main panels: an ``input'' panel located along the left side of the application where users can provide design parameters for the SW-CRT they wish to calculate power or sample size for, and a ``display'' panel occupying the center of the application where the results of calculations will be displayed. The display panel also features three tabs: the default ``results'' tab that displays results of the power and sample size calculations, the ``design matrix'' tab which creates a trial schematic to visualize the treatment sequence timing, and a ``references and resources'' tab that provides contact information for the application authors and directions to additional resources such as the code repository.\\

\begin{figure}
    \centering
    \includegraphics[width=\textwidth]{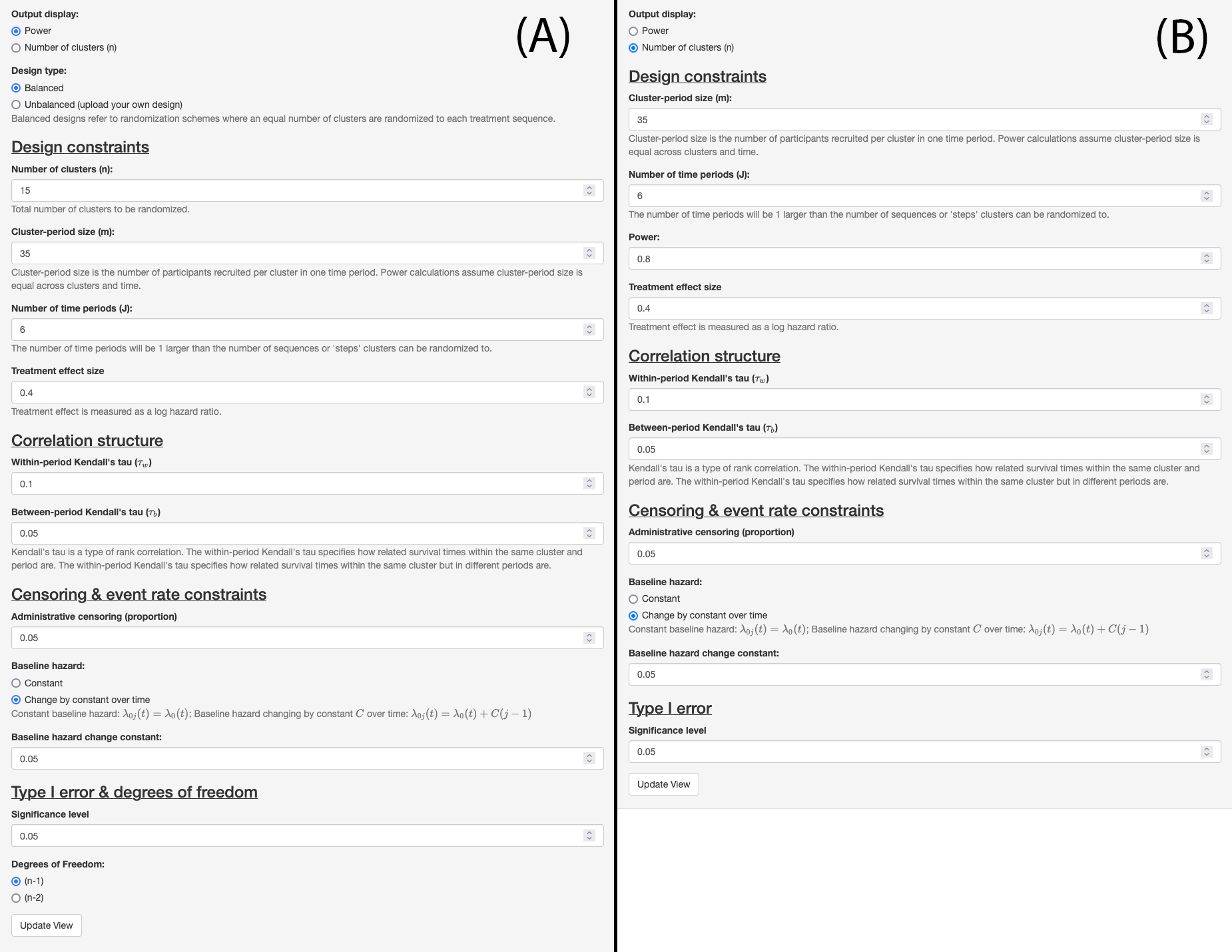}%{figures/mitchell-power-sensitivity-tau-02.png}
    \caption{\label{fig:input-panel} Screenshots of Shiny application input panel when the ``Power'' display option is chosen (A), and when the ``Number of clusters ($n$)'' option is chosen (B). Inputs for panel (A) are specified as: Output display - ``Power''; Design type - ``Balanced''; Number of clusters ($n$) - $15$; Cluster-period size ($m$) -  $35$; Number of time periods ($J$) - $6$; Power - $0.8$; Treatment effect size - $0.4$; Within-period Kendall's tau ($\tau_w$) - $0.1$; Between-period Kendall's tau ($\tau_b$)- $0.05$; Administrative censoring (proportion) - $0.05$; Baseline hazard  ``Change by constant over time''; Baseline hazard change constant - $0.05$; Significance level - $0.05$; Degrees of freedom - $(n-1)$. Inputs for panel (B): Output display - ``Number of clusters ($n$)''; Cluster-period size ($m$) -  $35$; Number of time periods ($J$) - $6$; Power - $0.8$; Treatment effect size - $0.4$; Within-period Kendall's tau ($\tau_w$) - $0.1$; Between-period Kendall's tau ($\tau_b$)- $0.05$; Administrative censoring (proportion) - $0.05$; Baseline hazard  ``Change by constant over time''; Baseline hazard change constant - $0.05$; Significance level - $0.05$.}
\end{figure}

\noindent Within the input panel users are asked for a variety of study design information to populate the power and sample size calculations on the application back-end. The ``output display'' option determines what design parameters the user is prompted to supply. If the ``power'' display is chosen, users are prompted for the design type (balanced or unbalanced/upload your own), total number of clusters ($n$) to randomize, cluster-period size ($m$), and number of time periods $J$. If the ``number of clusters ($n$)'' is chosen, users are only asked for cluster-period size ($m$), and number of time periods $J$, as well as the target power for the study. After these options are provided, users must input the anticipated treatment effect sizes on the log hazard ratio scale, measures of within- and between-period correlation (as measured by Kendall's tau), and the proportion of observation times that will be administratively censoring. Next, because we consider a Cox model with baseline hazards stratified by study period, the ``baseline hazard'' option asks users to consider whether the baseline hazard will remain constant across all trial periods (``constant''), or whether it will additively increase/decrease by some constant $C$ as the study progresses from one period to the next (``change by constant over time''). If ``change by constant over time'' is chosen, users will then need to specify the value of the constant in the ``baseline hazard change constant'' option. Finally, users are asked to input their significance level or type I error rate. If users chose to calculate power, they will also be asked how many degrees of freedom they would like to use for the $t$-distribution in their power calculation ($(n-1)$ or $(n-2)$); if they chose to estimate the number of clusters needed to achieve a particular power, a standard normal distribution will be used and users will not be asked to specify degrees of freedom. Once users have input all the requested information, they can launch the calculations by pressing the ``update view'' button at the bottom of the input panel. Examples of how the input panel is laid out are shown in Figure \ref{fig:input-panel}.\\

\noindent We will use the CATH TAG example from Section $5$ to demonstrate how to use the application. We are interested in estimating how many clusters would be necessary to achieve $80$\% power, so we will select ``number of clusters ($n$)'' under the output display option. We can then input the cluster-period size ($35$), the number of time periods ($6$), the power ($0.8$), and the targeted treatment effect size ($0.4$, as this needs to be input on the log hazard ratio scale; this is equal to a hazard ratio of $1.5$). Next, we need to supply information about the dependence between survival times in the same and different periods for individuals belonging to the same cluster; a variety of Kendall's tau combinations were explored in Section $5$ but for demonstration we will use the first set -- a within-period Kendall's tau of $0.1$ and a constant between-period Kendall's tau of $0.05$. Next we can input the anticipated proportion of observations that will be administratively censored, which will be $0.05$ since we specified a $5$\% administrative censoring rate in Section $5$. Concerning the form of the baseline hazard, we first considered one that increased at a minimal rate of $5$\% with each period, so we will select ``change by constant over time`` in the baseline hazard option and then input $0.05$ for the baseline hazard change constant option. Finally we specify a $5$\% type I error rate by inputting 0.05 under the ``significance level'' option and, since we are estimating the number of clusters needed, do not need to specify degrees of freedom.\\

\begin{figure}
    \centering
    \includegraphics[width=\textwidth]{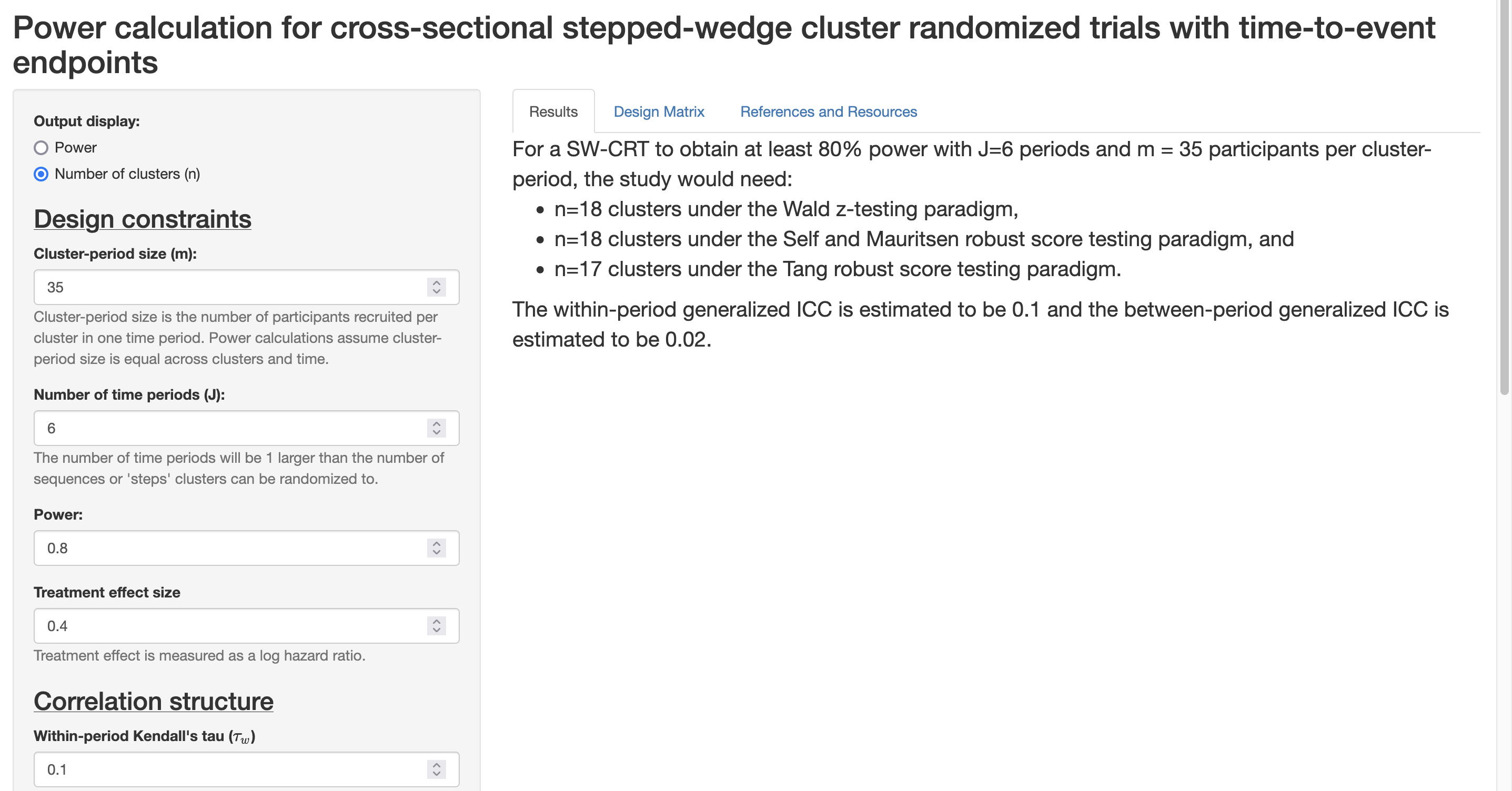}%{figures/mitchell-power-sensitivity-tau-02.png}
    \caption{\label{fig:main-display} Screenshot of Shiny application on the ``Results'' tab after design parameters have been input. Input selections are specified as: Output display - ``Number of clusters ($n$)''; Cluster-period size ($m$) -  $35$; Number of time periods ($J$) - $6$; Power - $0.8$; Treatment effect size - $0.4$; Within-period Kendall's tau ($\tau_w$) - $0.1$; Between-period Kendall's tau ($\tau_b$)- $0.05$; Administrative censoring (proportion) - $0.05$; Baseline hazard  ``Change by constant over time''; Baseline hazard change constant - $0.05$; Significance level - $0.05$.}
\end{figure}

\noindent Pressing the ``update view button'', a green ``loading'' box will display while the calculation is being run. Once the calculations are complete, the message box will disappear and text will populate the main display panel. For the inputs we provided above, the text will read: ``For a SW-CRT to obtain at least $80$\% power with $J=6$ periods and $m = 35$ participants per cluster-period, the study would need: $n=18$ clusters under the Wald $z$-testing paradigm, $n=18$ clusters under the Self and Mauritsen robust score testing paradigm, and $n=17$ clusters under the Tang robust score testing paradigm. The within-period generalized ICC is estimated to be $0.1$ and the between-period generalized ICC is estimated to be $0.02$'' (Figure \ref{fig:main-display}).\\

\begin{figure}
    \centering
    \includegraphics[width=\textwidth]{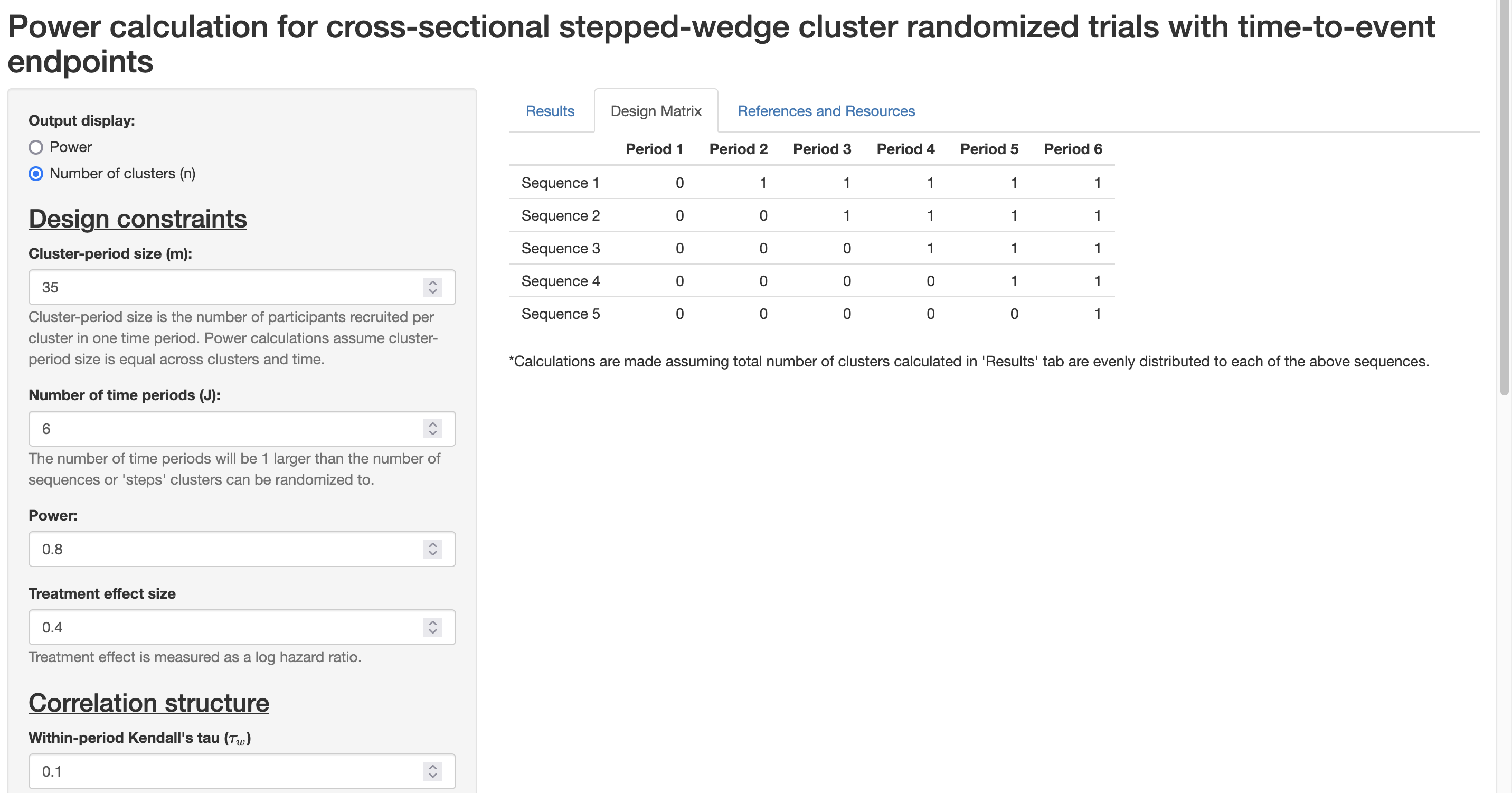}%{figures/mitchell-power-sensitivity-tau-02.png}
    \caption{\label{fig:design-display} Screenshot of Shiny application on the ``Design Matrix'' tab after design parameters have been input. Input selections are specified as: Output display - ``Number of clusters ($n$)''; Cluster-period size ($m$) -  $35$; Number of time periods ($J$) - $6$; Power - $0.8$; Treatment effect size - $0.4$; Within-period Kendall's tau ($\tau_w$) - $0.1$; Between-period Kendall's tau ($\tau_b$)- $0.05$; Administrative censoring (proportion) - $0.05$; Baseline hazard  ``Change by constant over time''; Baseline hazard change constant - $0.05$; Significance level - $0.05$.}
\end{figure}

\noindent If we wanted to see visual representation of this design, we could go to the ``design matrix'' tab. The main display window then changes to show a $5\times 6$ design schematic with $0$s representing the control condition and $1s$ representing the treatment condition (Figure \ref{fig:design-display}). Instead of illustrating a row for each cluster, this display only illustrates the timing of the $5$ treatment sequences; a note appears below that reads: ``*Calculations are made assuming total number of clusters calculated in `Results' tab are evenly distributed to each of the above sequences.'' This is meant to account for the fact that, when back-solving the power equation for number of clusters, you may end up with a number of clusters that is not evenly divisible by the number of treatment sequences. If we wanted to investigate the trial's power when the number of clusters is unevenly distributed among the treatment sequences, we could use the ``Unbalanced (upload your own design)'' option; in general, greatest power will be obtained if more clusters are assigned to ``outer'' sequences (first/last) rather than ``inner''/middle sequences. If the design matrix tab is selected when output display is set to ``power'', this matrix will illustrate treatment timing on the cluster level since the user will have either chosen a balanced design (such that the number of clusters is evenly distributed among the treatment sequences) or have uploaded their own design schematic from which the tab may pull from.\\

\noindent In addition, you will observe differences in the Wald power under the ``power'' and ``number of clusters ($n$)'' output display options; there are two causes for this. First would be due to differences in cluster allocation (equal, fractional allocation versus unequal, integer allocation); the second may be attributed to the use of the Normal distribution when calculating number of clusters for a given power, versus a $t$-distribution when calculating power for a given sample size. In trials will small numbers of clusters, power estimation via a $t$-distribution with $(n-2)$ degrees of freedom is recommended. In the case where sample size is unknown, we suggest an iterative workflow. First, estimate the number of clusters needed given a fixed sample size using the ``number of clusters ($n$)'' option. Next, using the ``power'' option, input the same information as previously, as well as the estimated number of clusters obtained in the last step. If the number of clusters indicates an unbalanced design,  use the ``Unbalanced (upload your own design)'' option to specify which treatment sequences will receive more/fewer clusters; you may also increase the number of clusters to acheive a balanced design. If the power obtained under the Wald $t$-testing paradigm is below your threshold, repeat this step by increasing the number of clusters by $1$ until the estimated power threshold is reached. In the CATH TAG setting, $18$ clusters does not divide evenly across $5$ treatment sequences, so we must use an unbalanced design. If we place $4$ clusters on sequences $1$, $3$, and $5$, and $3$ on all others, we predict $76$\% power under the Wald $t$-test, $82$\% power under S\&M, and $83$\% power under Tang. If we put $4$ clusters on sequences $2 - 4$ and $3$ clusters on sequences $1$ and $5$, we predict $75$\% power under Wald, $81$\% power under S\&M, and $82$\% power under Tang. Increasing the number of clusters to $20$ will given us a balanced design, and subsequently will put us above our $80$\% power threshold under the Wald $t$-testing paradigm: we estimate $80.8$\% power under Wald, $85.5$\% power using the S\&M score method, and $86.3$\% power using the Tang score method.

\end{document}

\maketitle
\vspace{-1cm}
\tableofcontents
\doublespacing
\setstackgap{L}{.6\baselineskip}
\input{A}
\input{B}
\input{C}
\input{D}
\newpage
\input{E}
\input{F}
\input{G}
\input{H}
%\input{I}
% \input{M_WebFig}
% \input{N_WebTab}

%%%%%%%%%%%%%%%%%%%%%%%%%%%%%%%%%%%%%%%%%%%%%%%%%%%%%%%%%%%%%%
% Bibliography
\newpage
\singlespace
%\linespread{1.3} %1.3 = 1.5 spaced, 1.6 = double spaced
%\bibliographystyle{apalike}
\bibliographystyle{biom}
\bibliography{ref_supl}